\documentclass[a4paper,11pt]{article}
\usepackage{jheppub} % for details on the use of the package, please see the JINST-author-manual
\usepackage[table]{xcolor}
\usepackage{tabularx}
\usepackage{multirow}
\usepackage{graphicx}
\graphicspath{{figs/}}
\usepackage{amsmath,amssymb}
\usepackage[capitalize]{cleveref}
\usepackage{bbold}
\usepackage{subcaption}
\makeatletter
\def\maketag@@@#1{\hbox{\m@th\normalfont\normalsize#1}}
\makeatother

\DeclareMathOperator{\Tr}{Tr}

\newcolumntype{C}{>{\Centering\arraybackslash}X} % centered "X" column

\newcommand{\Op}{\mathcal{O}}
\newcommand{\bO}{\mathrm{O}}
\newcommand{\dd}{\mathrm{d}}
\newcommand{\GeV}{\mathrm{GeV}}
\newcommand{\cC}{\mathcal{C}}

\newcommand{\fourpi}{\left(4\pi\right)}

\newcommand{\Tab}[1]{\cref{#1}}

\newcommand{\ifl}{\mathcal{I}}
\newcommand{\itfl}{\mathcal{I}_3}
\newcommand{\yfl}{\mathcal{Y}}

\newcommand{\db}{\mathbf{d}}

\usepackage{tikz}
\usepackage[compat=1.1.0]{tikz-feynhand}
\usepackage{ragged2e}

\newcommand{\currsep}{0.2}

\newcommand{\threepointsep}{0.5}
\newcommand{\currentmarker}[1]{
  \filldraw[fill=white,line width=1pt](#1)circle(.12);
  \draw[line width=.6pt] (#1) +(-135:.12) -- +(45:.12) +(-45:.12) -- +(135:.12);
}

\newcommand{\angleone}{45}
\newcommand{\angletwo}{-135}
\newcommand{\anglethree}{-45}
\tikzset{iso0style/.style={circle,fill,inner sep=2pt}}
\tikzset{isohalfstyle/.style={circle,draw=black,fill=white,inner sep=2pt}}
\tikzset{iso1style/.style={rectangle,fill,inner sep=2pt}}
\tikzset{iso3halfstyle/.style={rectangle,draw=black,fill=white,inner sep=2pt}}
\tikzset{iso2style/.style={star,star points=5,fill,inner sep=2pt}}
\tikzset{labelstyle/.style={text=red,font=\footnotesize,label distance=-5pt}}

\DeclareRobustCommand{\isozerokey}{\begin{tikzpicture}[baseline=-0.65ex] \node[iso0style] (A) at (0,0) {}; \end{tikzpicture}}
\DeclareRobustCommand{\isohalfkey}{\begin{tikzpicture}[baseline=-0.65ex] \node[isohalfstyle] (A) at (0,0) {}; \end{tikzpicture}}
\DeclareRobustCommand{\isoonekey}{\begin{tikzpicture}[baseline=-0.65ex] \node[iso1style] (A) at (0,0) {}; \end{tikzpicture}}
\DeclareRobustCommand{\isothreehalfkey}{\begin{tikzpicture}[baseline=-0.65ex] \node[iso3halfstyle] (A) at (0,0) {}; \end{tikzpicture}}
\DeclareRobustCommand{\isotwokey}{\begin{tikzpicture}[baseline=-0.65ex] \node[iso2style] (A) at (0,0) {}; \end{tikzpicture}}

\makeatletter\g@addto@macro\bfseries\boldmath\makeatother

\title{Diagonalising the LEFT}

\author[]{Sophie Renner,}
\emailAdd{sophie.renner@glasgow.ac.uk}

\author[]{Benjamin Smith,}
\emailAdd{b.smith.4@research.gla.ac.uk}

\author[]{and Dave Sutherland}
\emailAdd{david.w.sutherland@glasgow.ac.uk}

\affiliation[]{School of Physics and Astronomy, University of Glasgow, Glasgow G12 8QQ, United Kingdom}

\abstract{We organise the four-fermion vector current interactions below the weak scale --- i.e., in the low energy effective field theory (LEFT) --- into irreps of definite parity and $SU(N)$ flavour symmetry. Their coefficients are thus arranged into small subsets with distinct phenomenology, which are significantly smaller than traditional groupings of operators by individual fermion number. 
As these small subsets only mix among themselves, we show that the renormalisation group evolution is soluble semi-analytically, and examine the resulting eigenvalues and eigenvectors of the one- and two-loop running. This offers phenomenological insights, for example into the radiative stability of lepton flavour non-universality. We use these to study model-independent implications for $b\to s \tau \tau$ decays, as well as setting indirect bounds on flavour changing four-quark interactions.
}

\begin{document}
\maketitle
\flushbottom

\section{Introduction\label{sec:intro}}

Symmetries are a powerful organisational and calculational tool in particle physics. This is especially so within effective field theories where we are unsure of aspects of the UV physics. The archetype is the $SU(3)_V$ symmetry among the light $uds$ quarks. Knowledge of symmetry breaking spurions (such as the quark masses) combined with group theory machinery makes many useful predictions for composite operators built from the quarks, and for the hadrons they interpolate, despite the complexities of QCD dynamics.

The situation at energies significantly above the QCD confinement scale is strikingly similar. There exist non-Abelian generational symmetries among the up-type quarks $uc$, down-type quarks $dsb$, and leptons $e\mu\tau$, broken by the fermion masses, and whatever UV physics generates them. However, we rarely use the non-Abelian symmetries to their fullest extent. (Perhaps because one compelling weakly-coupled UV completion, the Standard Model (SM), is more amenable to brute force calculation.) This is despite the symmetries being as powerful a calculational tool as ever, and the only sensible means to organise the `zoo' of thousands of composite fermion operators that can be generated by SM, or unknown BSM, physics. In this paper, we explore the practicalities and benefits of using the non-Abelian symmetries.

We study the four-fermion vector current operators in the LEFT --- the EFT of SM particles lighter than the $W$ --- and provide their explicit decomposition into operators with definite flavour and parity charges in \cref{sec:decomp}. This partitions the degrees of freedom in the Wilson coefficients into small groups with distinct phenomenology, the largest of which contains $23$ real degrees of freedom. We stress that the resulting groups are considerably smaller than the more familiar partitioning obtained by considering just the Abelian charges of individual fermion number (for example, all operators mediating a single $b\to s$ transition). 

Calculations and predictions in one block can be related to many others by symmetry, and using the flavour symmetry to its fullest extent can effectively factorise flavour considerations out of kinematic, colour, and other pieces of a calculation within the LEFT, akin to the relations between strong matrix elements in chiral perturbation theory. This offers considerable simplifications. A case in point is the renormalisation group (RG) evolution in the LEFT, which we study using our decomposition in the rest of the paper. The RG is a key component of many predictions, and was first given at one loop for all the operators up to dimension 6 in \cite{Jenkins:2017dyc}, and which has recently appeared for large subsets of these operators at two loops in \cite{Aebischer:2025hsx,Naterop:2025cwg,Naterop:2025lzc} (see also \cite{Buras:1991jm,Buras:1992tc,Buras:1992zv,Ciuchini:1993vr,Ciuchini:1993ks,Chetyrkin:1996vx,Chetyrkin:1997gb,Buras:2000if,Gambino:2003zm,Bobeth:2003at,Gorbahn:2004my,Huber:2005ig,Aebischer:2024xnf}). The anomalous dimensions of operators arise from QED and QCD interactions, which are completely parity and flavour symmetric to all orders. Therefore, the RG only mixes operators within the small blocks we identify in \cref{sec:decomp} via a handful of distinct kinematic and colour factors, distributed predictably across the blocks by the flavour symmetry. We discuss this in \cref{sec:blockgamma}.

Having decomposed the running into small atomic blocks, it is relatively easy to construct semi-analytically the evolution matrix for the LEFT operators to per-mille accuracy, which we do explicitly for the operators that mediate $\Delta F_d = 1$ transitions in the LEFT in \cref{sec:82block}. We study some basis-independent properties of the RG flow in this sector via the eigenvectors and eigenvalues of this matrix, including where two-loop effects are important.

In \cref{sec:pheno} we consider the phenomenology of the $b \to s$ sector, making use of the eigenvectors and eigenvalues found previously, and the fact that new physics effects will generically contribute to more than one eigenvector of the running. In particular, the Wilson coefficients of four-fermion operators in $bsll$ decays ($l=e,\mu$) are comparatively well measured. Absent cancellation between the tree-level matching and RG contributions to these coefficients, any $b \to s$ process with the same symmetry charges unavoidably leaks into the $bsll$ operators at the per-mille level between the electroweak and $b$ mass scales. We use this behaviour to put indirect constraints on tauonic new physics, and four-fermion operators. 
These general arguments are enabled by the symmetry decomposition of the LEFT operators. We provide this decomposition, along with the one-loop beta functions for flavour and parity decomposed Wilson coefficients and a precomputed evolution matrix, in the ancillary files.

\section{Flavour and parity decomposition of the LEFT\label{sec:decomp}}

The field content of the LEFT consists of 2 up-type quarks ($u,c$), 3 down-type quarks ($d,s,b$), 3 charged leptons ($e,\mu,\tau$) and 3 left-handed neutrinos ($\nu_e,\nu_\mu,\nu_\tau$), as well as the gluon and the photon. The Lagrangian can be written schematically as the sum of the QED and QCD kinetic terms, the fermion mass terms, and a series of all higher dimensional operators, which are compatible with an $SU(3)_c \times U(1)_Q$ gauge symmetry:
\begin{equation}
\mathcal{L}_\text{LEFT} = \mathcal{L}_\text{gauge} + \mathcal{L}_\text{mass} + \sum_k c^{(D)}_{k}\Op^{(D)}_{k} \, , \label{eq:leftLag}
\end{equation}
where $c^{(D)}_{k}$ and $\Op^{(D)}_{k}$ are respectively Wilson coefficients and operators of dimension $D>4$. $c^{(D)}_{k}$ have an implicit suppression of at least $O\left(v^{4-D}\right)$. In this work, we ignore the QCD theta term.

The gauge part of the Lagrangian is parity and flavour symmetric. Up to $U(1)$ charges, the symmetry group is given by independent special unitary rotations among the different generations of left- and right-handed fields, together with a $\mathbb{Z}_2$ parity symmetry that exchanges the left- and right-handed charged fields, and which therefore interchanges the action of the corresponding left- and right-handed flavour groups. The symmetry group is formally 
\begin{align}
    &\left( SU(3)_{d_L} \times SU(3)_{d_R} \right) \, \times \, \left( SU(3)_{e_L} \times SU(3)_{e_R} \right) \notag \\[5pt]
    & \hspace{5ex}
      \times \left( SU(2)_{u_L} \times SU(2)_{u_R} \right) \, \times \, SU(3)_{\nu_L} \rtimes \mathbb{Z}_{2}  \, ,
    \label{eq:fullsymgroup}
\end{align}
where `$\rtimes$' denotes a semi-direct product, due to the exchange action of the $\mathbb{Z}_2$ symmetry.

The large symmetry group \cref{eq:fullsymgroup} is hierarchically broken by the mass terms. It is also broken by the higher dimension terms; whether these come from matching to the SM, or indeed to most phenomenologically viable models of BSM physics, the breaking pattern also tends to be hierarchical. It is therefore a powerful organisational tool to decompose the many Wilson coefficients in terms of their irreps under \cref{eq:fullsymgroup}. We stress that we do this without imposing any flavour assumptions on the Wilson coefficients themselves. The Wilson coefficients are effectively a fully general set of symmetry breaking spurions.

In this paper, we will restrict to studying the four-fermion vector current operators under the vectorial subgroup of \cref{eq:fullsymgroup}, which contains the bulk of the phenomenology in which we are interested. However, our treatment can be generalised. The reader not interested in the specifics of the decomposition can skip ahead to \cref{sec:flavour_parity_charges}.

\subsection{The four-fermion vector current operators}

The four-fermion vector current operators of the LEFT can be written as the product of two neutral  vector currents.\footnote{This excludes vectorial operators of the form $\overline{\nu}_L \gamma e_L \overline{d} \gamma u$ and $\overline{u}_L \gamma d_L \overline{d}_R \gamma u_R$, which only mix under RG among themselves due to their flavour charges.} A full list is given in the `San Diego' basis in \cite{Jenkins:2017jig}. In this section we outline how to decompose these operators into a basis with definite charges under the vectorial subgroup of \cref{eq:fullsymgroup}, namely
\begin{align}
    SU(3)_{d} \times SU(3)_{e} \times SU(2)_{u} \times \mathbb{Z}_{2}  \, .
    \label{eq:vecsymgroup}
\end{align}
In the vectorial subgroup both left- and right-handed species of $e$, $u$ and $d$ transform equivalently, and parity is an independent $\mathbb{Z}_2$ symmetry. Note that we ignore the flavour symmetry group of the neutrinos --- being uncharged under QCD and QED their contribution to operators' running is trivial, see \cref{sec:blockgamma}.
We choose to work with the subgroup \cref{eq:vecsymgroup} as its irreps are considerably simpler and more familiar than those of the full group. However, working with the full group would yield more relations between operators, we discuss this point in \cref{sec:fullflav}. 

We construct the operators starting with the uncontracted products of two vector currents 
\begin{equation}
    [L_\mu(f)]_\alpha^\beta [L^\mu(f^\prime)]_\gamma^\delta , \,
    [L_\mu(f)]_\alpha^\beta [R^\mu(f^\prime)]_\gamma^\delta , \, 
    [R_\mu(f)]_\alpha^\beta [L^\mu(f^\prime)]_\gamma^\delta , \,
    [R_\mu(f)]_\alpha^\beta [R^\mu(f^\prime)]_\gamma^\delta
    \label{eq:uncontractedOps}
\end{equation}
where
\begin{equation}
    \begin{aligned}
        \relax
        [L_\mu(f)]_\alpha^\beta =& \bar{f}_\alpha \gamma_\mu P_L f^\beta, \\
        [R_\mu(f)]_\alpha^\beta =& \bar{f}_\alpha \gamma_\mu P_R f^\beta, \\
    \end{aligned}
\label{eq:LR_notation}
\end{equation}
for a given $f, f^\prime \in \{u,d,e,\nu\}$. $\alpha, \beta, \ldots$ label fermion flavour and colour for quarks, and just flavour for leptons. We contract the colour indices in a later step; we first arrange \cref{eq:uncontractedOps} into parity eigenstates that, if $f=f^\prime$, also have a definite symmetry under exchange of the flavour/colour indices $\alpha, \beta, \ldots$. The manifest exchange symmetry will make it easier to identify the irreps under the $SU(N)$ flavour groups within operator's Wilson coefficients.

For operators composed entirely of electrically charged fermion fields ($u,d,e$), operators with definite parity charges can be constructed through
\begin{subequations}
\begin{align}
    [\Op_A^{+}]_{\alpha \gamma}^{\beta \delta} =& \frac{1}{\sqrt{2}} \left( [R_\mu(f)]_\alpha^\beta [R^\mu(f^\prime)]_\gamma^\delta + [L_\mu(f)]_\alpha^\beta [L^\mu(f^\prime)]_\gamma^\delta \right), \\
    [\Op_{B}^{+}]_{\alpha \gamma}^{\beta \delta} =& \frac{1}{\sqrt{2}} \left( [L_\mu(f)]_\alpha^\beta [R^\mu(f^\prime)]_\gamma^\delta + [R_\mu(f)]_\alpha^\beta [L^\mu(f^\prime)]_\gamma^\delta \right), \\
    [\Op_{A}^{-}]_{\alpha \gamma}^{\beta \delta} =& \frac{1}{\sqrt{2}} \left( [R_\mu(f)]_\alpha^\beta [R^\mu(f^\prime)]_\gamma^\delta - [L_\mu(f)]_\alpha^\beta [L^\mu(f^\prime)]_\gamma^\delta \right), \\
    [\Op_{B}^{-}]_{\alpha \gamma}^{\beta \delta} =& \frac{1}{\sqrt{2}} \left( [L_\mu(f)]_\alpha^\beta [R^\mu(f^\prime)]_\gamma^\delta - [R_\mu(f)]_\alpha^\beta [L^\mu(f^\prime)]_\gamma^\delta \right), 
\end{align}%
\label{eq:AandB_op_def}%
\end{subequations}%
where $\Op_{A}^{\pm}$ and $\Op_B^{\pm}$ have parity charges $\pm1$. For $\Op_{\nu f}$ type operators we define the parity operation as swapping the chirality of the charged fermions only. Combinations of operators with definite parity charges $\pm$ are formed through
\begin{equation}
    [\Op_{\nu}^{\pm}]_{\alpha \gamma}^{\beta \delta} = \frac{1}{\sqrt{2}} [L_\mu(\nu)]_\alpha^\beta \left([L^\mu(f)]_\gamma^\delta \pm [R^\mu(f)]_\gamma^\delta \right) \, .
\label{eq:nu_ops_parity_decomp}
\end{equation}

Of course, only operators which are colour singlets appear in \cref{eq:leftLag}. When either none or two of the fermions are coloured, there is only one way to make a colour singlet out of the $A$ and $B$ operators of \cref{eq:AandB_op_def}. When all four fermions are coloured, there are two distinct ways to make a colour singlet, which we choose to write as
\begin{subequations}
  \begin{align}
    [\Op^{\cdot,+}]^{q s}_{p r} = \delta^{(IK)}_{(JL)} [\Op^{\cdot}]_{\alpha \gamma}^{\beta \delta} \, , \\
    [\Op^{\cdot,-}]^{q s}_{p r} = \delta^{[IK]}_{[JL]} [\Op^{\cdot}]_{\alpha \gamma}^{\beta \delta} \, ,
  \end{align}
  \label{eq:op_colour_contractions}%
\end{subequations}
where the `$\cdot$' is a placeholder for the parity charge. In \cref{eq:op_colour_contractions} the Greek indices are split into a flavour and colour index ($\alpha=p,I$, $\beta=q,J$, and so on), and the colour indices are contracted with the structures
\begin{subequations} 
\begin{align}
  \delta^{(IK)}_{(JL)} \equiv \frac12\left(\delta^{I}_{J}\delta^{K}_{L}+\delta^{I}_{L}\delta^{K}_{J}\right) \, , \\
\label{eq:symmetric_colour_struc}
  \delta^{[IK]}_{[JL]} \equiv \frac12\left(\delta^{I}_{J}\delta^{K}_{L}-\delta^{I}_{L}\delta^{K}_{J}\right) \, .
\end{align}%
\end{subequations}
The colour-contracted operators \cref{eq:op_colour_contractions} can be expressed in terms of the more usual colour contractions
\begin{subequations}
  \begin{align}
      \Op^{\cdot,+} =&  \frac{4}{3}\Op^{\cdot,(1)} + \Op^{\cdot,(8)} \, ,\\
      \Op^{\cdot,-} =&  \frac{2}{3}\Op^{\cdot,(1)} - \Op^{\cdot,(8)} \, ,
  \end{align}
  \label{eq:colour_sym_ops}%
\end{subequations}
where in `$(1)$' each fermion bilinear is contracted to form a singlet, and in `$(8)$' each fermion bilinear is contracted with an $SU(3)$ generator $T^A$ to form an octet, normalisation $\Tr \left[ T^A T^B\right] = \frac12 \delta^{AB}$. However, we prefer to work in a basis of (anti)symmetric colour structures, as when $f=f^\prime$ this yields a definite exchange symmetry for the remaining flavour indices, see \cref{sec:flav_decomp_indist}.

Similarly, when $f \neq f^\prime$, we could have constructed definite-parity-charge operators from vector and axial currents, instead of the combinations in \cref{eq:AandB_op_def}. However, in the case where $f=f^\prime$, the non-trivial interactions between the flavour and parity symmetries force us to use the combinations \cref{eq:AandB_op_def}, and so we use them everywhere in the interest of uniformity.

\subsection{Decomposition for distinguishable currents}

The Wilson coefficients of operators involving distinguishable currents, where $f \neq f^\prime$, decompose into direct products of irreps under two of the flavour groups $SU(2)_u$, $SU(3)_d$, and $SU(3)_e$. Specifically,
\begin{align}
    [c_{ud}^{\pm,\pm}]^{pr}_{qs} \, :& \,  {\mathbf{2}}_{u}\otimes\mathbf{2}_{u}\otimes\overline{\mathbf{3}}_{d}\otimes\mathbf{3}_{d} = (\mathbf{1}_{u}\otimes\mathbf{1}_{d})\oplus(\mathbf{1}_{u}\otimes\mathbf{8}_{d})\oplus(\mathbf{3}_{u}\otimes\mathbf{1}_{d})\oplus(\mathbf{3}_{u}\otimes\mathbf{8}_{d}) \, , \\
    [c_{eu}^{\pm}]^{pr}_{qs} \, :& \,  \overline{\mathbf{3}}_{e}\otimes\mathbf{3}_{e}\otimes{\mathbf{2}}_{u}\otimes\mathbf{2}_{u} = (\mathbf{1}_{e}\otimes\mathbf{1}_{u})\oplus(\mathbf{1}_{e}\otimes\mathbf{3}_{u})\oplus(\mathbf{8}_{e}\otimes\mathbf{1}_{u})\oplus(\mathbf{8}_{e}\otimes\mathbf{3}_{u}) \, , \\
    [c_{ed}^{\pm}]^{pr}_{qs} \, :& \,  \overline{\mathbf{3}}_{e}\otimes\mathbf{3}_{e}\otimes\overline{\mathbf{3}}_{d}\otimes\mathbf{3}_{d} = (\mathbf{1}_{e}\otimes\mathbf{1}_{d})\oplus(\mathbf{1}_{e}\otimes\mathbf{8}_{d})\oplus(\mathbf{8}_{e}\otimes\mathbf{1}_{d})\oplus(\mathbf{8}_{e}\otimes\mathbf{8}_{d}) \, .
\label{eqn:ed_decomp_eqn}
\end{align}

Since neutrino currents are not charged under the gauge interactions of the LEFT, we decompose operators coupling charged fermions to neutrinos into irreps of the charged fermion flavour group only. Consequently, the Wilson coefficients decompose as 
\begin{align}
  [c_{\nu d}^{\pm}]^{pr}_{qs} \, :& \,   \overline{\mathbf{3}}_{d}\otimes\mathbf{3}_{d} = \mathbf{1}_{d}\oplus\mathbf{8}_{d} \, , \\
  [c_{\nu e}^{\pm}]^{pr}_{qs} \, :& \,   \overline{\mathbf{3}}_{e}\otimes\mathbf{3}_{e} = \mathbf{1}_{e}\oplus\mathbf{8}_{e} \, , \\
  [c_{\nu u}^{\pm}]^{pr}_{qs} \, :& \,   \mathbf{2}_{u}\otimes\mathbf{2}_{u} = \mathbf{1}_{u}\oplus\mathbf{3}_{u} \, .
\end{align}

\paragraph{Notation}
We denote parity and flavour-decomposed Wilson coefficients, for operators with distinguishable fermion currents, as 
\begin{equation}
[c^{\pm,\pm}_{ff^\prime, A/B}]_{\{\db, i\}_f, \{\db, j\}_{f^\prime}} \, ,
\end{equation}
where the first $\pm$ denotes parity charge, the second $\pm$ denotes colour symmetry for coefficients of $\Op_{ud}$ operators but will otherwise be omitted, $\db$ is the dimension of the flavour irrep under the relevant flavour group and $i$ and $j$ label irrep components. For the coefficients of $\Op_{\nu f}$ operators we write
\begin{equation}
[c^{\pm}_{\nu f}]_{\{\db, i\}_f \, ,}
\end{equation}
since we only decompose under the charged fermion flavour group.

\subsection{Decomposition for indistinguishable currents\label{sec:flav_decomp_indist}}

When $f=f^\prime$, Fierz identities show that the A-type operators ($\Op_{A}^\pm$, \cref{eq:AandB_op_def}) are intrinsically symmetric under the exchange of either just the fermion labels ($\beta,\delta$), just the antifermion labels ($\alpha,\gamma$), or both simultaneously. However, B-type operators ($\Op_{B}^\pm$, \cref{eq:AandB_op_def}) only have definite symmetries under the simultaneous exchange of both fermion and antifermion labels. Additionally, the parity charge of these operators affects the sign of the exchange symmetry; $\Op_{B}^+$ is symmetric while $\Op_{B}^-$ is antisymmetric. In sum, when $f=f^\prime$ the operators satisfy
\begin{align}
    [\Op_{A}^{\pm}]^{\alpha \gamma}_{\beta \delta} = [\Op_{A}^{\pm}]^{\gamma \alpha}_{\beta \delta} = [\Op_{A}^{\pm}]^{\alpha \gamma}_{\delta \beta} = [\Op_{A}^{\pm}]^{\gamma \alpha}_{\delta \beta} \, , \label{eq:A_type_op_sym} \\
    [\Op_B^{\pm}]^{\alpha \gamma}_{\beta \delta} = \pm[\Op_B^{\pm}]^{\gamma \alpha}_{\delta \beta} \, . \label{eq:B_type_op_sym}
\end{align}
When we define the various colour contractions, this reduces to the following symmetry in just the flavour indices ($p$, $q$, $r$, and $s$)
\newcommand{\bpm}{{\color{blue}\pm}}
\newcommand{\opm}{{\color{orange}\pm}}
\begin{align}
  [\Op_{A}^{\bpm,\opm}]^{q s}_{p r} = (\opm 1) [\Op_{A}^{\bpm,\opm}]^{s q}_{p r} = (\opm 1) [\Op_{A}^{\bpm,\opm}]^{q s}_{r p} = [\Op_{A}^{\bpm,\opm}]^{s q}_{r p} \, , \label{eq:A_type_flav_sym} \\
  [\Op_B^{\bpm,\opm}]^{q s}_{p r} = (\bpm 1) [\Op_B^{\bpm,\opm}]^{s q}_{r p} \, . \label{eq:B_type_flav_sym}
\end{align}
Above, the parity charge of the operator is shown in blue, and the exchange symmetry of the colour invariant is shown in orange. The resulting definite exchange symmetry of the flavour indices is the key to decomposing each operator's Wilson coefficients into irreps under the $SU(N)$ flavour groups.

To satisfy \cref{eq:A_type_flav_sym}, the coefficients of A-type operators with symmetric or antisymmetric colour structures satisfy
\begin{align}
    [c_A^{\cdot,+}]^{pr}_{qs}=& \, [c_A^{\cdot,+,+}]^{(pr)}_{(qs)} \, ,\\
    [c_A^{\cdot,-}]^{pr}_{qs}=& \, [c_A^{\cdot,-,-}]^{[pr]}_{[qs]} \, ,
\label{eq:A_type_decomp}
\end{align}
where `$\cdot$' is a placeholder for parity charge, $p,r,q,s$ label flavour, and we have added a third $\pm$ on the right-hand side to show the Wilson coefficient's symmetry under exchange of the fermion (i.e.\ upper) indices.\footnote{The symmetrisations of the indices are normalised such that
\begin{subequations}
  \begin{align}
    c^{(pr)}_{(qs)} \equiv& \, \frac14\left(c^{pr}_{qs}+c^{rp}_{qs}+c^{pr}_{sq}+c^{rp}_{sq}\right) \, ,\\
    c^{[pr]}_{[qs]} \equiv& \,  \frac14 \left(c^{pr}_{qs}-c^{rp}_{qs} -c^{pr}_{sq} +c^{rp}_{sq} \right) \, , \\
    c^{(pr)}_{[qs]} \equiv& \, \frac14\left(c^{pr}_{qs}+c^{rp}_{qs}-c^{pr}_{sq}-c^{rp}_{sq}\right) \, , \\
    c^{[pr]}_{(qs)} \equiv& \, \frac14 \left(c^{pr}_{qs}-c^{rp}_{qs}+c^{pr}_{sq} -c^{rp}_{sq} \right) \, .
 \end{align}
\end{subequations}}
Coefficients of B-type operators must satisfy \cref{eq:B_type_flav_sym}, where the parity charge of the corresponding operator determines the sign of this symmetry. The coefficients of parity-even operators decompose as 
\begin{equation}
[c_B^{+,\pm}]^{pr}_{qs}=[c_B^{+,\pm,+}]^{(pr)}_{(qs)}+[c_B^{+,\pm,-}]^{[pr]}_{[qs]} \, ,
\end{equation}
whereas the coefficients of parity-odd operators decompose as
\begin{equation}
  [c_B^{-,\pm}]^{pr}_{qs}=[c_B^{-,\pm,+}]^{(pr)}_{[qs]}+[c_B^{-,\pm,-}]^{[pr]}_{(qs)} \, .
\end{equation}
The above is independent of the symmetry of the colour structure. The (anti)symmetrised flavour components can now be mapped onto $SU(3)_{d,e}$ and $SU(2)_u$ flavour irreps.

When $f=d,e$ the various symmetrised components decompose as:
\begin{align}
  c^{(pr)}_{(qs)} \, :& \,  \left(\mathbf{3}_{f}\otimes\mathbf{3}_{f}\right)_{\text{sym}}\otimes\left(\overline{\mathbf{3}}_{f}\otimes\overline{\mathbf{3}}_{f}\right)_{\text{sym}} = \mathbf{6}_{f}\otimes \overline{\mathbf{6}}_{f}= \mathbf{1}_{f}\oplus \mathbf{8}_{f}\oplus \mathbf{27}_{f} \, , \\
  c^{[pr]}_{[qs]} \, :& \,  \left(\mathbf{3}_{f}\otimes\mathbf{3}_{f}\right)_{\text{antisym}}\otimes\left(\overline{\mathbf{3}}_{f}\otimes\overline{\mathbf{3}}_{f}\right)_{\text{antisym}} =\overline{\mathbf{3}}_{f}  \otimes\mathbf{3}_{f}= \mathbf{1}_{f}\oplus \mathbf{8}_{f} \, , \\
  c^{[pr]}_{(qs)} \, :& \,  \left(\mathbf{3}_{f}\otimes\mathbf{3}_{f}\right)_{\text{antisym}} \otimes \left(\overline{\mathbf{3}}_{f}\otimes\overline{\mathbf{3}}_{f}\right)_{\text{sym}}= \overline{\mathbf{3}}_{f}\otimes \overline{\mathbf{6}}_{f} = \mathbf{8}_{f} \oplus \overline{\mathbf{10}}_{f} \, , \\
  c^{(pr)}_{[qs]} \, :& \,  \left(\mathbf{3}_{f}\otimes\mathbf{3}_{f}\right)_{\text{sym}} \otimes \left(\overline{\mathbf{3}}_{f}\otimes\overline{\mathbf{3}}_{f}\right)_{\text{antisym}}= \mathbf{6}_{f}\otimes \mathbf{3}_{f} = \mathbf{8}_{f} \oplus \mathbf{10}_{f} \, ,
\end{align}
These different flavour irreps, and the colour structures that they are multiplied by, are shown in \cref{tab:dd_decomp_table}. Since leptons are uncharged under QCD, the $\Op_{e e}$ operators only populate the left-hand side of the table, while $\Op_{dd}$ operators populate both sides depending on their colour structure.

When $f=u$, the various symmetrised components decompose as:
\begin{align}
  c^{(pr)}_{(qs)} \, :& \,  \left(\mathbf{2}_{u}\otimes\mathbf{2}_{u}\right)_{\text{sym}} \otimes\left(\mathbf{2}_{u}\otimes\mathbf{2}_{u}\right)_{\text{sym}} = \mathbf{3}_{u}\otimes \mathbf{3}_{u}= \mathbf{1}_{u}\oplus \mathbf{3}_{u}\oplus \mathbf{5}_{u} \, , \\
  c^{[pr]}_{[qs]} \, :& \,  \left(\mathbf{2}_{u}\otimes\mathbf{2}_{u}\right)_{\text{antisym}}\otimes\left(\mathbf{2}_{u}\otimes\mathbf{2}_{u}\right)_{\text{antisym}} =\mathbf{1}_{u}  \otimes\mathbf{1}_{u}= \mathbf{1}_{u} \, , \\
  c^{[pr]}_{(qs)},c^{(pr)}_{[qs]} \, :& \,  \left(\mathbf{2}_{u}\otimes\mathbf{2}_{u}\right)_{\text{sym}} \otimes \left(\mathbf{2}_{u}\otimes\mathbf{2}_{u}\right)_{\text{antisym}}= \mathbf{3}_{u}  \otimes\mathbf{1}_{u}= \mathbf{3}_{u} \, .
\end{align}
These different flavour irreps, and the parity and colour structures that they are multiplied by, are shown in \cref{tab:uu_table}.

\paragraph{Notation}
In \cref{tab:dd_decomp_table,tab:uu_table} we introduce a notation to label the Wilson coefficients of operators, involving indistinguishable fermion currents, after decomposition into flavour and parity eigenstates. We use the notation
\begin{align}
    [c^{\pm,\pm,\pm}_{ff,A/B}]_{\{\db, i\}} \label{eq:indistinguishable_coeff_notation}
\end{align}
to label the flavour-decomposed coefficient of an A- or B-type combination of $\Op_{ff}$ operators. The superscript `$\pm$' labels denote, in order, the symmetry under parity, the symmetry of the colour structure, and the symmetry of the flavour structure under exchange of the fermion (i.e.\ upper) indices. The `$\{\db, i\}$' label indicates the dimensionality and component of the representation of the coefficient under $SU(3)_f$ or $SU(2)_u$.

\begin{table}
\begin{centering}
\resizebox{\textwidth}{!}{%
\begin{tabularx}{1.24\textwidth}{|c|C|C|C|C|C|C|}
\hline
{$\Op$} & \multicolumn{3}{ c |}{$\delta^{(..)}_{(..)}$ colour structure} & \multicolumn{3}{ c |}{$\delta^{[..]}_{[..]}$ colour structure}\\ 
\cline{2-7}
& {Flavour structure} & {$SU(3)_{d,e}$ irrep} & Notation $(f=d,e)$ & {Flavour structure} & {$SU(3)_d$ irrep} & Notation\\
\hline
$\Op^{+}_{A}$ & $c^{(..)}_{(..)}$ & $\mathbf{1}\oplus\mathbf{8}\oplus\mathbf{27}$ & $[c^{+,+,+}_{ff,A}]_{\{\db, i\}}$ & $c^{[..]}_{[..]}$ & $\mathbf{1}\oplus\mathbf{8}$ & $[c^{+,-,-}_{dd,A}]_{\{\db, i\}}$ \\ \hline
$\Op^{-}_{A}$ & $c^{(..)}_{(..)}$ & $\mathbf{1}\oplus\mathbf{8}\oplus\mathbf{27}$ & $[c^{-,+,+}_{ff,A}]_{\{\db, i\}}$ & $c^{[..]}_{[..]}$ & $\mathbf{1}\oplus\mathbf{8}$ & $[c^{-,-,-}_{dd,A}]_{\{\db, i\}}$ \\ \hline
\multirow{ 2}{*}{$\Op^{+}_{B}$} & $c^{(..)}_{(..)}$ & $\mathbf{1}\oplus\mathbf{8}\oplus\mathbf{27}$ &$[c^{+,+,+}_{ff, B}]_{\{\db, i\}}$ & $c^{(..)}_{(..)}$ & $\mathbf{1}\oplus\mathbf{8}\oplus\mathbf{27}$ &$[c^{+,-,+}_{dd, B}]_{\{\db, i\}}$ \\ 
& $c^{[..]}_{[..]}$ & $\mathbf{1}\oplus\mathbf{8}$ &$[c^{+, +, -}_{ff,B}]_{\{\db, i\}}
$ & $c^{[..]}_{[..]}$ & $\mathbf{1}\oplus\mathbf{8}$ &$[c^{+,-,-}_{dd, B}]_{\{\db, i\}}$ \\ \hline
\multirow{ 2}{*}{$\Op^{-}_{B}$} & $c^{(..)}_{[..]}$ & $\mathbf{10}\oplus\mathbf{8}$ &$[c_{ff, B}^{-,+,+}]_{\{\db, i\}}$ & $c^{(..)}_{[..]}$ & $\mathbf{10}\oplus\mathbf{8}$ &$[c^{-,-,+}_{dd, B}]_{\{\db, i\}}$\\
& $c^{[..]}_{(..)}$ & $\overline{\mathbf{10}}\oplus\mathbf{8}$ &$[c^{-,+,-}_{ff, B}]_{\{\db, i\}}$ & $c^{[..]}_{(..)}$ & $\overline{\mathbf{10}}\oplus\mathbf{8}$&$[c^{-,-,-}_{dd,B}]_{\{\db, i\}}$ \\ \hline
\end{tabularx}
}
\caption{The flavour irrep decomposition of the Wilson coefficients of each parity basis operator for $\Op_{d d}$ and $\Op_{e e}$ operators. Wilson coefficients of $\Op_{e e}$ only decompose into the $\delta^{(..)}_{(..)}$ column since they cannot support a non-trivial colour structure.}
\label{tab:dd_decomp_table}
\end{centering}
\end{table}

\begin{table}
\begin{centering}
\resizebox{\textwidth}{!}{%
\begin{tabularx}{1.24\textwidth}{|c|C|C|C|C|C|C|}
\hline
{$\Op$} & \multicolumn{3}{ c |}{$\delta^{(..)}_{(..)}$ colour structure} & \multicolumn{3}{ c |}{$\delta^{[..]}_{[..]}$ colour structure}\\ 
\cline{2-7}
& {Flavour structure} & {$SU(2)_u$ irreps} & Notation & {Flavour structure} & {$SU(2)_u$ irreps} & Notation\\
\hline
$\Op_{A}^+$ & $c^{(..)}_{(..)}$ & $\mathbf{1}\oplus\mathbf{3}\oplus\mathbf{5}$ & $[c^{+,+,+}_{uu,A}]_{\{\db, i\}}$& $c^{[..]}_{[..]}$ & $\mathbf{1}$ & $[c^{+,-,-}_{uu,A}]_{\{\db, i\}}$\\ \hline
$\Op_{A}^-$ & $c^{[..]}_{[..]}$  & $\mathbf{1}$ & $[c^{-,+,-}_{uu,A}]_{\{\db, i\}}$ & $c^{(..)}_{(..)}$ & $\mathbf{1}\oplus\mathbf{3}\oplus\mathbf{5}$ & $[c^{-,-,+}_{uu,A}]_{\{\db, i\}}$\\ \hline
\multirow{2}{*}{$\Op_{B}^+$} & $c^{(..)}_{(..)}$  & $\mathbf{1}\oplus\mathbf{3}\oplus\mathbf{5}$ & $[c^{+,+,+}_{uu,B}]_{\{\db, i\}}$ & $c^{(..)}_{(..)}$ & $\mathbf{1}\oplus\mathbf{3}\oplus\mathbf{5}$ & $[c^{+,-,+}_{uu,B}]_{\{\db,i\}}$\\
& $c^{[..]}_{[..]}$ & $\mathbf{1}$ & $[c^{+,+,-}_{uu,B}]_{\{\db,i\}}$ & $c^{[..]}_{[..]}$ & $\mathbf{1}$ & $[c^{+,-,-}_{uu,B}]_{\{\db,i\}}$\\ \hline
\multirow{2}{*}{$\Op_{B}^-$} & $c^{(..)}_{[..]}$ & $\mathbf{3}$ & $[c^{-,+,+}_{uu,B}]_{\{\db,i\}}$&  $c^{(..)}_{[..]}$ & $\mathbf{3}$ & $[c^{-,-,+}_{uu,B}]_{\{\db,i\}}$\\
& $c^{[..]}_{(..)}$ & $\mathbf{3}$ & $[c^{-,+,-}_{uu,B}]_{\{\db,i\}}$ & $c^{[..]}_{(..)}$ & $\mathbf{3}$ & $[c^{-,-,-}_{uu,B}]_{\{\db,i\}}$\\ \hline
\end{tabularx}
}
\caption{Flavour irrep decomposition of Wilson coefficients for $\Op_{u u}$ parity basis operators.}
\label{tab:uu_table}
\end{centering}
\end{table}

\subsection{Flavour and parity charges\label{sec:flavour_parity_charges}}

Each Wilson coefficient of a four-fermion vector current operator carries a set of charges under the group \cref{eq:vecsymgroup}, namely: the parity charge; its 3 irreps under $SU(3)_d$, $SU(3)_e$, and $SU(2)_u$, together with the specific components of the irreps to which the operator belongs. We use the results of this Section as well as the conventions of \cite{deSwart:1963pdg,Kaeding:1995vq} to calculate the Clebsch-Gordan decomposition of the relevant Wilson coefficients in the San Diego basis \cite{Jenkins:2017jig} into coefficients with a definite set of flavour and parity charges. The linear mapping is provided in the ancillary files accompanying this paper, which also specify the conjugation properties of the components, their charges as specified in \cref{eq:flavour_charges}, and their one-loop beta functions.

The component of an $SU(3)$ irrep is uniquely identified by an index that runs between $1$ and the dimension of the irrep. Alternatively, we can uniquely identify an irrep component by its total isospin $\ifl$, its third component of isospin $\itfl$, and its hypercharge $\yfl$. These are measured respectively by operators proportional to $T_1^2+T_2^2+T_3^2$, $T_3$, and $T_8$, where $T$ is an $SU(3)$ generator. The weight diagrams of \cref{fig:weight_diagrams} show the correspondence between the component index and the aforementioned charges in the conventions of \cite{deSwart:1963pdg}. Similarly, we can uniquely identify a component of an $SU(2)$ irrep by its third component of isospin $\itfl$, as measured by the third $SU(2)$ generator. In our conventions $\itfl=j,j-1,\ldots,-j$ for the $SU(2)$ irrep of dimension $2j+1$, with the components indexed from $1$ to $2j+1$ in order of descending $\itfl$.

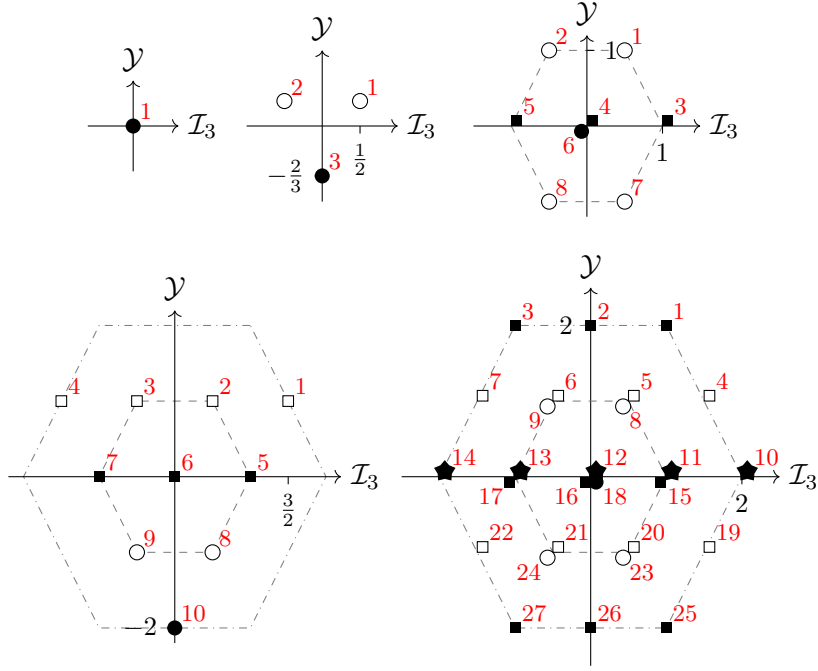
\begin{figure}
  \centering
  \begin{tikzpicture}[scale=1]
    \draw[->] (-0.6,0) -- (0.6,0) node[right] {$\itfl$};
    \draw[->] (0,-0.6) -- (0,0.6) node[above] {$\yfl$};
    \node at (0,0) [iso0style,label={[labelstyle]\angleone:$1$}]{}; 
    \begin{scope}[shift = {(2.5,0)}]
    \draw[->] (-1,0) -- (1,0) node[right] {$\itfl$};
    \draw[->] (0,-1) -- (0,1) node[above] {$\yfl$};
    \draw (0.5,0) -- (0.5,-0.1) node[below] {\small$\frac12$};
    \draw (0,-0.66) -- (-0.1,-0.66) node[left] {\small$-\frac23$};
    \node at (0.5,0.33) [isohalfstyle,label={[labelstyle]\angleone:$1$}]{};
    \node at (-0.5,0.33) [isohalfstyle,label={[labelstyle]\angleone:$2$}]{};
    \node at (0,-0.66) [iso0style,label={[labelstyle]\angleone:$3$}]{};
  \end{scope}
    \begin{scope}[shift = {(6,0)}]
      \coordinate (e1) at (0.5,1);
      \coordinate (e2) at (-0.5,1);
      \coordinate (e3) at (1,0);
      \coordinate (e4) at (0,0);
      \coordinate (e5) at (-1,0);
      \coordinate (e6) at (0,0);
      \coordinate (e7) at (0.5,-1);
      \coordinate (e8) at (-0.5,-1);
      \draw[gray,dashed] (e1) -- (e2) -- (e5) -- (e8) -- (e7) -- (e3) -- cycle;
      \draw[->] (-1.5,0) -- (1.5,0) node[right] {$\itfl$};
      \draw[->] (0,-1.2) -- (0,1.2) node[above] {$\yfl$};
      \draw (1,0) -- (1,-0.1) node[below] {\small$1$};
      \draw (0,1) -- (0.1,1) node[right] {\small$1$};
      \node at (e1) [isohalfstyle,label={[labelstyle]\angleone:$1$}]{};
      \node at (e2) [isohalfstyle,label={[labelstyle]\angleone:$2$}]{};
      \node at ($(e3) + (\angleone:0.1)$) [iso1style,label={[labelstyle]\angleone:$3$}]{};
      \node at ($(e4) + (\angleone:0.1)$) [iso1style,label={[labelstyle]\angleone:$4$}]{};
      \node at ($(e5) + (\angleone:0.1)$) [iso1style,label={[labelstyle]\angleone:$5$}]{};
      \node at ($(e6) + (\angletwo:0.1)$) [iso0style,label={[labelstyle]\angletwo:$6$}]{};
      \node at (e7) [isohalfstyle,label={[labelstyle]\angleone:$7$}]{};
      \node at (e8) [isohalfstyle,label={[labelstyle]\angleone:$8$}]{};
        \end{scope}
  \end{tikzpicture}

  \vspace{2ex}

  \begin{tikzpicture}[scale=1]
    \coordinate (e1) at (1.5,1);
    \coordinate (e2) at (0.5,1);
    \coordinate (e3) at (-0.5,1);
    \coordinate (e4) at (-1.5,1);
    \coordinate (e5) at (1,0);
    \coordinate (e6) at (0,0);
    \coordinate (e7) at (-1,0);
    \coordinate (e8) at (0.5,-1);
    \coordinate (e9) at (-0.5,-1);
    \coordinate (e10) at (0,-2);
    \draw[gray,dash dot] (1,2) -- (-1,2) -- (-2,0) -- (-1,-2) -- (1,-2) -- (2,0) -- cycle;
    \draw[gray,dashed] (0.5,1) -- (-0.5,1) -- (-1,0) -- (-0.5,-1) -- (0.5,-1) -- (1,0) -- cycle;
    \draw[->] (-2.2,0) -- (2.2,0) node[right] {$\itfl$};
    \draw[->] (0,-2.2) -- (0,2.2) node[above] {$\yfl$};
    \draw (1.5,0) -- +(0,-0.1) node[below] {\small$\frac32$};
    \draw (0,-2) -- +(-0.1,0) node[left] {\small$-2$};
    \node at (e1) [iso3halfstyle,label={[labelstyle]\angleone:$1$}]{};
    \node at (e2) [iso3halfstyle,label={[labelstyle]\angleone:$2$}]{};
    \node at (e3) [iso3halfstyle,label={[labelstyle]\angleone:$3$}]{};
    \node at (e4) [iso3halfstyle,label={[labelstyle]\angleone:$4$}]{};
    \node at (e5) [iso1style,label={[labelstyle]\angleone:$5$}]{};
    \node at (e6) [iso1style,label={[labelstyle]\angleone:$6$}]{};
    \node at (e7) [iso1style,label={[labelstyle]\angleone:$7$}]{};
    \node at (e8) [isohalfstyle,label={[labelstyle]\angleone:$8$}]{};
    \node at (e9) [isohalfstyle,label={[labelstyle]\angleone:$9$}]{};
    \node at (e10) [iso0style,label={[labelstyle]\angleone:$10$}]{};
 \begin{scope}[shift = {(5.5,0)}]
   \draw[->] (-2.5,0) -- (2.5,0) node[right] {$\itfl$};
   \draw[->] (0,-2.5) -- (0,2.5) node[above] {$\yfl$};
   \draw (2,0) -- (2,-0.1) node[below] {\small$2$};
   \draw (0,2) -- (-0.1,2) node[left] {\small$2$};
   \draw[gray,dash dot] (1,2) -- (-1,2) -- (-2,0) -- (-1,-2) -- (1,-2) -- (2,0) -- cycle;
   \draw[gray,dashed] (0.5,1) -- (-0.5,1) -- (-1,0) -- (-0.5,-1) -- (0.5,-1) -- (1,0) -- cycle;
   \node at (1,2) [iso1style,label={[labelstyle]\angleone:$1$}]{};
   \node at (0,2) [iso1style,label={[labelstyle]\angleone:$2$}]{};
   \node at (-1,2) [iso1style,label={[labelstyle]\angleone:$3$}]{};

   \node at ($(1.5,1) + (\angleone:0.1)$) [iso3halfstyle,label={[labelstyle]\angleone:$4$}]{};
   \node at ($(0.5,1) + (\angleone:0.1)$) [iso3halfstyle,label={[labelstyle]\angleone:$5$}]{};
   \node at ($(-0.5,1) + (\angleone:0.1)$) [iso3halfstyle,label={[labelstyle]\angleone:$6$}]{};
   \node at ($(-1.5,1) + (\angleone:0.1)$) [iso3halfstyle,label={[labelstyle]\angleone:$7$}]{};
   \node at ($(0.5,1) + (\angletwo:0.1)$) [isohalfstyle,label={[labelstyle]\anglethree:$8$}]{};
   \node at ($(-0.5,1) + (\angletwo:0.1)$) [isohalfstyle,label={[labelstyle]\angletwo:$9$}]{};

   \node at ($(2,0) + (\angleone:0.1)$) [iso2style,label={[labelstyle]\angleone:$10$}]{};
   \node at ($(1,0) + (\angleone:0.1)$) [iso2style,label={[labelstyle]\angleone:$11$}]{};
   \node at ($(0,0) + (\angleone:0.1)$) [iso2style,label={[labelstyle]\angleone:$12$}]{};
   \node at ($(-1,0) + (\angleone:0.1)$) [iso2style,label={[labelstyle]\angleone:$13$}]{};
   \node at ($(-2,0) + (\angleone:0.1)$) [iso2style,label={[labelstyle]\angleone:$14$}]{};

   \node at ($(1,0) + (\angletwo:0.1)$) [iso1style,label={[labelstyle]\anglethree:$15$}]{};
   \node at ($(0,0) + (\angletwo:0.1)$) [iso1style,label={[labelstyle]\angletwo:$16$}]{};
   \node at ($(-1,0) + (\angletwo:0.1)$) [iso1style,label={[labelstyle]\angletwo:$17$}]{};
   \node at ($(0,0) + (\anglethree:0.1)$) [iso0style,label={[labelstyle]\anglethree:$18$}]{};

   \node at ($(1.5,-1) + (\angleone:0.1)$) [iso3halfstyle,label={[labelstyle]\angleone:$19$}]{};
   \node at ($(0.5,-1) + (\angleone:0.1)$) [iso3halfstyle,label={[labelstyle]\angleone:$20$}]{};
   \node at ($(-0.5,-1) + (\angleone:0.1)$) [iso3halfstyle,label={[labelstyle]\angleone:$21$}]{};
   \node at ($(-1.5,-1) + (\angleone:0.1)$) [iso3halfstyle,label={[labelstyle]\angleone:$22$}]{};
   \node at ($(0.5,-1) + (\angletwo:0.1)$) [isohalfstyle,label={[labelstyle]\anglethree:$23$}]{};
   \node at ($(-0.5,-1) + (\angletwo:0.1)$) [isohalfstyle,label={[labelstyle]\angletwo:$24$}]{};

   \node at (1,-2) [iso1style,label={[labelstyle]\angleone:$25$}]{};
   \node at (0,-2) [iso1style,label={[labelstyle]\angleone:$26$}]{};
   \node at (-1,-2) [iso1style,label={[labelstyle]\angleone:$27$}]{};
         \end{scope}
 \end{tikzpicture}
 \caption{The mapping of the components (labelled in red) of the irreps $\mathbf{1}$, $\mathbf{3}$, $\mathbf{8}$, $\mathbf{10}$, and $\mathbf{27}$ onto their flavour charges --- the generational analogues of isospin $\ifl$, the third component of isospin $\itfl$, and the hypercharge $\yfl$ --- in the conventions of \cite{deSwart:1963pdg}. `Isospin' ($\ifl$) key: \isozerokey$\,=0$, \isohalfkey$\,=\frac12$, \isoonekey$\,=1$, \isothreehalfkey$\,=\frac32$, \isotwokey$\,=2$. The dashed hexagon intersects components that contribute to $\Delta F = 1$ processes, the dot-dashed hexagon intersects those that contribute to $\Delta F = 2$ processes. (Note that the conventions of \cite{Aebischer:2025hsx} distinguish the points on the vertices of the dot-dashed hexagon from those in the middle of the edges, respectively labelling them $\Delta F = 2$ and $\Delta F = 1.5$.) The diagrams for $\overline{\mathbf{3}}$ and $\overline{\mathbf{10}}$ are obtained by inverting the diagrams for the corresponding unconjugated irrep about the origin.\label{fig:weight_diagrams}}
\end{figure}

The Wilson coefficients of the LEFT can therefore be partitioned according to the charges
\begin{equation}
  \{\db,\itfl\}_u, \{\db,\ifl,\itfl,\yfl\}_d, \{\db,\ifl,\itfl,\yfl\}_e, P ,
  \label{eq:flavour_charges}
\end{equation}
where $P$ is the parity charge and $\db$ the dimension of a given irrep. As parity and flavour are violated in nature in a very non-generic pattern, it is helpful to organise the space of Wilson coefficients into these blocks, which will each have a distinct phenomenology. Not least because the blocks do not mix under running, as the gauge interactions are flavour and parity symmetric --- see \cref{sec:blockgamma}.

It is worth pausing to consider what the full Clebsch-Gordan decomposition of the special unitary flavour groups $SU(3)^2\times SU(2)$ adds here. A more typical approach to organise the Wilson coefficients of the LEFT is by the $U(1)^8$ of individual fermion number (see e.g.~\cite{Buchalla:1995vs}). In the case of the $\mathcal{O}_{dd}$ operator, for example, this amounts to considering the $U(1)^3$ charge by counting the number of $d$s (minus the number of $\bar{d}$s), and similarly for $s$s and $b$s, with the other five $U(1)$ charges being trivial. Given that all vector current operators that we consider have zero net fermion number overall, this is equivalent to counting the charges of just the $U(1)^2 \subset SU(3)_d$. These two charges are w.l.o.g.\ given by $\ifl_{3,d}$ and $\yfl_d$. Therefore, the more typical approach is to partition the Wilson coefficients of the LEFT by just the Abelian charges
\begin{equation}
  \{\itfl\}_u, \{\itfl,\yfl\}_d, \{\itfl,\yfl\}_e \, .
  \label{eq:flavour_charges_abelian}
\end{equation}
Doing so for the vector current operators gives the block sizes and degeneracies in \cref{tab:magnetic_blocks}, which we have checked agrees with the definition of these blocks in the \texttt{wcxf} format \cite{Aebischer:2017ugx}, as implemented by \texttt{wilson} \cite{Aebischer:2018bkb}. \cref{tab:large_blocks_table,tab:small_blocks_table} document the much smaller blocks obtained by adding the parity and non-Abelian information contained in \cref{eq:flavour_charges}.

To give an example, the block which contributes to lepton-flavour-conserving $\Delta F_d = 1$ transitions has 52 real degrees of freedom (d.o.f.) when arranged according to the Abelian charges \cref{eq:flavour_charges_abelian}. However, when arranged by the full set of charges \cref{eq:flavour_charges}, this decomposes into $2$ blocks with $12$ d.o.f., $8$ blocks with $2$ d.o.f., and $12$ blocks with $1$ d.o.f., each with its own distinct phenomenology, and none of which mix under RG.

In the next Section we consider the individual behaviour of these blocks under RG.

\begin{table}
  \centering
  \begin{tabular}{|c | c | c |}
      \hline
      Block quantum numbers & Block size & \# of blocks \\ \hline\hline
      - & $203\times203$ & $1$ \\[5pt]\hline
      ${\itfl}_d^2 + \frac34 \yfl_d^2 = 1$ & $52\times52$ & $6$ \\[5pt]\hline
      ${\itfl}_u^2 = 1$ & $48\times48$ & $2$ \\[5pt]\hline
      ${\itfl}_e^2 + \frac34 \yfl_e^2 = 1$ & $34\times34$ & $6$ \\[5pt]\hline
      ${\itfl}_d^2 + \frac34 \yfl_d^2 = 1, {\itfl}_u^2 = 1$ & $8\times8$ & $12$ \\[5pt]\hline
      ${\itfl}_d^2 + \frac34 \yfl_d^2 = 3$ & $6\times6$ & $6$ \\[5pt]\hline
      ${\itfl}_d^2 + \frac34 \yfl_d^2 = 1,{\itfl}_e^2 + \frac34 \yfl_e^2 = 1$ & $4\times4$ & $36$ \\[5pt]\hline
      ${\itfl}_u^2 = 1,{\itfl}_e^2 + \frac34 \yfl_e^2 = 1$ & $4\times4$ & $12$ \\[5pt]\hline
      ${\itfl}_e^2 + \frac34 \yfl_e^2 = 3$ & $4\times4$ & $6$ \\[5pt]\hline
      ${\itfl}_d^2 + \frac34 \yfl_d^2 = 4$ & $4\times4$ & $6$ \\[5pt]\hline
      ${\itfl}_u^2 = 2$ & $4\times4$ & $2$ \\[5pt]\hline
      ${\itfl}_e^2 + \frac34 \yfl_e^2 = 4$ & $3\times3$ & $6$ \\[5pt]\hline
  \end{tabular}
  \caption{The size of the blocks of vector current operators, in terms of real d.o.f., when decomposed according to just the Abelian flavour quantum numbers ${\itfl}_u,{\itfl}_d,\yfl_d,{\itfl}_e,\yfl_e$. For each block, the quantum numbers are zero unless stated in the left-hand column.}
  \label{tab:magnetic_blocks}
\end{table}

\section{The decomposed anomalous dimension matrix\label{sec:blockgamma}}

In this Section, we consider the schematic form of the RG in the LEFT, and how it simplifies under the flavour and parity decomposition of \cref{sec:decomp}.

\subsection{The schematic form of the LEFT RG}

Consider the parameters of the LEFT Lagrangian up to dimension 6. At dimension 3, $M$ denote mass matrices for the quarks and charged leptons. At dimension 4, we denote as $e$ and $g$ the gauge couplings for QED and QCD respectively, and ignore the QCD theta term. The dimension 5 coefficients, $c_{\psi X}$, multiply dipole terms. Barring the operators built from three gluon field strengths, which we neglect, the dimension 6 terms are four-fermion operators. We classify the coefficients into vector ($c_V$), scalar ($c_S$), and tensor ($c_T$), according to the spinor contractions in the corresponding operators.

The one-loop running was calculated in \cite{Jenkins:2017dyc}. Schematically, up to the insertion of $\mathcal{O}(1)$ rational coefficients, it takes the form of the coupled differential equations\footnote{Here we assume $M$ is the coefficient of an operator structure $\overline{f}_L f_R$ (in contrast to the convention of \cite{Jenkins:2017dyc}), and $c_{\psi X}$ is the coefficient of an operator structure $\overline{f}_L \sigma f_R F$. $M^*$ and $c_{\psi X}^*$ are associated to the conjugate operators.}
\newcommand{\nglct}{\color{gray}}
\begin{subequations}
\begin{align}   
\left(4\pi\right)^2 \dot{M}  & = (e^2+ g^2)M \nglct +(ec_{\psi X}+gc_{\psi X})M M^* \notag\\
&\hspace{8ex} \nglct +(c_S +c_{\psi X}^2) M (M^*)^2  + (c_V +c_{\psi X}c_{\psi X}^*) M^2 M^*\, ,\\
\left(4\pi\right)^2 \dot{e} &= e^3 \nglct +e^2(c_{\psi X}M^* +c_{\psi X}^* M) +c_{\psi X}^2 (M^*)^2 + (c_{\psi X}^*)^2 M^2 \, ,\\
\left(4\pi\right)^2 \dot{g} &= g^3 \nglct +g^2(c_{\psi X}M^* +c_{\psi X}^* M) +c_{\psi X}^2 (M^*)^2 + (c_{\psi X}^*)^2 M^2  \, ,\\
\left(4\pi\right)^2 \dot{c}_{\psi X} & = (e^2+g^2+eg)c_{\psi X} +(e+g)M^* (c_S + c_T) \nglct +(e+g) Mc_{\psi X}c_{\psi X}^* \, ,\label{eq:dipoleRGs}\\
\left(4\pi\right)^2 \dot{c}_S & = (e^2+ g^2) (c_S + c_T) +(e^2+ g^2 + eg)c_{\psi X}^2 \, ,\\
\left(4\pi\right)^2 \dot{c}_T & = (e^2+ g^2) (c_S + c_T) \, , \\
\left(4\pi\right)^2 \dot{c}_V &= (e^2+ g^2) c_V + (e^2+ g^2 + eg)c_{\psi X}c_{\psi X}^* \, , \label{eq:vectorialRGE}
\end{align}\label{eq:schematicRG}%
\end{subequations}
where a dot represents the action of $\mu \frac{\dd}{\dd \mu} \equiv \frac{\dd}{\dd \ln \mu} \equiv \frac{\dd}{\dd t}$; we define $t \equiv \ln \mu$ henceforth. We wish to solve these equations to percent-level accuracy when running across the typical range of application of the five-flavour LEFT: from the pole $W$ mass to the $\overline{\mathrm{MS}}$ $b$ mass
\begin{equation}
    \mu_W = 80.4 \, \GeV \, ; \quad \mu_b = 4.2 \, \GeV \, ,
\end{equation}
with initial conditions \cite{ParticleDataGroup:2024cfk}
    \begin{equation}
        e(m_Z) = 0.313 \, ; \quad g(m_Z) = 1.217 \, .
    \end{equation}
We assume that the scale of the higher dimension coefficients is bounded from below by the electroweak vev, i.e., $c_{\psi X}^2 , c \lesssim \frac{1}{v^2}$.

In \cref{eq:schematicRG}, we show the terms that we neglect in the subsequent analysis in grey. First, we neglect any mass-suppressed term of the form
\begin{equation}
    \left(4\pi\right)^2 \dot x = g \, M^* \,c_{\psi X} \, x \text{ OR } g \, M \,c_{\psi X}^* \, x\, ,
\end{equation}
where $x$ is an arbitrary parameter. At leading order this gives a fractional change in the low scale value of $x$ of
\begin{equation}
    \frac{\Delta x}{x} \sim g \frac{1}{\left(4\pi\right)^2} \frac{m_b}{v} \ln \frac{\mu_b}{\mu_W} \sim 10^{-3} g \, ,
\end{equation}
which is beyond our required precision, even without accounting for the $\frac{1}{\left(4\pi\right)^2}$ suppression of $c_{\psi X}$ which arises when matching to the SM or any other weakly coupled theory \cite{Jenkins:2013fya,Grojean:2024tcw}. On similar grounds we also neglect all $\bO(M^2)$ terms. This significantly decouples \cref{eq:schematicRG}; the equations can now be solved sequentially in the sets $\{e\}$, $\{g\}$, $\{M\}$, $\{c_{\psi X},c_S,c_T\}$, and $\{c_V\}$.

Effectively we decouple the equations by neglecting certain mass insertions. The terms without mass insertions can be understood by considering the helicity selection rules in amplitudes of massless particles, which we describe briefly in the next section, before returning to the effect of the flavour and parity decomposition on the running of the vector operators.

\subsection{Helicity selection rules}
If we take all particles to be massless, we can define SM states of definite helicity as
\begin{equation}
    \psi^+, \psi^-, V^+, V^-,
\end{equation}
where $\psi^{\pm}$ represent any fermion and $V^{\pm}$ represent any vector, with respective helicities
\begin{equation}
 h=\frac{1}{2}, -\frac{1}{2}, 1, -1,
\end{equation}
defined for \emph{incoming} particles without loss of generality. The various classes of LEFT operators can then be labelled by coordinates $(n,\sum_i h_i)$, corresponding to the number of legs $n$ and total helicity $\sum_i h_i$ of the amplitudes they induce at tree level.\footnote{$(n,\sum_i h_i)$ are linear combinations of the holomorphic coordinates introduced in \cite{Cheung:2015aba}.} So, we can refer to e.g.~the vectorial 4-fermion operators $O_V\sim\psi^2\bar{\psi}^2$, as the \emph{(4,0) operators}, since they induce 4-point amplitudes with zero total helicity.

This classification of operators in terms of their helicity structure leads to `non-renormalisation' theorems \cite{Cheung:2015aba}. 
These arise due to the fact that if 2-cuts of a loop amplitude vanish, then the loop is finite, and cannot contribute to anomalous dimensions.\footnote{Caveat: if the loop has IR divergences, then the 2-cut may vanish in dimensional regularisation even though there may be a compensating UV divergence. But IR divergences can only arise in self-renormalisation diagrams, so this issue will not change the arguments of this section, which are focused on mixing between different classes of operators.} This leads to the following relation between tree-level amplitudes and the singularities that they can generate at one loop \cite{Cheung:2015aba}:
\begin{equation}
\label{eq:nonrenormalizationhelicity}
\begin{tikzpicture}[baseline=-13ex]
\node[draw,circle,fill=gray,inner sep=8pt] (amp1) at (0,0) {$a$};
\node[draw,circle,fill=gray,inner sep=8pt] (amp2) at (3,0) {$b$};
\node[draw,circle,fill=gray,inner sep=8pt] (amp3) at (6,0) {$c$};
\node at (-0.1,-2.0) {$\begin{pmatrix} n_a \\ \sum h_a \end{pmatrix}$};
\node at (1.5,-2.0) {$+ \! \begin{pmatrix} -4 \\ 0 \end{pmatrix} \! +$};
\node at (3.1,-2.0) {$\begin{pmatrix} n_b \\ \sum h_b \end{pmatrix}$};
\node at (6,-2.0) {$\begin{pmatrix} n_c \\ \sum h_c \end{pmatrix}$\,.};
\node at (4.5,-2.0) {$=$};
\node at (4.5,0) {$=$};
\foreach \ang in {100,140,...,260}
  \draw[thick] (amp1) -- (\ang:1);
\foreach \ang in {-80,-40,...,80}
  \draw[thick] (amp2) -- ++(\ang:1);
\foreach \ang in {0,36,...,324}
  \draw[thick] (amp3) -- ++(\ang:1);
\draw[thick] (amp1) to [out=45,in=180] (1,0.8) node[label=$\pm$] {};
\draw[thick] (amp1) to [out=-45,in=180] (1,-0.8) node[label=$\pm$] {};
\draw[thick] (amp2) to [out=135,in=0] (2,0.8) node[label=$\mp$] {};
\draw[thick] (amp2) to [out=-135,in=0] (2,-0.8) node[label=$\mp$] {};
\draw[dashed] (1.5,-1) -- (1.5,1);
\end{tikzpicture}
\end{equation}
The helicity of the legs on either side of the cut are equal and opposite, since the momenta of the legs of the tree-level amplitudes are all defined incoming. If we take `$c$' to be a LEFT amplitude, then we can use this relation to find constraints on which `$c$' operators can be renormalised by which `$a$' and `$b$' operators. Specifically, if we assume that `$c$' corresponds to an insertion of a dimension-six operator, then we have two possible situations: \textit{(i)} `$a$' is a dimension-six LEFT amplitude and `$b$' is an SM amplitude \textit{(ii)} `$a$' and `$b$' are each dimension-five LEFT amplitudes. Focussing first on case \textit{(i)}, the $(n,\sum_i h_i)$ coordinates of the tree-level dimension-four SM `$b$' amplitudes below the EW scale \emph{always} obey the rule\footnote{Above the EW scale there are a few exceptions to this, proportional to Yukawa couplings, but with the Higgs integrated out of the theory these exceptions do not exist at dimension four.}
\begin{equation}
\label{eq:SMhelicity}
\left| \sum h_{SM} \right|\leq n_{SM}-4.
\end{equation}
\begin{figure}
    \centering
    \includegraphics[width=0.3\linewidth]{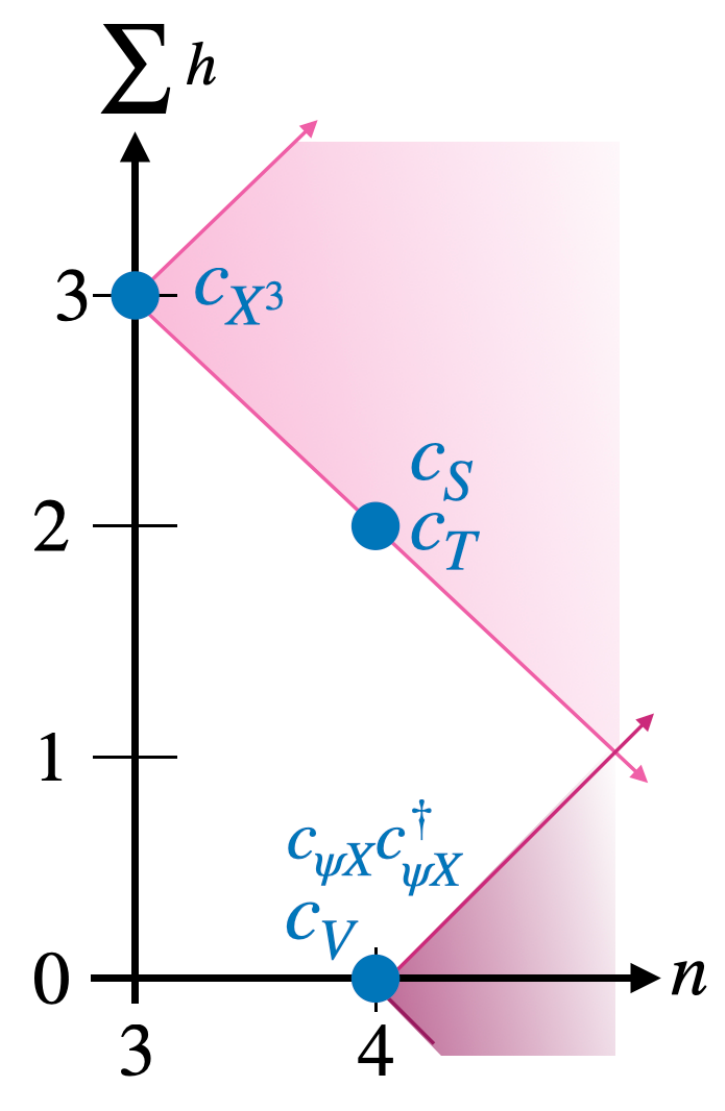}
    \caption{LEFT dimension five and six operator coefficients, arranged by the total helicity $(\sum h)$ and the number of legs ($n$) of the tree level amplitudes they generate. Conjugation reflects the graph in the $n$ axis.\label{fig:helicity}}
\end{figure}
So when the relations in \cref{eq:nonrenormalizationhelicity,eq:SMhelicity} are taken in combination, we find constraints on which `$c$' operators can be renormalised by which `$a$' operators, in terms of their respective helicities and number of legs. This is shown in \cref{fig:helicity}: only operators whose $(n,\sum_i h_i)$ coordinates lie on or within any given pink cone can be renormalised by the operators at the apex of the cone. For example, this means scalar $O_S$ and tensor $O_T$ operators cannot renormalise vector $O_V$ operators, and vice versa. In the LEFT, a dimension-six amplitude `$a$' could contain one dimension-six vertex, or two dimension-five vertices, and both options are shown in \cref{fig:helicity}. This latter case arises in an amplitude containing a dipole operator and a conjugate dipole operator, shown schematically as $c_{\psi X} c_{\psi X}^\dagger$, and in the first two diagrams of \cref{fig:dipoleamps}. This produces an amplitude with $(n,\sum_i h_i)=(4,0)$, and can renormalise the vector operators $O_V$.

Now if we return to case \textit{(ii)}, in which both `$a$' and `$b$' are dimension five four-point amplitudes, the relevant amplitudes are shown in the right two diagrams of \cref{fig:dipoleamps}. We see from \cref{eq:nonrenormalizationhelicity} that if both amplitudes contain $O_{\psi X}$, then the resulting `$c$' amplitude is a $(4,2)$ amplitude, so $O_S$ and $O_T$ can be renormalised in this case. Note that in all non-zero cuts the two dipole insertions must appear on different fermion lines, so the dipoles cannot have flavour indices contracted with each other when renormalising an $O_S$ or $O_T$ operator. If one amplitude contains $O_{\psi X}$ and the other $O_{\psi X}^\dagger$, then the resulting `$c$' amplitude is a $(4,0)$ amplitude, so $O_V$ can be renormalised; this is similar to our findings in case \textit{(i)}. In the case where the insertions of the dipole and its conjugate appear on different fermion lines, the helicities are such that it can only renormalise the operators $\Op^V_{uddu}$ and $\Op^{V,LR}_{ff}$ where $f \in \{e,d,u\}$. All other contributions of the dipole operators to $(4,0)$ have a flavour index contracted between the two dipole coefficients. Thus there are a limited number of flavour structures within the current-current operators which have dipole source terms in their renormalisation group equations (RGEs), as shown in \cref{tab:large_blocks_table,tab:small_blocks_table}. It is also worth highlighting here that the dipole coefficients themselves are not renormalised at one loop by the vector coefficients, but only by scalar and tensor coefficients with a mass insertion, as seen from \cref{eq:dipoleRGs}.

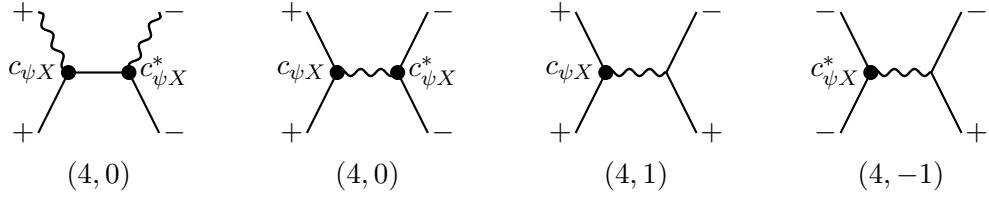
\begin{figure}
\begin{centering}
\begin{tikzpicture}[baseline=-0.65ex,scale=0.8]
 \draw[thick] (-1,-1)  -- (-\threepointsep,0) node[left] {$c_{\psi X}$} -- (\threepointsep,0) node[right] {$c_{\psi X}^*$} -- (1,-1);
  \draw[/tikzfeynhand/photon] (\threepointsep,0) -- (1,1);
  \draw[/tikzfeynhand/photon] (-\threepointsep,0) -- (-1,1);
  \node[circle,fill,inner sep=2pt] at (-\threepointsep,0) {};
  \node[circle,fill,inner sep=2pt] at (\threepointsep,0) {};
  \node at (-1.25,-1.0) {$+$};
  \node at (-1.25,1.0) {$+$};
    \node at (1.25,-1.0) {$-$};
  \node at (1.25,1.0) {$-$};
    \node at (0.0,-1.7) {$(4,0)$};
\end{tikzpicture}~~~~~~
\begin{tikzpicture}[baseline=-0.65ex,scale=0.8]
 \draw[thick] (-1,1)  -- (-\threepointsep,0) node[left] {$c_{\psi X}$}-- (-1,-1);
  \draw[thick] (1,-1) -- (\threepointsep,0) node[right] {$c_{\psi X}^*$} -- (1,1);
  \draw[/tikzfeynhand/photon] (\threepointsep,0) -- (-\threepointsep,0);
  \node[circle,fill,inner sep=2pt] at (-\threepointsep,0) {};
  \node[circle,fill,inner sep=2pt] at (\threepointsep,0) {};
  \node at (-1.25,-1.0) {$+$};
  \node at (-1.25,1.0) {$+$};
    \node at (1.25,-1.0) {$-$};
  \node at (1.25,1.0) {$-$};
    \node at (0.0,-1.7) {$(4,0)$};
\end{tikzpicture}~~~~~~
\begin{tikzpicture}[baseline=-0.65ex,scale=0.8]
 \draw[thick] (-1,1)  -- (-\threepointsep,0) node[left] {$c_{\psi X}$}-- (-1,-1);
  \draw[thick] (1,-1) -- (\threepointsep,0) -- (1,1);
  \draw[/tikzfeynhand/photon] (\threepointsep,0) -- (-\threepointsep,0);
  \node[circle,fill,inner sep=2pt] at (-\threepointsep,0) {};
  \node at (-1.25,-1.0) {$+$};
  \node at (-1.25,1.0) {$+$};
    \node at (1.25,-1.0) {$+$};
  \node at (1.25,1.0) {$-$};
 \node at (0.0,-1.7) {$(4,1)$};
\end{tikzpicture}~~~~~~
\begin{tikzpicture}[baseline=-0.65ex,scale=0.8]
 \draw[thick] (-1,1)  -- (-\threepointsep,0) node[left] {$c_{\psi X}^*$}-- (-1,-1);
  \draw[thick] (1,-1) -- (\threepointsep,0) -- (1,1);
  \draw[/tikzfeynhand/photon] (\threepointsep,0) -- (-\threepointsep,0);
  \node[circle,fill,inner sep=2pt] at (-\threepointsep,0) {};
  \node at (-1.25,-1.0) {$-$};
  \node at (-1.25,1.0) {$-$};
    \node at (1.25,-1.0) {$+$};
  \node at (1.25,1.0) {$-$};
   \node at (0.0,-1.7) {$(4,-1)$};
\end{tikzpicture}
\caption{\label{fig:dipoleamps} Four-point amplitudes built from dipole operators (drawn as black circles) and SM gauge vertices. The helicities of (ingoing) legs are shown as $+$ or $-$, and the number of legs and total helicity of the amplitudes are given in $(n,\sum h)$ form below the amplitudes.}
\end{centering}
\end{figure}

\subsection{Block diagonalisation for vector current operators\label{sec:blockdiag}}

In this Section, we block-diagonalise the anomalous dimension matrix (ADM) of the four-fermion vector operators through the flavour and parity basis of \cref{sec:decomp}. The anomalous dimension matrix is built entirely from QED and QCD interactions, and can be divided into three diagrammatic classes, which we consider explicitly at one loop.

The first class is the diagrams where the gauge boson interacts only with a single current, i.e.,
\begin{equation}
  \mathord{\begin{tikzpicture}[baseline=-0.65ex,scale=0.8]
    \draw[thick] (-1,1) -- (-\currsep,0) -- (-1,-1);
    \draw[thick] (1,1) -- (\currsep,0) node[right] {} -- (1,-1);
    \draw[/tikzfeynhand/photon] (0.85,0.81) -- (0.85,-0.81);
    \currentmarker{\currsep,0}
    \currentmarker{-\currsep,0}
  \end{tikzpicture}}+
  \mathord{\begin{tikzpicture}[baseline=-0.65ex,scale=0.8]
    \draw[thick] (-1,1) -- (-\currsep,0) -- (-1,-1);
    \draw[thick] (1,1) --  (\currsep,0) node[right] {} node[pos=0.1,inner sep=0pt] (v1) {} node[pos=0.75,inner sep=0pt] (v2) {} -- (1,-1);
    \coordinate (v3) at (0.85,0.81); 
    \coordinate (v4) at (0.36, 0.20);
        \draw[/tikzfeynhand/photon,/tikzfeynhand/half left] (v3) to (v4);
    \currentmarker{\currsep,0}
    \currentmarker{-\currsep,0}
  \end{tikzpicture}} +
  \mathord{\begin{tikzpicture}[baseline=-0.65ex,scale=0.8]
    \draw[thick] (-1,1) -- (-\currsep,0) -- (-1,-1);
    \draw[thick] (1,1) --  (\currsep,0) node[right] {} -- (1,-1) node[pos=0.4,inner sep=0pt] (v1) {} node[pos=0.8,inner sep=0pt] (v2) {};
    \coordinate (v3) at (0.85,-0.81); 
    \coordinate (v4) at (0.36, -0.20);
    \draw[/tikzfeynhand/photon,/tikzfeynhand/half right] (v3) to (v4);
        \currentmarker{\currsep,0}
    \currentmarker{-\currsep,0}
  \end{tikzpicture}}
  = 0 \, ,
\end{equation}
where each `$\mathord{\begin{tikzpicture}[baseline=-0.65ex,scale=0.8]
  \currentmarker{0,0}
\end{tikzpicture}}$' denotes a vector current. The second and third diagrams denote the wavefunction renormalisation corrections to the legs of the current. The sum of all three diagrams is zero, due to the conservation of fermion number current \cite{Collins:2005nj}.

The second class is the diagrams where the gauge boson interacts with both currents, i.e.,
\begin{equation}
\mathord{\begin{tikzpicture}[baseline=-0.65ex,scale=0.8]
  \draw[thick] (-1,1) -- (-\currsep,0) -- (-1,-1);
  \draw[thick] (1,1) -- (\currsep,0) node[right] {} -- (1,-1);
  \draw[/tikzfeynhand/photon] (-0.80,0.75) -- (0.80,0.75);
    \currentmarker{\currsep,0}
  \currentmarker{-\currsep,0}
\end{tikzpicture}}
+
\mathord{\begin{tikzpicture}[baseline=-0.65ex,scale=0.8]
  \draw[thick] (-1,1) -- (-\currsep,0) -- (-1,-1);
  \draw[thick] (1,1) -- (\currsep,0) node[right] {} -- (1,-1);
  \draw[/tikzfeynhand/photon] (0.80,-0.75) -- (-0.80,-0.75);
    \currentmarker{\currsep,0}
  \currentmarker{-\currsep,0}
\end{tikzpicture}}
+
\mathord{\begin{tikzpicture}[baseline=-0.65ex,scale=0.8]
  \draw[thick] (-1,1) -- (-\currsep,0) -- (-1,-1);
  \draw[thick] (1,1) -- (\currsep,0) node[right] {} -- (1,-1);
  \draw[/tikzfeynhand/photon,/tikzfeynhand/half left] (-0.85,0.80) to (0.813, -0.78);
      \currentmarker{\currsep,0}
  \currentmarker{-\currsep,0}
\end{tikzpicture}}
+
\mathord{\begin{tikzpicture}[baseline=-0.65ex,scale=0.8]
  \draw[thick] (-1,1) -- (-\currsep,0) -- (-1,-1);
  \draw[thick] (1,1) -- (\currsep,0) node[right] {} -- (1,-1);
  \draw[/tikzfeynhand/photon,/tikzfeynhand/half left] (0.85,0.80) to (-0.84, -0.79);
      \currentmarker{\currsep,0}
  \currentmarker{-\currsep,0}
\end{tikzpicture}}
. \label{eq:softDiagrams}
\end{equation}
Up to effects on the colour structures of the operators, these are self-renormalisation diagrams, in the sense that the gauge bosons do not change the species of the fermion in the current, nor its helicity. They also affect all flavour states equally.

The third class is the diagrams where a fermion loop is closed, e.g.,
\begin{equation}
  \mathord{\begin{tikzpicture}[baseline=-0.65ex,scale=0.8]
    \draw[thick] (-1,1) -- (-\currsep,0) -- (-1,-1);
    \draw[thick] (1,1) -- (\currsep,0) node[right] {$c^{pv}_{qw}$} -- (1,-1);
    \currentmarker{\currsep,0}
    \currentmarker{-\currsep,0}
  \end{tikzpicture}}
  \mathord{\begin{tikzpicture}[baseline=-0.65ex,scale=0.8]
    \draw[thick] (-1,1) -- (-\threepointsep,0) node[left] {$\delta^w_v$}-- (-1,-1);
    \draw[thick] (1,-1) -- (\threepointsep,0) node[right] {$\delta^r_s$} -- (1,1);
    \draw[/tikzfeynhand/photon] (\threepointsep,0) -- (-\threepointsep,0);
      \end{tikzpicture}} \propto c^{pw}_{qw} \delta^r_s
  \, . \label{eq:hardDiagrams}
\end{equation}
Such a diagram can change the fermion species, and also treats flavour states differently. The only affected irreps within the Wilson coefficients are those which contribute to the trace over the flavour indices of the fermions in the loop.

Because the neutrinos are uncharged, the anomalous dimension matrix trivially does not mix the operators with different numbers of neutrinos. The operators with two neutrino currents do not run at all. The operators with one neutrino current do not feature any diagrams of the type \cref{eq:softDiagrams}, and so the only ones that run are in a singlet in the charged fermion irrep --- because only then are the remaining diagrams of the type \cref{eq:hardDiagrams} are non-zero. Moreover, these singlet operators must be parity-even, as for a parity odd operator (proportional to $L-R$ in the charged fermion current), the contributions of the left and right-handed fermions in the closed fermion loop in \cref{eq:hardDiagrams} are equal and opposite. This leaves three real coefficients among the neutrino operators that run at all in the decomposed basis.

\begin{figure}
  \centering
  \includegraphics[width=0.5\linewidth]{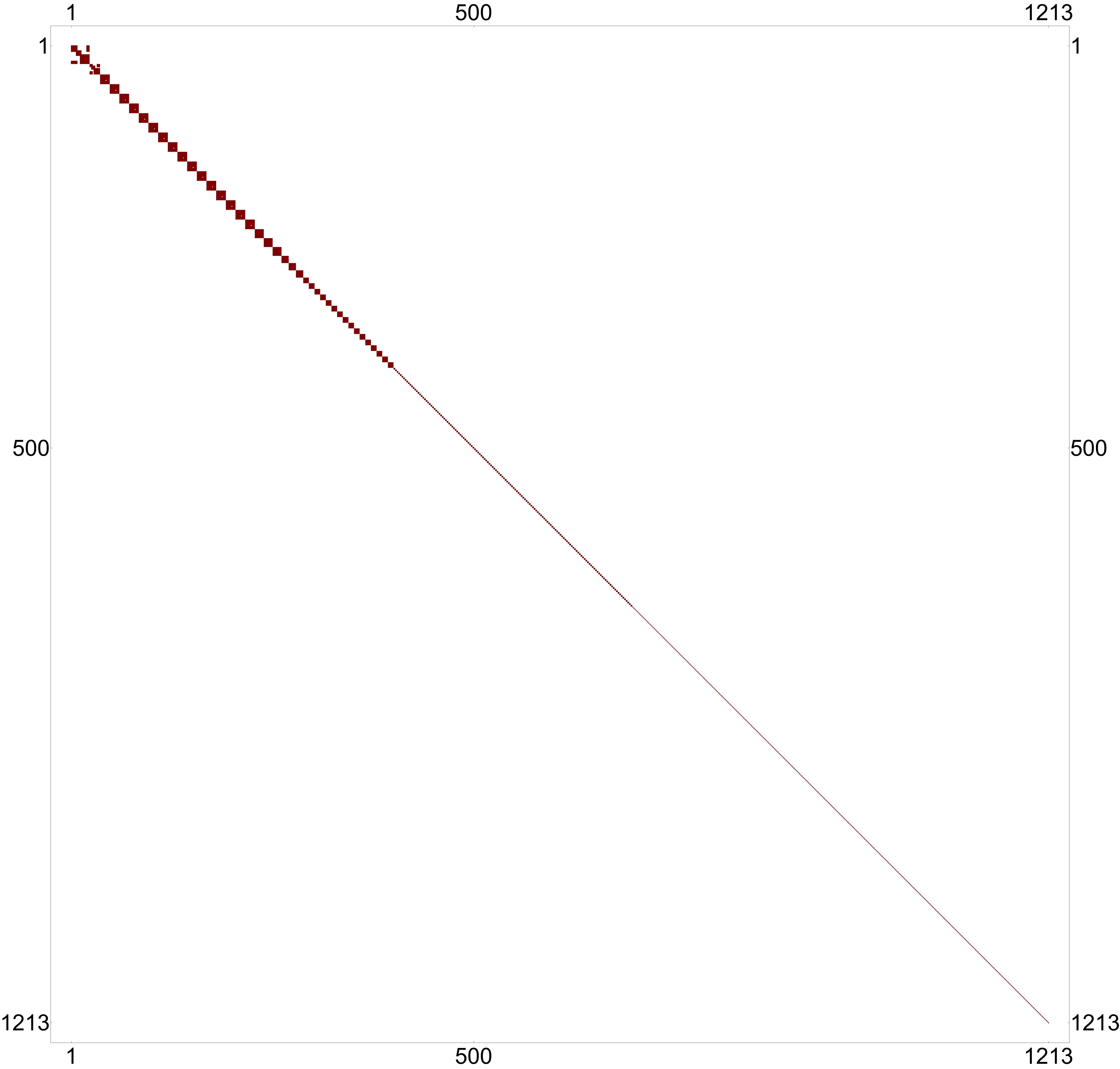}
  \caption{The ADM of the charged four-fermion vector operators in the basis of definite parity and flavour charges. Shaded blocks indicate non-zero entries. Diagonal blocks are ordered by decreasing dimensions to match \Tab{tab:large_blocks_table} followed by \Tab{tab:small_blocks_table}.}  \label{fig:full_matrix_plot}
\end{figure}

Regarding the operators built from two charged fermion currents, which comprise 1213 real d.o.f.s, a flavour and parity decomposition arranges the anomalous dimension matrix into a block diagonal form built from the blocks given in \cref{tab:large_blocks_table,tab:small_blocks_table} and shown in \cref{fig:full_matrix_plot}, of which the largest is the parity-even singlet block of 23 d.o.f.s. The smaller $2 \times 2$ and $1 \times 1$ blocks of \cref{tab:small_blocks_table} only self-renormalise (and possibly mix colour structures), as they are composed of entirely traceless flavour irreps. Moreover, they separate into blocks of operators of `type A' (built from $LL$ and $RR$ currents) and operators of `type B' (built from $LR$ and $RL$ currents), even if they have the same flavour and parity charges. This is because the graphs of the type \cref{eq:softDiagrams} do not change the chirality of the current.

All structures discussed above are, in principle, robust to higher loop order, although in practice can be spoilt by choice of scheme, see \cref{sec:importanceTwoLoop}. Explicitly at one-loop order, the anomalous dimension matrices are dense within the confines of the blocks discussed above. There are a few zeroes in the larger blocks of \cref{tab:large_blocks_table} where there is explicitly no diagram; for example, an operator built out of two down currents cannot run into one built from two lepton currents at one loop. The only zeroes not explainable in this fashion are the entries corresponding to the self-renormalisation of B-type $\Op_{ed}$ operators, which are present in both the parity-even and parity-odd $\{\db_{d} = 8, \db_{u,e}=1\}$ blocks. This is due to a cancellation between the QED diagrams of \cref{eq:hardDiagrams} and \cref{eq:softDiagrams}, which is specific to the $U(1)_Q$ charges of down-type quarks and leptons and does not occur for other combinations of fields. The zeroes discussed in this paragraph will be `filled in' at higher-order, be that in the two-loop ADM (see \cref{sec:importanceTwoLoop}) or by resumming the one-loop matrix.

\begin{table}
\centering
\begin{tabular}{|c | c | c | c | c|}
  \hline
  Quantum numbers & Size & \# of blocks & \# of zero entries & Source \\ \hline\hline
  $\db_{u,d,e} = 1, P=+1$ & $23\times23$ & $1$ & 216 & Y \\[5pt]\hline
  $\db_{u,d,e} = 1, P=-1$ & $13\times13$ & $1$ & 40 & Y \\[5pt]\hline
  \cellcolor{blue!30} $\db_{d} = 8, \db_{u,e}=1, P=+1$ & $12\times12$ & $8$ & 1 & Y \\[5pt]\hline
  \cellcolor{blue!30} $\db_{d} = 8, \db_{u,e}=1, P=-1$ & $12\times12$ & $8$ & 1 & Y \\[5pt]\hline
  $\db_{u} = 3, \db_{d,e}=1, P=-1$ & $11\times11$ & $3$ & 0 & Y \\[5pt]\hline
  $\db_{u} = 3, \db_{d,e}=1, P=+1$ & $9\times9$ & $3$ & 0 & Y \\[5pt]\hline
  $\db_{e} = 8, \db_{u,d}=1, P=+1$ & $7\times7$ & $8$ & 0 & Y \\[5pt]\hline
  $\db_{e} = 8, \db_{u,d}=1, P=-1$ & $7\times7$ & $8$ & 0 & Y \\[5pt]\hline
\end{tabular}
\caption{The largest blocks from the block-diagonalised ADM for the LEFT. $\db_f$ quantum numbers specify the dimension of the irrep under $SU(3)_f$ or $SU(2)_f$. The $P$ quantum numbers label the parity charge of coefficients. A `Y' entry in the final column denotes the presence of a source term in \cref{eq:mainRG}. Where the quantum numbers have a blue background, they appear in the decomposition of the $52 \times 52$ block based purely on Abelian charges that mediates $\Delta F_d = 1$ transitions, as discussed at the end of \cref{sec:flavour_parity_charges}.}
\label{tab:large_blocks_table}
\end{table}

\begin{table}
\centering
\scalebox{0.90}{
\begin{tabular}{|c | c | c | c | c | c|}
  \hline
  Type & Quantum numbers & Size & \# & Source & ADM \\ \hline\hline
  A & \cellcolor{blue!30} $\db_{u} = 3, \db_{d}=8, P=\pm1$ & $2\times2$ & $48$  &  & $\left(
\begin{array}{cc}
12q_uq_de^2 + \frac{16}3g^2 & \frac{8 g^2}{3} \\
-\frac{20 g^2}{3} &  12q_uq_de^2-\frac{28}{3}g^2 \\
\end{array}
\right)$ \\[5pt]\hline
 B & \cellcolor{blue!30} $\db_{u} = 3, \db_{d}=8, P=\pm1$ & $2\times2$ & $48$  &  & $\left(
\begin{array}{cc}
-12q_uq_de^2-\frac{34}{3}g^2 & \frac{10 g^2}{3} \\
\frac{56 g^2}{3} & -12q_uq_de^2-\frac{8}{3}g^2 \\
\end{array}
\right)$\\[5pt]\hline
  B & $\db_{u} = 5, P=+1$ & $2\times2$ & $5$  & Y & $\left(
\begin{array}{cc}
-12q_u^2e^2-\frac{34}{3} g^2 & \frac{10 g^2}{3} \\
\frac{56 g^2}{3} & -12q_u^2e^2-\frac{8}{3}g^2\\
\end{array}
\right)$\\[5pt]\hline
 B & \cellcolor{blue!30} $\db_{d} = 27, P=+1$ & $2\times2$ & $27$  & Y & $\left(
\begin{array}{cc}
-12q_d^2e^2-\frac{34}{3} g^2 & \frac{10 g^2}{3} \\
\frac{56 g^2}{3} &  -12q_d^2e^2-\frac{8}{3} g^2 \\
\end{array}
\right)$\\[5pt]\hline
  B & \cellcolor{blue!30} $\db_{d} = 10, P=-1$ & $2\times2$ & $10$  & Y & $\left(
\begin{array}{cc}
-12q_d^2e^2-\frac{34}{3} g^2 & \frac{10 g^2}{3} \\
\frac{56 g^2}{3} &  -12q_d^2e^2-\frac{8}{3} g^2 \\
\end{array}
\right)$\\[5pt]\hline
  B & \cellcolor{blue!30} $\db_{d} = \overline{10}, P=-1$ & $2\times2$ & $10$ & Y & $\left(
\begin{array}{cc}
-12q_d^2e^2-\frac{34}{3} g^2 & \frac{10 g^2}{3} \\
\frac{56 g^2}{3} &  -12q_d^2e^2-\frac{8}{3} g^2 \\
\end{array}
\right)$\\[5pt]\hline
  A & $\db_{u} = 3, \db_{e}=8, P=\pm1$ & $1\times1$ & $48$ &  & $\begin{array}{c}
12q_uq_ee^2 \\
\end{array}$\\[5pt]\hline
 B & $\db_{u} = 3, \db_{e}=8, P=\pm1$ & $1\times1$ & $48$ &  & $\begin{array}{c}
-12q_uq_ee^2 \\
\end{array}$\\[5pt]\hline
  A & \cellcolor{blue!30} $\db_{d} = 8, \db_{e}=8, P=\pm1$ & $1\times1$ & $128$  &  & $\begin{array}{c}
12q_dq_ee^2 \\
\end{array}$\\[5pt]\hline
  B & \cellcolor{blue!30} $\db_{d} = 8, \db_{e}=8, P=\pm1$ & $1\times1$ & $128$  &  & $\begin{array}{c}
-12q_dq_ee^2 \\
\end{array}$\\[5pt]\hline
  A & $\db_{u} = 5, P=\pm1$ & $1\times1$ & $10$  & Y & $\begin{array}{c}
12q_u^2e^2+4 g^2 \\
\end{array}$\\[5pt]\hline
 A & \cellcolor{blue!30} $\db_{d} = 27, P=\pm1$ & $1\times1$ & $54$  & Y & $
\begin{array}{c}
12q_d^2e^2+4g^2 \\
\end{array}
$\\[5pt]\hline
 A & $\db_{e} = 27, P=+1$ & $1\times1$ & $27$ & Y & $
\begin{array}{c}
12 q_e^2e^2 \\
\end{array}
$\\[5pt]\hline
  A & $\db_{e} = 27, P=-1$ & $1\times1$ & $27$ & Y & $
\begin{array}{c}
12 q_e^2e^2 \\
\end{array}
$\\[5pt]\hline
 B & $\db_{e} = 27, P=+1$ & $1\times1$ & $27$ & Y & $
\begin{array}{c}
-12 q_e^2e^2 \\
\end{array}
$\\[5pt]\hline
  B & $\db_{e} = 10, P=-1$ & $1\times1$ & $10$ & Y & $
\begin{array}{c}
-12 q_e^2e^2 \\
\end{array}
$\\[5pt]\hline
  B & $\db_{e} = \overline{10}, P=-1$ & $1\times1$ & $10$ & Y & $
\begin{array}{c}
-12 q_e^2e^2 \\
\end{array}
$\\[5pt]\hline
\end{tabular}
}
\caption{Characterisation of smaller blocks in the full block-diagonalised LEFT ADM. Notation of block quantum numbers matches \cref{tab:large_blocks_table}; however, dimensions of representations under flavour groups not explicitly written should now be understood as singlets. Blocks with `Y' entries in the source column have a non-zero source term contribution $s(t)$, as defined in \cref{eq:mainRG}. The ADM is given explicitly in terms of fermion charges ($q_d, q_u, q_e$) and the QED and QCD gauge couplings ($e, g$). $2\times2$ blocks for coloured coefficients are shown with the coefficient ordering $(c^{\cdot,+}, c^{\cdot,-})$ following the notation of \cref{eq:indistinguishable_coeff_notation}. Where the quantum numbers have a blue background, they appear in the decomposition of the $52 \times 52$ magnetic block that mediates $\Delta F_d = 1$ transitions, possibly multiple times.}
\label{tab:small_blocks_table}
\end{table}

\section{A case study of the down-type octet operators\label{sec:82block}}

Having reduced the anomalous dimension matrices into small tractable blocks via a flavour and parity decomposition, we now consider how to solve the RG flow in the LEFT and extract physically meaningful, basis-independent quantities from it. We apply this to the examples of the parity-even and parity-odd $12 \times 12$ blocks that mediate lepton flavour universal $\Delta F_d = 1$ transitions; these contain operators which are octets of down flavour ($\db_d=8$), and singlets of up and lepton flavour ($\db_{u,e}=1$). Hereafter these $12 \times 12$ blocks are referred to as the `down-octet' blocks. Without loss of generality, we consider $b \to s$ transitions here. This was first studied in its entirety in \cite{Aebischer:2017gaw} as part of the larger block constructed by Abelian charges alone; we derive further insights into the RG flow from the smaller non-Abelian blocks, shown with blue background in \cref{tab:large_blocks_table} and \cref{tab:small_blocks_table}.

\subsection{Solution of the renormalisation group equations\label{sec:solution_RGs}}

The $c_V$ RG equation can be written in terms of a vector of dimension-6 Wilson coefficients $c$, an anomalous dimension matrix $\gamma$, and a vector of source terms, $s$, whose one-loop elements are proportional to $\{e^2,g^2,eg\}\times c_{\psi X}c_{\psi X}^*$:
\begin{equation}
    \fourpi^2 \frac{\dd c}{\dd t} = \gamma(t) \,  c(t) + s(t)\, .\label{eq:mainRG}
\end{equation}
The $t$ dependence of the anomalous dimension matrix comes from that of the gauge couplings, which we parameterise up to two loop as
\begin{equation}
  \gamma(t) = e^2(t) \hat{\gamma}^{(1)}_e + g^2(t) \hat{\gamma}^{(1)}_g + \frac{e^4(t)}{\fourpi^2} \hat{\gamma}^{(2)}_{e} + \frac{e^2(t) g^2(t)}{\fourpi^2} \hat{\gamma}^{(2)}_{eg} + \frac{g^4(t)}{\fourpi^2} \hat{\gamma}^{(2)}_g + \ldots \, , \label{eq:tDependenceGamma}
\end{equation}
in terms of some constant matrices $\hat{\gamma}$ whose entries are products of order-one rational coefficients, $N_f$, $N_c$, and order-one surds arising from the Clebsch-Gordan decomposition into flavour irreps. For the largest parity-even down-octet block these matrices are given in \cref{app:onetwoloopexplicit}, which are constructed from the results of \cite{Jenkins:2017dyc,Naterop:2025cwg} with the help of \texttt{DSixTools} \cite{Celis:2017hod,Fuentes-Martin:2020zaz}.

Formally, \cref{eq:mainRG} is solved by the use of an integrating factor $U(t,t^\prime)$ to give the coefficients at the $b$-scale
\begin{equation}
    c(t_b) = U(t_b,t_W) c(t_W) + \int_{t_W}^{t_b} \dd t \, U(t_b,t) s(t) \, .
\label{eq:solution_eq}
\end{equation}
$U$ is the solution of the matrix equations
\begin{equation}
    \fourpi^2 \frac{\dd}{\dd t} U(t,t^\prime) = \gamma(t) \,  U(t,t^\prime) \, ; \quad U(t,t) = \mathbb{1} \, ; \quad U(t,t^\prime) U(t^\prime,t^{\prime\prime}) = U(t,t^{\prime\prime}) \quad \forall t,t^\prime,t^{\prime\prime} \, .
\end{equation}

\begin{figure}
  \centering
  \includegraphics[width=0.83\linewidth]{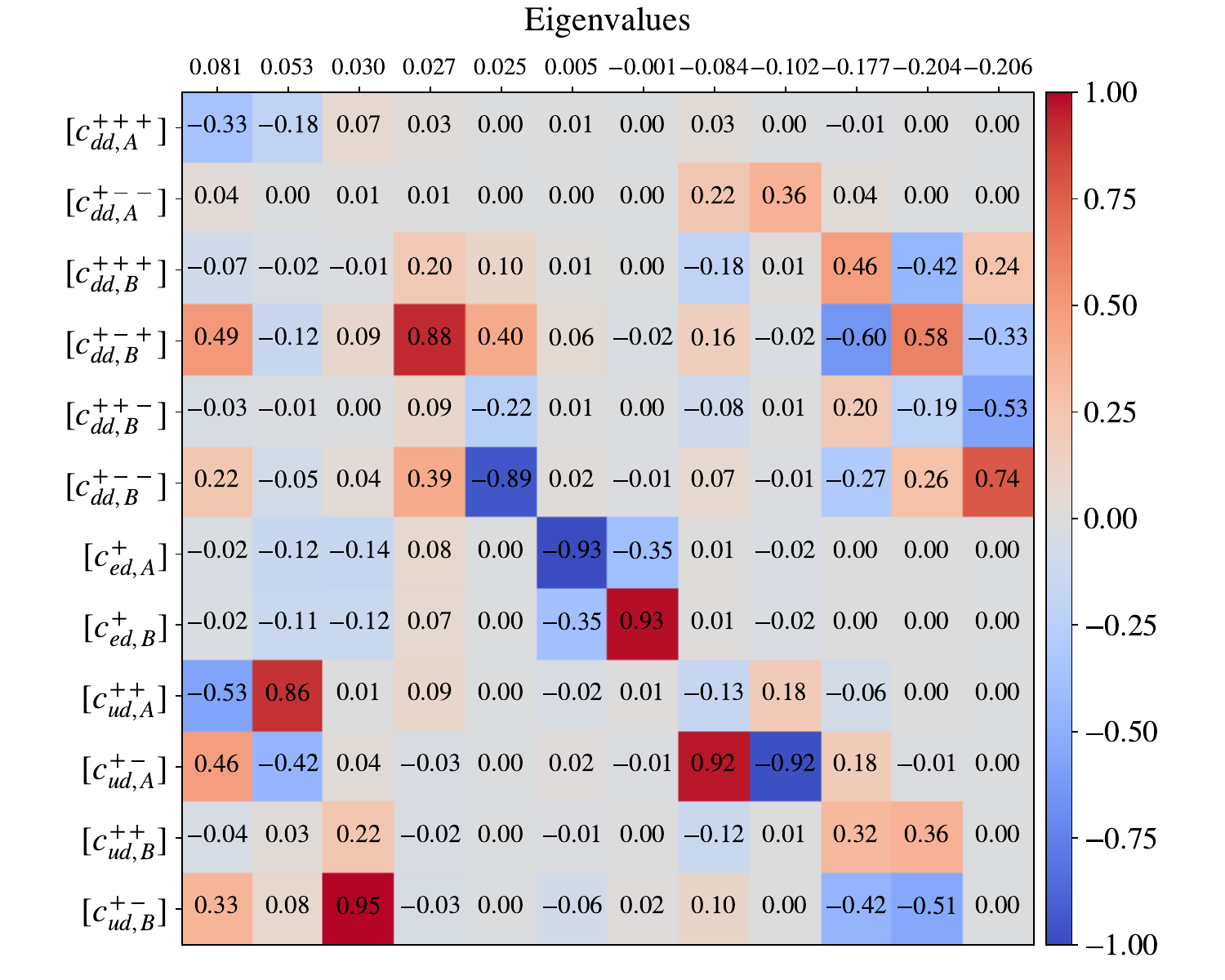}\\[-3pt]
  \includegraphics[width=0.83\linewidth]{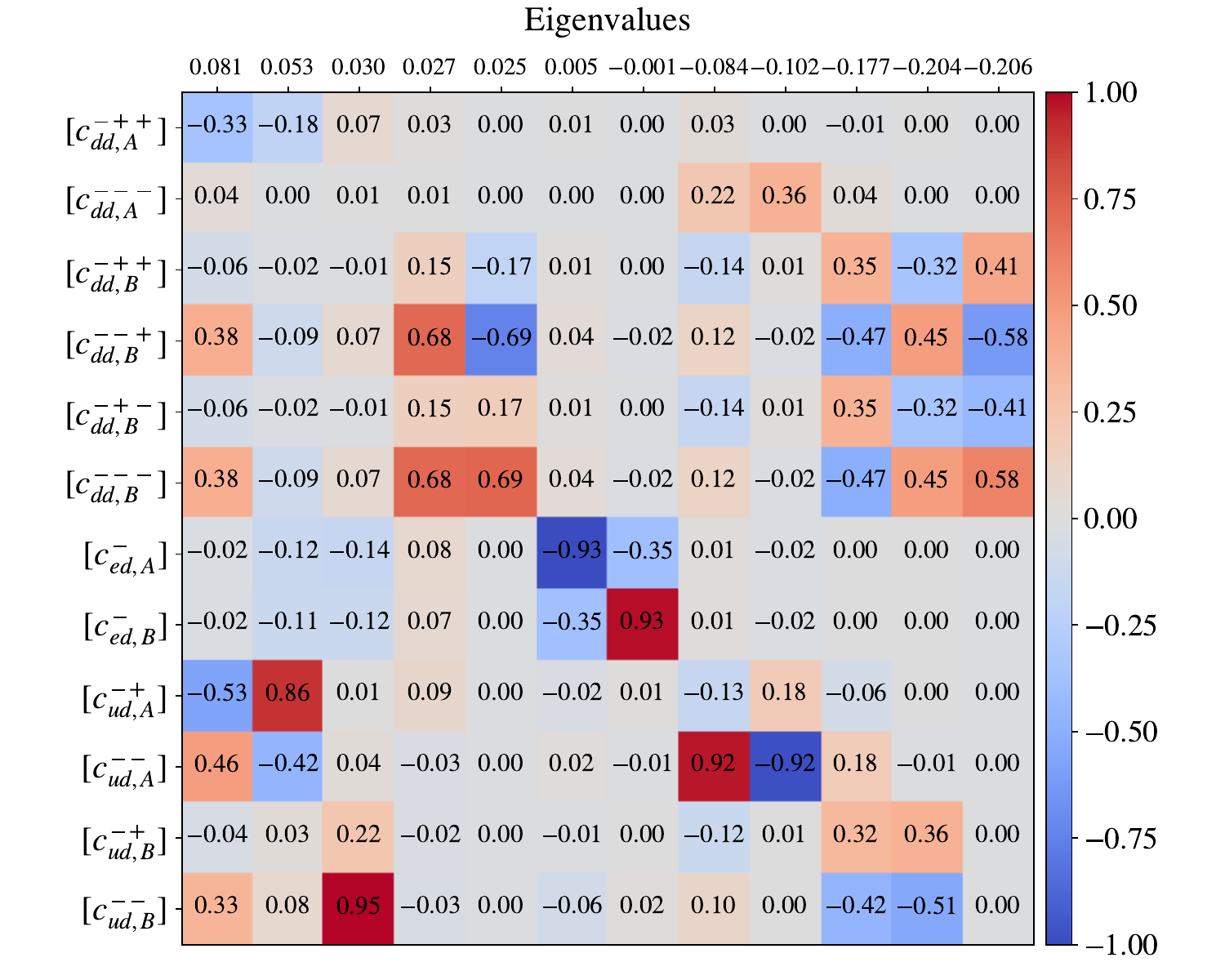}
  \caption{The eigenvectors and eigenvalues $d_i$, defined in \cref{eq:effectiveAnomalousDims}, of the parity-even (top) and parity-odd (bottom) $\db_{d} = 8, \db_{u,e}=1$ blocks. Normalised eigenvectors, in terms of their components in the Wilson coefficients listed at the side, are arranged vertically with their corresponding eigenvalues at the top.\label{fig:eigenvector_plot}}
\end{figure} 

To extract basis-independent information from the map between the $W$ and $b$ scales, $U(t_b,t_W)$, we compute its eigenvalues $u_i$:
\begin{equation}
  U(t_b,t_W) = S \cdot \mathrm{diag} \left(u_1,u_2,\ldots\right) \cdot S^{-1} \, .
\end{equation}
\cref{fig:eigenvector_plot} gives the eigenvalues for the down-octet blocks in the form of the effective anomalous dimensions
\begin{equation}
  d_i = \frac{\ln u_i}{\ln \left(\frac{\mu_b}{\mu_W}\right)} \, ,
  \label{eq:effectiveAnomalousDims}
\end{equation}
where the Wilson coefficients of the corresponding eigendirections scale as
\begin{equation}
  c_i(\mu_b) = \left(\frac{\mu_b}{\mu_W}\right)^{d_i} c_i(\mu_W) \,.
  \label{eq:eigencoefficient_scaling}
\end{equation}
\cref{fig:eigenvector_plot} also gives the entries of the matrix $S$, whose columns are the normalised eigenvectors of $U(t_b,t_W)$.

\cref{fig:eigenvector_plot} shows which directions in Wilson coefficient space are relevant ($d_i < 0$) or irrelevant ($d_i>0$) in the IR, relative to the classical expectation. The eigenvalues between the parity-even and parity-odd blocks are the same; however, the eigenvectors are configured differently among the `$B$-type' (that is, built from $LR$ and $RL$) indistinguishable current $dd$ operators. The largest eigenvalue in absolute size is $d_i = -0.2$ --- note that the eigenvalues $d_i$ can be an order-of-magnitude size larger than a typical anomalous dimension coefficient, which is of order
\begin{equation}
  \frac{\gamma^{(1)}_{jk}}{\fourpi^2} \approx \frac{g^2}{\fourpi^2} \left( N_c \text{ OR } N_f \right) \approx 0.03 \, ,
  \label{eq:one_loop_element_estimate}
\end{equation}
simply because the effect of many of these elements in the dense $12 \times 12$ matrix `add up' in some eigenvalues. Unsurprisingly, the four-quark operators run more strongly, and mix little with the $ed$ operators, whose running is a purely QED effect. However, the little mixing of the $ed$ operators with the rest prefers to make them irrelevant (which introduces a model-independent difference between operators in a leptonic singlet vs.\ non-singlet, see \cref{sec:LFUV}). Also, given the strong experimental bounds on $ed$ operators involving muons and the relatively weak bounds on some $\Delta F_d=1$ four-quark operators, the mixing of four-quark operators into $ed$ operators, although very small, can still be phenomenologically relevant (see \cref{sec:fourquarkpheno}).

Interestingly, in both parity-even and parity-odd blocks, there are also two eigendirections built entirely out of $B$-type $dd$ operators, with \emph{exactly} no admixture of $ed$ and $ud$. They are eigendirections of both the QCD and QED matrices individually, $\hat{\gamma}^{(1)}_g$ and $\hat{\gamma}^{(1)}_e$. In the parity-even block, these correspond to the San Diego basis directions
\begin{equation}
\label{eq:justdd1}
  -\frac12 [c_{dd}^{V1,LR}]^{12}_{13} = \frac13 [c_{dd}^{V1,LR}]^{21}_{13} = [c_{dd}^{V1,LR}]^{22}_{23} = [c_{dd}^{V1,LR}]^{32}_{33} \, ,
\end{equation}
with the same values given to the parity-flipped coefficients. The $d=-0.206$ ($d=0.025$) eigenvalue has the $[c_{dd}^{V8,LR}]$ coefficients $6$ ($-\frac34$) times as large as the corresponding $[c_{dd}^{V1,LR}]$. In the parity odd block, the eigendirections correspond to
\begin{equation}
\label{eq:justdd2}
  [c_{dd}^{V1,LR}]^{21}_{13} = [c_{dd}^{V1,LR}]^{22}_{23} = -[c_{dd}^{V1,LR}]^{32}_{33} \, ,
\end{equation}
with the opposite values given to the parity-flipped coefficients, and the same ratios of $V8$ to $V1$ coefficients as the parity-even case. In other words, there are four directions which will never mix into $ed$ and $ud$ in the one-loop resummed RG, despite being (na\"ively) allowed to by symmetry, although this can be explained by studying the irreps of the full flavour group \cref{eq:fullsymgroup}, see \cref{sec:fullflav}. Notably, the two different colour structures corresponding to the different eigenvalues --- $1:6$ and $4:-3$ in the $V1$ versus $V8$ operators --- arise respectively from integrating out a field with the same charges as the Higgs doublet, and a colour octet version of a Higgs doublet~\cite{deBlas:2017xtg}.

The effects of dipoles, via the second term in \cref{eq:solution_eq}, can often be safely neglected. For example, if the $c_{\psi X}$ coefficients are calculated by matching to the SM, they are GIM and loop suppressed, and rotating to the eigenbasis of the down-octet block, the additive contribution of the dipoles to the RHS of \cref{eq:eigencoefficient_scaling} is
\begin{equation}
  \frac{10^{-13}}{m_W^2} \times \left(2.1, -2.4, 
  5.5, -0.04, -2.5, 0.03, 
  0.31, -0.36, -1.9, -0.95, \
 -0.66, 0.36 \right) \, .
\end{equation}
We list the contribution to each eigencoefficient in the order the eigenvalue appears in \cref{fig:eigenvector_plot}. Beyond the SM, the dipoles are always loop suppressed for any weakly coupled new physics (see e.g.~\cite{Jenkins:2013fya,Grojean:2024tcw}).

\subsection{An efficient semi-analytic solution\label{sec:magnus}}

Given the relatively simple $t$-dependence of the anomalous dimension matrix, \cref{eq:tDependenceGamma}, it is convenient to solve for $U$ using a Magnus expansion\footnote{To write the expression compactly, we define $\int \equiv \int_{t_W}^{t_b} \dd t_1$, $\int_{t_1<t_2} \equiv \int_{t_W}^{t_b} \dd t_1 \int_{t_W}^{t_1} \dd t_2$, $\int_{t_1<t_2<t_3} \equiv \int_{t_W}^{t_b} \dd t_1 \int_{t_W}^{t_1} \dd t_2 \int_{t_W}^{t_2} \dd t_3$.}
\begin{align}
                                                \ln U(t_b,t_W) =&  \frac{1}{\fourpi^2} \int \gamma(t_1) + 
    \frac{1}{2\fourpi^4} \int_{t_1<t_2} [\gamma(t_1),\gamma(t_2)] \notag\\
    &\hspace{1ex}+ 
    \frac{1}{6\fourpi^6} \int_{t_1<t_2<t_3} \left( [\gamma(t_1),[\gamma(t_2),\gamma(t_3)]] + [\gamma(t_3),[\gamma(t_2),\gamma(t_1)]]\right) \notag\\[5pt] 
    &\hspace{1ex}+ \mathrm{O}\left(\gamma^4\right) \, .
\end{align}
Restricting to the one-loop anomalous dimension matrix, $U$ has a semi-analytic expression
\begin{align}
  \ln U(t_b,t_W) =& \, \hat{\gamma}_e \int \frac{e^2(t_1)}{\fourpi^2} + \hat{\gamma}_g \int \frac{g^2(t_1)}{\fourpi^2}
          +\frac12 [\hat{\gamma}_e ,\hat{\gamma}_g]\int_{t_1<t_2} \frac{e^2(t_1)g^2(t_2) - g^2(t_1)e^2(t_2)}{\fourpi^4}  \notag\\
        &\hspace{1ex}+ 
    \frac16 \Bigg\{ [\hat{\gamma}_e,[\hat{\gamma}_e ,\hat{\gamma}_g]] \int_{t_1<t_2<t_3} \frac{1}{\fourpi^6}  \Big( e^2(t_1) e^2(t_2) g^2(t_3) + g^2(t_1) e^2(t_2) e^2(t_3)  \notag\\
    &\hspace{7ex} - 2 e^2(t_1) g^2(t_2) e^2(t_3)\Big) + \{e \leftrightarrow g\}\Bigg\} + \mathrm{O}\left(\gamma^4\right) \, , \notag\\
    =&  -\hat{\gamma}_e \times 1.803 \times 10^{-3} - \hat{\gamma}_g \times 3.783 \times 10^{-2} -\frac12 [\hat{\gamma}_e ,\hat{\gamma}_g] \times 7.379 \times 10^{-6}   \notag\\
    &\hspace{1ex}- 
    \frac16 [\hat{\gamma}_e,[\hat{\gamma}_e ,\hat{\gamma}_g]] \times 8.70 \times 10^{-10} - 
    \frac16 [\hat{\gamma}_g,[\hat{\gamma}_g ,\hat{\gamma}_e]] \times 8.94 \times 10^{-9} \notag\\[5pt] 
    &\hspace{1ex}+ \mathrm{O}\left(\gamma^4\right)\, .
\label{eq:magnusexp}
\end{align}
In the last equality, we have numerically solved the running of the gauge couplings, and used this to evaluate the coupling integrals. To do so, we use the \texttt{DSixTools} implementation \cite{Celis:2017hod,Fuentes-Martin:2020zaz} of the one-loop beta function for $e$, and the four-loop beta function for $g$ \cite{vanRitbergen:1997va}, \emph{with all higher order Wilson coefficients $d, c_S, c_T, c_V$ turned off}.

In the Magnus expansion, the first order terms $\propto \hat{\gamma}_e$ or $\propto \hat{\gamma}_g$ resum respectively the pure QED and QCD contributions to the running of the dimension 6 coefficients. If neither $e$ nor $g$ had any scale dependence, which is not a terrible approximation between $\mu_W$ and $\mu_b$, then all remaining higher order terms would vanish. Therefore, the remaining higher order terms are suppressed by orders of magnitude, and rapidly converge on the resummation of the mixed QED/QCD terms. Compare a more typical method, where the pure QCD terms are first resummed as
\begin{equation}
    U_0(t_b,t_W) = \exp \left( \hat{\gamma}_g \int \frac{g^2(t_1)}{\fourpi^2} \right) \, ,
\end{equation}
and then the QED pieces are included as a fixed-order perturbation to approximate the full evolution matrix
\begin{equation}
    U(t_b,t_W) = U_0(t_b,t_W) + \int \frac{e^2(t_1)}{\fourpi^2} \,  U_0(t_b,t_1) \,  \hat{\gamma}_e \, U_0(t_1,t_W) + \mathrm{O}(e^4) \, .
    \label{eq:fixed_qed_order_exp}
\end{equation}
The higher order terms are less suppressed.

As the Magnus expansion also resums the pure QED pieces, we find it to be more accurate, especially for the blocks containing semi-leptonic operators where the QED running induces non-trivial effects. To illustrate this, we take the $12 \times 12$ down-octet operator block, and calculate its evolution matrix, $U_\text{num}$, fully numerically. Its eigenvalues and eigenvectors respectively populate the $u_{i,\text{num}}$ and matrix $S_\text{num}$ in the decomposition
\begin{equation}
    U_\text{num} = S_\text{num} \mathrm{diag} \left(u_1,u_2,\ldots\right)_\text{num} S_\text{num}^{-1} \, .
\end{equation}
For each approximation of the evolution matrix $U_\text{approx}$, we calculate the fractional deviation in the $i$th eigenvalue compared to the fully numerical solution,
\begin{equation}
    \delta_i = \frac{u_{i,\text{approx}} - u_{i,\text{num}}}{u_{i,\text{num}}} \, ,
    \label{eq:delta_i_definition}
\end{equation}
as well as the deviation in the alignment of the $i$th eigenvector,
\begin{equation}
    \epsilon_i = \left(S_\text{num}S_\text{approx}^{-1}\right)_{ii} - 1  \, .
    \label{eq:epsilon_i_definition}
\end{equation}

\cref{fig:approximationAccuracy} shows the accuracy of the eigenvalues and eigenvectors for the $12\times 12$ down-octet operator block at the first two fixed orders in the QED expansion, and the first 2 orders of the Magnus expansion. Perhaps unsurprisingly, the first order of the Magnus expansion, which resums the pure QCD and pure QED pieces, performs better than the fixed order approach. This gives a one-part-in-a-million accuracy, which translates into a per-mille accurate solution of the RG if mass effects are neglected. Adding the dominant mixed QED and QCD terms via the 2\textsuperscript{nd} order of the Magnus expansion has little effect on the eigenvalues (in fact it makes the accuracy of the eigenvalues slightly worse, we believe that it is reaching an accuracy comparable to our numerical solution), but does improve the eigenvectors slightly. Due to the two eigendirections built entirely from $dd$ $B$-type operators (numbers $5$ and $12$ in \cref{fig:approximationAccuracy}) being simultaneous eigendirections of both the QCD and QED anomalous dimension matrices, all higher order terms in the Magnus expansion vanish. The first-order Magnus expansion is exact for these two eigendirections, and the non-zero $\delta_i$ and $\epsilon_i$ are indicative of the size of our numerical error.

We have computed the evolution matrices from $\mu_W$ to $\mu_b$ for all blocks up to second order in the Magnus expansion; we provide these in the ancillary files in the form of replacement rules.

\begin{figure}
    \centering
    \includegraphics[height=0.25\textwidth]{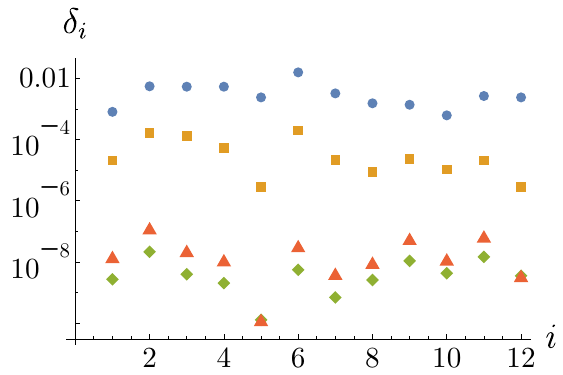}\hspace{10pt}\includegraphics[height=0.25\textwidth]{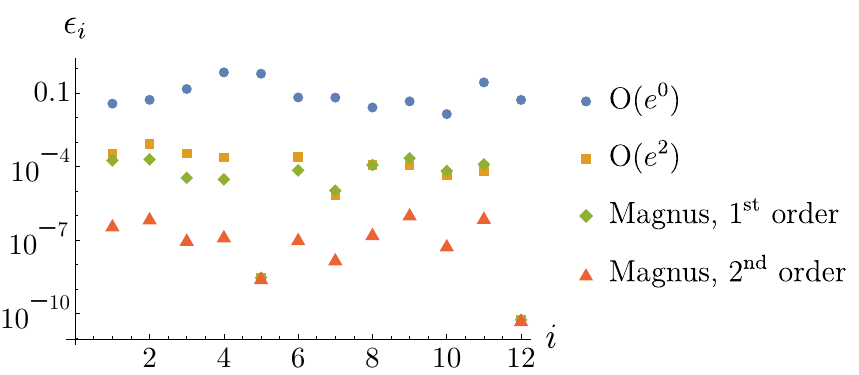}
    \caption{The fractional error $\delta_i$, \cref{eq:delta_i_definition}, in the eigenvalues of the one-loop evolution matrix (left) and the misalignment of the eigenvectors, $\epsilon_i$, \cref{eq:epsilon_i_definition}, (right), in the $12 \times 12$ down-octet block when using fixed QED order \cref{eq:fixed_qed_order_exp} and Magnus expansion \cref{eq:magnusexp} solutions.\label{fig:approximationAccuracy}}
\end{figure}

\subsection{The importance of two loop\label{sec:importanceTwoLoop}}

At one loop, the anomalous dimension matrix manifests the physical flavour and parity invariance of the gauge interactions by separating into blocks based on the flavour and parity charges of the operators, see \cref{sec:blockgamma}. Importantly, the decomposition relies on the exchange symmetry of flavour indices (e.g.\ \cref{eq:A_type_flav_sym}) that results from 4d Fierz identities.

At two loops and higher the manifest flavour and chiral symmetries in the anomalous dimension matrix can therefore be broken by the treatment of evanescent operators. Consequently, we do not necessarily expect that the two-loop matrices for the four-fermion operators reported in \cite{Buras:1991jm,Buras:1992tc,Buras:1992zv,Ciuchini:1993vr,Ciuchini:1993ks,Chetyrkin:1996vx,Chetyrkin:1997gb,Buras:2000if,Gambino:2003zm,Bobeth:2003at,Gorbahn:2004my,Huber:2005ig,Aebischer:2025hsx} preserve the flavour symmetry. We have explicitly checked that the two-loop matrix of \cite{Aebischer:2025hsx} does not have the same block-diagonal form of the one-loop matrix; it is possible that the chosen scheme violates the flavour symmetries, which could explain the lack of block diagonalisation.

However, it is possible to judiciously choose a set of finite subtractions that preserves the symmetries \cite{Naterop:2023dek,Naterop:2025lzc}. Recent work \cite{Naterop:2025cwg} uses a set of subtractions in the 't Hooft-Veltman scheme to do so, and we have explicitly checked that the reported two-loop matrix takes on the same block-diagonal form as the one-loop matrix for the four-fermion vector operators that we consider.

This two-loop matrix is effectively as dense as the one-loop matrix. The `accidental' zero in the running of the $ed$ operators in the down-octet block (see \cref{sec:blockdiag}) is filled in at two loop, as expected. In the singlet blocks, at one loop there are zeroes due to the lack of a one-loop diagram that can change the species of both currents in one operator. In the two-loop ADM these are not filled in, because they are entirely accounted for by one-loop insertions of the one-loop counterterms. At two loop in the scheme of \cite{Naterop:2025cwg}, there are also non-zero contributions to the self-renormalisation of operators with neutrinos, see \cite[\S 4.1]{Naterop:2025cwg}.

Using the eigenvalues and eigenvectors of the one-loop matrix we can estimate where two-loop running effects are significant. Approximately (i.e., to first order in the Magnus expansion) the evolution matrix is
\begin{equation}
  \frac{\ln U(t_b,t_W)}{\ln \left(\frac{\mu_b}{\mu_W}\right)} = \frac{\left\langle \gamma^{(1)} \right\rangle}{\fourpi^2}  + \frac{\left\langle \gamma^{(2)} \right\rangle}{\fourpi^4}  \, ,
  \label{eq:perturbed_evolution_matrix}
\end{equation}
where $\left\langle \gamma^{(1)} \right\rangle$ is the average of the one-loop matrix over the logarithmic scale interval,
\begin{equation}
  \left\langle \gamma^{(1)} \right\rangle = \frac{\int \dd \ln \mu \, \gamma^{(1)}}{\int \dd \ln \mu} \, ,
\end{equation}
and similarly for the two-loop matrix $\gamma^{(2)}$. The effect of the second term in \cref{eq:perturbed_evolution_matrix} is a small perturbation on the first term; we expect the approximate size of an element of the two-loop term to be \cite{Buras:1992tc}
\begin{equation}
  \frac{\left\langle \gamma^{(2)} \right\rangle_{jk}}{\fourpi^4} \approx \frac{\left\langle g^4 \right\rangle}{\fourpi^4} \times \left( N_c^2 \text{ OR } N_c N_f \right) \approx 0.001 \, ,
  \label{eq:two_loop_element_estimate}
\end{equation}
cf.\ \cref{eq:one_loop_element_estimate} for the one-loop term.

Considering only the first term, the eigenvalues and eigenvectors of \cref{eq:perturbed_evolution_matrix} are given by $d_i$ and $S_{ij}$, defined in \cref{sec:solution_RGs}. By matrix perturbation theory, the leading effect of the second term on the eigenvalues,
\begin{equation}
  \Delta d_i = S^{-1}_{ij} \frac{\left\langle \gamma^{(2)} \right\rangle_{jk}}{\fourpi^4} S_{ki} \, , \label{eq:delta_d}
\end{equation}
and eigenvectors,
\begin{equation}
  \Delta S_{ij} = \sum_{i \neq j} \frac{S^{-1}_{ik} \frac{\left\langle \gamma^{(2)} \right\rangle_{kl}}{\fourpi^4} S_{lj}}{d_j - d_i} \, . \label{eq:delta_s}
\end{equation}
For the small blocks that we consider, the size of \cref{eq:delta_d} and the numerator \cref{eq:delta_s} could easily be $\mathrm{O}(10^{-2})$, an order of magnitude higher than an individual element \cref{eq:two_loop_element_estimate}, due the neglected rational coefficients, or the `constructive interference' of many elements in the sum. This is just like how some eigenvalues of the one-loop matrix are an order of magnitude bigger than the individual elements.

Therefore, we estimate that the RG flow will be significantly affected by the two-loop contributions in particular eigendirections: one, those whose one-loop eigenvalue $d_i$ is less than $\mathrm{O}(10^{-2})$, where the comparable two-loop contribution \cref{eq:delta_d} may reverse the sign of the eigenvalue and therefore reverse the direction of the RG flow; two, sets of eigendirections whose eigenvalues differ by $\mathrm{O}(10^{-2})$ or less, and which could therefore be significantly mixed by the two-loop contribution in \cref{eq:delta_s}.

\begin{figure}
  \centering
  \includegraphics[width=0.83\textwidth]{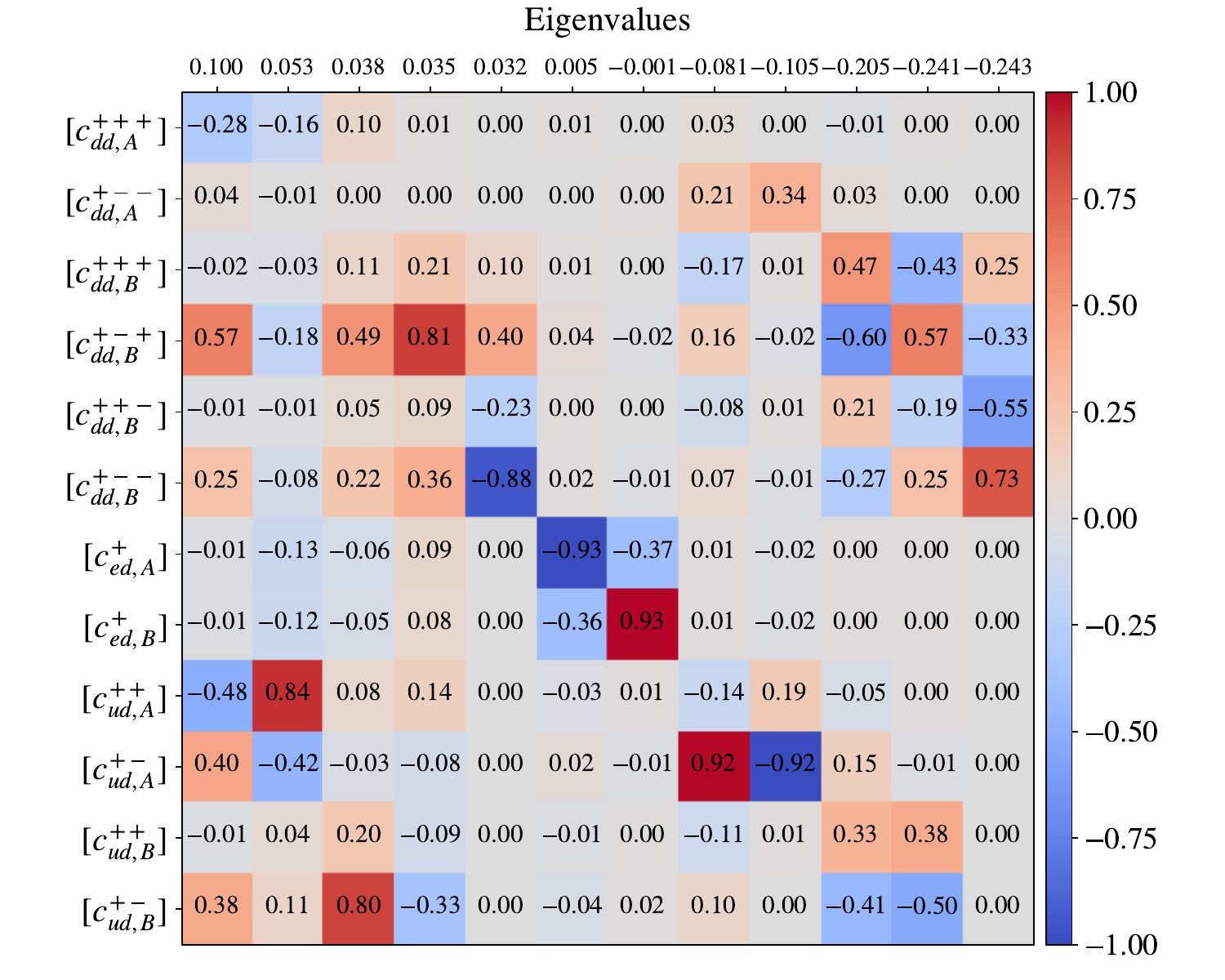}
  \caption{The eigenvectors and eigenvalues (defined in \cref{eq:effectiveAnomalousDims}) of the parity-even $\db_{d} = 8, \db_{u,e}=1$ block, including two-loop effects. Normalised eigenvectors, in terms of their components in the Wilson coefficients listed at the side, are arranged vertically with their corresponding eigenvalues at the top. Compared to the plot without two-loop effects, \cref{fig:eigenvector_plot} (top), the eigenvalues shift by up to $0.04$, and there is order-one mixing between nearly degenerate eigendirections. The $dd$-only directions are protected by symmetry.\label{fig:twoLoop82Block}}
\end{figure}

In the case of the parity-even $\Delta F_d = 1$ block, \cref{fig:twoLoop82Block} shows the eigenvectors and -values when including the two-loop pieces from \cite{Naterop:2025cwg}. In comparison to the same plot without the two-loop pieces, \cref{fig:eigenvector_plot} (top), the above rough estimates of the importance of the two-loop pieces are borne out. There are shifts in the eigenvalues up to $0.04$, or $20\%$. There is order-one mixing between the nearly degenerate eigenvectors in the 3rd and 4th columns. Note that the $dd$ only eigenvectors do not mix with adjacent eigenvectors, as they are protected by the full flavour symmetry, see \cref{sec:fullflav}.

\subsection{The full flavour symmetry\label{sec:fullflav}}

An obvious extension is to decompose the LEFT operators into irreps of the full flavour and parity group \cref{eq:fullsymgroup}. Being larger these irreps provide more symmetry relations between, e.g., coefficients of the anomalous dimension matrix.

The irreps of
\begin{equation}
  SU(N)_L \times SU(N)_R \rtimes \mathbb{Z}_2 \, ,
  \label{eq:schematicFullFlavGroup}
\end{equation}
where $\mathbb{Z}_2$ exchanges the actions of the $SU(N)$s, can be explicitly constructed analogously to the $SU(2)$ example in \cite{Gripaios:2014pqa}. For the purposes of this discussion we simply characterise the irreps by their decomposition to the parity-less subgroup $SU(N)_L \times SU(N)_R$ and the vectorial subgroup $SU(N)_V \times \mathbb{Z}_2$. We have effectively been using the vectorial subgroup in this paper.

For any irrep $\mathbf{A}$ of $SU(N)$, there will be two distinct irreps of \cref{eq:schematicFullFlavGroup} with dimension $A^2$. The two irreps are differentiated by a $\pm$, and decompose as
\begin{align}
  \mathbf{A^2}^\pm \overset{SU(N)_L \times SU(N)_R}{\longrightarrow}& \,\, \mathbf{A}_L \otimes \mathbf{A}_R \, , \\
  \overset{SU(N)_V \times \mathbb{Z}_2}{\longrightarrow}& \left( \mathbf{A}_V \otimes \mathbf{A}_V \right)^\pm \, ,
\end{align}
where the $\pm$ on the RHS gives the sign of the $\mathbb{Z}_2$ irrep. For any two \emph{inequivalent} irreps $\mathbf{A}$ and $\mathbf{B}$ of $SU(N)$, there will be an irrep of dimension $2AB$ that decomposes as
\begin{align}
  \mathbf{2AB} \overset{SU(N)_L \times SU(N)_R}{\longrightarrow}& \,\, \left(\mathbf{A}_L \otimes \mathbf{B}_R \right) \otimes \left( \mathbf{A}_R \otimes \mathbf{B}_L \right) \, , \\
  \overset{SU(N)_V \times \mathbb{Z}_2}{\longrightarrow}& \left( \mathbf{A}_V \otimes \mathbf{B}_V \right)^+ \otimes \left( \mathbf{A}_V \otimes \mathbf{B}_V \right)^- \, ,
\end{align}
The irreps of the full symmetry group will decompose the anomalous dimension matrices into smaller blocks by relating the many repeated entries in the parity-even and parity odd blocks in this paper.

In particular, the only four-fermion vector current operators that can populate the $\mathbf{A^2}^\pm$ irreps are indistinguishable currents of the form $\Op_{ff}^{V,LR}$. The eigendirections built purely out of $\Op_{dd}^{V,LR}$ in \cref{sec:solution_RGs} inhabit these irreps of the full flavour group, and therefore cannot mix with the other operators in the $12\times 12$ down-octet block, which are in different irreps.

\section{Phenomenological implications for BSM\label{sec:pheno}}

In this Section we make use of the results of previous sections to identify phenomenological messages applicable to flavoured heavy new physics. The non-Abelian nature of the lepton flavour symmetry explicitly separates lepton flavour universal operators (within singlet irreps of the lepton flavour group) from traceless lepton flavour non-universal operators (within higher irreps of the lepton flavour group), and these pieces run independently of each other. This allows us to study the radiative stability of lepton flavour non-universality in different operator types, which we do in the next subsection, before applying this to the phenomenology of $b\to s \tau \tau$ decays in \cref{sec:taupheno}. In \cref{sec:fourquarkpheno} we investigate indirect constraints on four-quark operators in the Standard Model Effective Field theory (SMEFT) from $b\to s ll$ processes which lie within the same flavour and parity blocks.

\subsection{General scaling of lepton non-universality}
\label{sec:LFUV}

Different flavour and parity irreps delineate phenomenologically significant sets of Wilson coefficients. As a simple example, lepton flavour singlets correspond to lepton flavour universal effects, and all other irreps to the non-universal effects. Moreover, different flavour and parity irreps run differently. For example, because of their non-trivial flavour charges, some operators generating lepton flavour non-universality simply do not mix among lepton generations below the electroweak scale at all: notably operators mediating $d_i \to u_j l \nu$ processes (tested in e.g.~$R(D)$ and $R(D)^*$) and $l_i \to l_j \nu_i \bar \nu_j$ (tested in e.g.~$\Gamma (\tau \to e \mu \mu)/\Gamma(\mu \to e \nu\nu)$). However, the irreps involved in measurements of $\Delta F_d=1$ decays involving charged leptons, such as $d_i \to d_j l^+_k l^-_k$, do mix in the LEFT. This makes the degree of non-universality scale dependent~\cite{Crivellin:2018yvo, Cornella:2021sby,Alguero:2022wkd}, which we classify in generality in this section using the results of \cref{sec:blockgamma}.

In the San Diego basis, and neglecting four-quark operators, the one-loop RGEs for $d_i \to d_j l^+_k l^-_k$ operators (with $i\neq j$) are \cite{Jenkins:2017dyc}
\begin{align}
     \fourpi^2 \frac{\dd [c_{ed}^{V,RR}]^{e_k d_i}_{e_k d_j} }{\dd\ln \mu} &= \frac{4}{3}e^2 q_e^2 \sum_{m}\left( [c_{ed}^{V,RR}]^{e_m d_i}_{e_m d_j} +[c_{ed}^{V,LR}]^{e_m d_i}_{e_m d_j}\right) + 12e^2q_e q_d [c_{ed}^{V,RR}]^{e_k d_i}_{e_k d_j}\,,\label{eq:LFUVrunningLEFT1}\\
     \fourpi^2 \frac{\dd [c_{ed}^{V,LR}]^{e_k d_i}_{e_k d_j} }{\dd\ln \mu} &= \frac{4}{3}e^2 q_e^2 \sum_{m}\left( [c_{ed}^{V,RR}]^{e_m d_i}_{e_m d_j} +[c_{ed}^{V,LR}]^{e_m d_i}_{e_m d_j}\right) - 12e^2q_e q_d [c_{ed}^{V,LR}]^{e_k d_i}_{e_k d_j}\,,\label{eq:LFUVrunningLEFT2}
\end{align}
and similar for the other quark chirality (with $R\leftrightarrow L$). If new physics contributes only to a lepton octet (i.e. traceless) flavour structure, then the sum over $m$ on the right-hand side of both equations is zero. In this case, each operator only self-renormalises. Instead, if new physics contains a lepton singlet piece, then the sum on the right-hand side contributes a flavour universal piece which is vectorial in the lepton current. 
In addition to lepton flavour universal theories that would only populate the lepton singlet components in the decomposition, we can identify possible UV completions which would only populate the lepton octet components: for example models based on the anomaly-free $U(1)$ gauge groups $L_\mu-L_\tau$, $L_e-L_\mu$ or $L_e-L_\tau$. However, any new physics which couples only to a single generation will contribute amounts of the same order to the octet structure and the singlet structure, and therefore will unavoidably generate effects in the other generations at a different scale. 

In the language of eigendirections of the ADMs, this can be understood by the fact that the singlet and octet pieces have different anomalous dimensions. In the basis of flavour and parity eigenstates of \cref{sec:decomp}, the $\Delta F=1$ $ed$ operators are decomposed in terms of the 12 coefficients
\begin{equation}
  [c_{ed,A/B}^\pm]_{\{ \mathbf{8} , 4 \}_e} \, , \,\, [c_{ed,A/B}^\pm]_{\{ \mathbf{8} , 6\}_e} \, , \,\, [c_{ed,A/B}^\pm]_{\{ \mathbf{1} , 1 \}_e} \, ,
\end{equation}
where we suppress the label for the $SU(3)_d$ charges, which is uniformly $\{ \mathbf{8} , n \}_d$ for some $n$, fixed by the flavours of the down quarks, $i,j$. These 12 coefficients are \emph{almost} an RG invariant subset. The octet coefficients are already eigendirections of the running. The singlet coefficients mix via QED diagrams with four-quark operators outside the stated subset, but only by a rather small amount: the two central eigendirections in \cref{fig:eigenvector_plot} only contain percent-level admixtures of the four-quark operators, meaning that a configuration with only $ed$ operators turned on at $\mu_W$ is still mostly $ed$ operators at the $\mu_b$ scale.
Ignoring the admixtures with four-quark operators, the approximate eigendirections of the running among the singlet pieces are
\begin{equation}
  \begin{pmatrix}
    [c_{ed,a}^\pm]_{\{ \mathbf{1} , 1 \}_e} \\ [c_{ed,b}^\pm]_{\{ \mathbf{1} , 1 \}_e}
  \end{pmatrix}
  =
  \begin{pmatrix}
    -c_\theta & -s_\theta \\
    -s_\theta & c_\theta
  \end{pmatrix}
  \begin{pmatrix}
    [c_{ed,A}^\pm]_{\{ \mathbf{1} , 1 \}_e} \\ [c_{ed,B}^\pm]_{\{ \mathbf{1} , 1 \}_e}
  \end{pmatrix} \, ,
\end{equation}
where the coefficients of the rotation matrix, $c_\theta = 0.93, s_\theta = 0.35$, can be read from \cref{fig:eigenvector_plot}.

For these eigendirections, the corresponding anomalous dimensions\footnote{as defined by the appropriate coefficient of the logarithm of the evolution matrix, see \cref{eq:effectiveAnomalousDims}} are given by
\begin{align}
   d_{\boldsymbol{8}A}&=- d_{\boldsymbol{8}B}=\frac{3q_d q_e \left\langle e^2 \right\rangle}{4\pi^2}=2.4 \times 10^{-3}; ~~d_{\boldsymbol{1}a}=5.3 \times 10^{-3}; ~~d_{\boldsymbol{1}b}=-1.1 \times 10^{-3}.
   \label{eq:bslleigenvalues}
\end{align}
The anomalous dimensions of the octet depend on the chirality structure of the operator and can be read from \cref{tab:small_blocks_table}. The anomalous dimensions of the singlet pieces are read from the eigenvalues of the mostly semileptonic eigenvectors in \cref{fig:eigenvector_plot}.

Without doing the full calculation, if the Wilson coefficient at the electroweak scale is purely within a single lepton flavour $c_{ed}^{l_i}$, we can generally expect that the mixing into the other flavours should go with the difference of singlet vs octet eigenvalues. Schematically: 
\begin{equation}
    c_{ed}^U(\mu_b) \sim {\Delta d} \log \left(\frac{\mu_b}{\mu_W} \right) c_{ed}^{l_i} (\mu_W) \gtrsim O(\Delta d)\,c_{ed}^{l_i}(\mu_b) .
    \label{eq:roughLFUV}
\end{equation}
From the eigenvalues in \cref{eq:bslleigenvalues}, we expect $\Delta d$ of order $10^{-3}$. Therefore a hierarchy between the coefficients of different lepton flavours of $O(10^{3})$ or above will not be radiatively stable, and will be diluted down to below the $O(10^3)$ level by running, assuming that the scale of new physics is near or above the electroweak scale. This is of course still a very large degree of lepton non-universality, but given the current hierarchies in precision between measurements of meson branching ratios involving different lepton flavours, this theoretical upper limit can still provide phenomenological information, see \cref{sec:taupheno}. Performing full running calculations, using the Magnus expansion method outlined in \cref{sec:magnus}, confirms the naive expectation of \cref{eq:roughLFUV}. For example, if we assume that only $[c_{e d}^{V,LL}]^{\tau d_i}_{\tau d_j}$ is non-zero at the scale $\mu_W$, then we find the following relation between coefficients at the scale $\mu_b$, where here $l=e,\mu$:
\begin{equation}
    [c_{e d}^{V,LL}]^{l d_i}_{l d_j}(\mu_b) = [c_{de}^{V,LR}]^{d_i l}_{ d_j l}(\mu_b) =-2.3\times 10^{-3}[c_{e d}^{V,LL}]^{\tau d_i}_{\tau d_j}(\mu_b)
\end{equation}
The situation for $u_i u_j l l$ operators is similar. The eigenvalues are instead 
\begin{align}
   &d_{\boldsymbol{3}A}^\prime =- d_{\boldsymbol{3}B}^\prime = \frac{3q_u q_e \left\langle e^2 \right\rangle}{4\pi^2}=-4.9 \times 10^{-3}; ~~d_{\boldsymbol{1}a}^\prime=7.3 \times 10^{-3}; ~~d_{\boldsymbol{1}b}^\prime=-3.2 \times 10^{-3},
   \label{eq:culleigenvalues}
\end{align}
meaning that we again expect that hierarchies above $O(10^3)$ in lepton flavours should not be radiatively stable. 

In summary, absent per-mille level fine-tuning at the electroweak scale, we do not expect heavy new physics to generate hierarchies larger than $O(10^3)$ between different lepton flavours in $d_id_jll$ or $u_iu_jll$ operators at the $b$ mass scale (or below). There are a couple of notable exceptions which were alluded to earlier:
\begin{itemize}
    \item \textit{Flavour-stabilised exception:} Any operator which lies entirely within the octet irrep for the leptons will be prevented by flavour symmetry from mixing into the singlet, and therefore will have a radiatively stable flavour structure. This applies to all coefficients which are traceless in the leptons, for example those arising from $L_i-L_j$ gauge models~\cite{Altmannshofer:2014cfa}. It is clear however that this exception cannot apply to new physics which is specific to just one lepton flavour, for example new physics coupling only to the third generation, since the traceless condition relies on the sum of contributions from two or more flavours. Experimentally, $\mu/e$ universality in $d_id_jll$ is relatively well established, and only $\tau$-containing observables have loose enough bounds that we could anticipate orders of magnitude enhancements over the SM prediction. So any non-universality which enhances two flavours at once over the third, although it could be radiatively stable, is not a phenomenologically viable option for achieving large degrees of lepton non-universality in $d_i\to d_j ll$ decays, since it would necessarily imply non-universality among the light leptons.
        \item \textit{Parity-stabilised exception:} Highly suppressed mixing into other flavours can happen for Wilson coefficients with an axial-vector structure in the lepton current, e.g. if $[c_{ed}^{V,RR}]^{e_k d_i}_{e_k d_j}=-[c_{ed}^{V,LR}]^{e_k d_i}_{e_k d_j}$. In this case the flavour universal pieces on the right-hand side of \cref{eq:LFUVrunningLEFT1,eq:LFUVrunningLEFT2} cancel.\footnote{The effect of resummation means that the mixing into the lepton-universal vectorial piece is not identically zero however, due to QED mixing into and then out of the $dd$ operators (which are not shown in the simplified RG equations \cref{eq:LFUVrunningLEFT1,eq:LFUVrunningLEFT2}). Among the two down-type currents in the $dd$ operators, there is a non-trivial interplay between flavour and parity for these operators which means that the parity-odd and parity-even eigenvectors are not identical, see \cref{fig:eigenvector_plot}.} This provides a genuine loophole in the above argument, but it is worth mentioning that this is not usually achieved by models of BSM physics arising from above the electroweak scale, most of which couple to particular chiralities of lepton. We comment more on this possibility, and its phrasing within the SMEFT, in the next subsection.
\end{itemize}

\subsection{Implications for tauonic heavy new physics}
\label{sec:taupheno}

New physics at the TeV scale may couple predominantly to the third generation of quarks and leptons. This is motivated by the lack of clear signs of new physics at the LHC --- which excludes flavour universal new physics to higher scales than if it is third generation-specific~\cite{Allwicher:2023shc} --- and by puzzles such as the hierarchy problem, the flavour problem, and experimental anomalies in $R_D^{(*)}$ observables, all of which are suggestive of a special role for the third generation.
Decays involving tau leptons, such as $B_s\to \tau \tau$ and $B\to K^{(*)}\tau\tau$ would be highly sensitive to these scenarios.
The precision of current and near-future projected limits on such branching ratios are however still several orders of magnitude above SM predictions, because of the experimental difficulties associated with tauonic final states, and are significantly less precise than corresponding measurements of semileptonic decays involving muons and electrons.

Here, we consider new physics coupled entirely to tauonic SMEFT operators at the EW scale, and investigate their inevitable mixing into lepton flavour-universal operators.
By starting from this maximal LFUV case, we can then set indirect upper bounds on the effects of new physics on branching ratios for processes involving $b\to s \tau^+ \tau^-$ transitions. Of course, tauonic heavy new physics will also undergo some running from the new physics scale down to the electroweak scale, which will work in the same direction to reduce the degree of LFUV, even before the matching onto the LEFT. In this sense, our approach is conservative, in that the true size of effects in flavour universal $b\to s ll$ processes will generally be larger than our estimates (always assuming no fine-tuned cancellations between tree level and RGE-induced contributions). This generation of flavour-universal effects from tauonic new physics has been demonstrated and studied previously in the context of combined explanations of $B$ anomalies~\cite{Crivellin:2018yvo, Cornella:2021sby,Alguero:2022wkd}. Here we take a more general view, with the aim of understanding what more we can hope to learn about tauonic new physics in $B$ decays at current and future experiments, regardless of UV completion.

Predictions for $b\to s ll$ decay observables are traditionally written in terms of the coefficients of the effective Hamiltonian
\begin{equation}
	\mathcal{H}_{\rm eff} \supset - \frac{4 G_F}{\sqrt{2}} \frac{\alpha}{4 \pi} V_{ts}^* V_{tb} \sum_i \mathcal{C}_i \Op_{i}, \label{eq:effHambsll}
\end{equation}
with
\begin{align}
	\Op_9^{(\prime)\ell} &= (\bar s \gamma_\alpha P_{L (R)} b) (\bar \ell \gamma^\alpha \ell), \\ 
	\Op_{10}^{(\prime)\ell} &= (\bar s \gamma_\alpha P_{L (R)} b) (\bar \ell \gamma^\alpha \gamma_5 \ell).
\end{align}
We express new physics at $\mu_W$ in terms of independent SMEFT operator coefficients $[\cC_{ld}]_{\tau s}^{\tau b}$, $[\cC_{ed}]_{\tau s}^{\tau b}$, $[\cC_{qe}]_{s\tau }^{b\tau }$, $[\cC_{lq}^{(1)}]_{\tau s}^{\tau b}$ and $[\cC_{lq}^{(3)}]_{\tau s}^{\tau b}$. Matching onto the effective Hamiltonian above, the BSM contributions to the tauonic Wilson coefficients are (where $\mathcal{C}_9^\ell = \mathcal{C}^\text{SM}_9+ \Delta\mathcal{C}_9^\ell$, etc)\footnote{More details of this matching and of the relations between the various operator bases can be found in \cref{sec:basisbsll}.}
\begin{align}
    \Delta\mathcal{C}_{9}^{\tau}(\mu_W)&=\frac{\pi v^2}{\alpha V_{ts}^*V_{tb}} \left( [\cC_{qe}]_{s\tau}^{b\tau}+[\cC^{(1)}_{lq}]_{\tau s}^{\tau b}+[\cC^{(3)}_{lq}]_{\tau s}^{\tau b} \right), \\
    \Delta\mathcal{C}_{10}^{\tau}(\mu_W)&=\frac{\pi v^2}{\alpha V_{ts}^*V_{tb}} \left( [\cC_{qe}]_{s\tau}^{b\tau}-[\cC^{(1)}_{lq}]_{\tau s}^{\tau b}-[\cC^{(3)}_{lq}]_{\tau s}^{\tau b} \right), \\
    \Delta\mathcal{C}_{9}^{\prime \tau}(\mu_W)&=\frac{\pi v^2}{\alpha V_{ts}^*V_{tb}} \left( [\cC_{ed}]_{\tau s}^{\tau b}+[\cC_{ld}]_{\tau s}^{\tau b} \right), \\
    \Delta\mathcal{C}_{10}^{\prime \tau}(\mu_W)&=\frac{\pi v^2}{\alpha V_{ts}^*V_{tb}} \left( [\cC_{ed}]_{\tau s}^{\tau b}-[\cC_{ld}]_{\tau s}^{\tau b} \right).
\end{align}
 The running from $\mu_W$ to $\mu_b$ then generates non-zero lepton flavour universal contributions, which we denote as $\mathcal{C}_i^U$. These are given by:
\begin{align}
\Delta\mathcal{C}_{9}^{U}(\mu_b) &= v^2( 47.4[\cC_{qe}]_{s\tau}^{b\tau}+47.1([\cC^{(1)}_{lq}]_{\tau s}^{\tau b}+[\cC^{(3)}_{lq}]_{\tau s}^{\tau b})), \label{eq:C9 equation}\\
\Delta\mathcal{C}_{10}^{U}(\mu_b) &= v^2(0.17[\cC_{qe}]_{s\tau}^{b\tau}+0.17([\cC^{(1)}_{lq}]_{\tau s}^{\tau b}+[\cC^{(3)}_{lq}]_{\tau s}^{\tau b})),\\
\Delta\mathcal{C}_{9}^{\prime U}(\mu_b) &= v^2(47.1[\cC_{ed}]_{\tau s}^{\tau b}+47.4[\cC_{ld}]_{\tau s}^{\tau b}),\\
\Delta\mathcal{C}_{10}^{\prime U}(\mu_b) &= -v^2(0.17[\cC_{ed}]_{\tau s}^{\tau b}+0.17[\cC_{ld}]_{\tau s}^{\tau b}) \label{eq:C10' equation},
\end{align}
where the tauonic SMEFT coefficients on the RHS are evaluated at the EW scale, and we use the second-order Magnus expansion to calculate the running.\footnote{We find very small but non-zero contributions to $\Delta\mathcal{C}_{10}^{U(\prime)}(\mu_b)$. These contributions arise from resummation of higher-loop effects from running into the $dd$ operators and back into $ed$. Specifically, they occur because the parity odd and parity-even parts of the ADM are not identical for $\Delta F=1$ $dd$ operators, see \cref{fig:eigenvector_plot} and discussions in \cref{sec:82block}.} Note that the total value of the $\cC_{9,10}^{\ell (\prime)}$ Wilson coefficients include also (flavour-universal) SM contributions, plus, for $\ell=\tau$, the tree-level matched contributions above. As expected from the results of the previous subsection, the hierarchy between tauonic and muonic effects at $\mu_b$ is $O(10^{3})$.  Under the assumption that new physics is entirely tauonic at the electroweak scale, it is therefore possible to relate the tauonic and light lepton Wilson coefficients at $\mu_b$, and hence interpret constraints from light leptons in the tauonic parameter space. To do this, we take the constraint on $\cC_9$ from \cite{Allwicher:2024ncl} (which is based on the analyses of \cite{Alguero:2022wkd,Bordone:2024hui,Isidori:2024lng}), while for $\cC_9^\prime$
we use \texttt{flavio (v2.5)} \cite{Straub:2018kue} to perform a global fit to all implemented $b\to s ll$ observables, obtaining the allowed regions:
\begin{equation}
    \label{eq:c9c9pregions}
    \Delta\cC_9^{U}(\mu_b)=-0.6\pm 0.2, ~~\Delta\cC_9^{\prime U}(\mu_b) =0.16\pm 0.23.
\end{equation}
The $\Delta\cC_9^{U}$ region does not contain zero at $2\sigma$; this is due to a pattern of discrepancies from SM predictions in $b\to s \mu \mu$ modes (see, e.g., \cite{Greljo:2022jac} and references therein). If confirmed, these could be explained either by new physics coupled to light leptons (i.e.~an additional small contribution on top of our tauonic assumptions), or the effect of the RGEs could generate them from purely tauonic NP (e.g., \cite{Crivellin:2018yvo,Cornella:2021sby,Alguero:2022wkd}). We remain agnostic about this point, taking the view that untuned tauonic new physics is constrained by the requirement not to generate too large an effect here, but that it is not required to fit the non-zero central value for $\Delta\cC_9^{\mu,e}$, which could instead be explained by a small direct coupling to muons (and electrons). 

The constraints from \cref{eq:c9c9pregions} are shown in planes of $\Delta\cC_{9,10}^{(\prime) \tau}$ in \cref{fig:c9c10taucomparison,fig:c9c10ptaucomparison}, along with current and projected constraints (listed in \cref{tab:bstautauobs}). We see that the requirement to not generate too large an effect in $\Delta\cC_9^{(\prime)U}$ already constrains $\Delta\cC_9^{(\prime)\tau}$ to smaller values than will be tested directly in the near future. On the other hand, the projected sensitivity of $\tau$ measurements at a Tera-Z run at FCC-ee \cite{Li:2020bvr,Miralles:2024,Allwicher:2025bub} will provide a stringent test of currently allowed parameter space; and, if the $b\to s l l$ fit persists at its current central value, will settle whether these deviations are due to tauonic heavy new physics.

\begin{figure}
    \centering
    \includegraphics[width=0.42\linewidth]{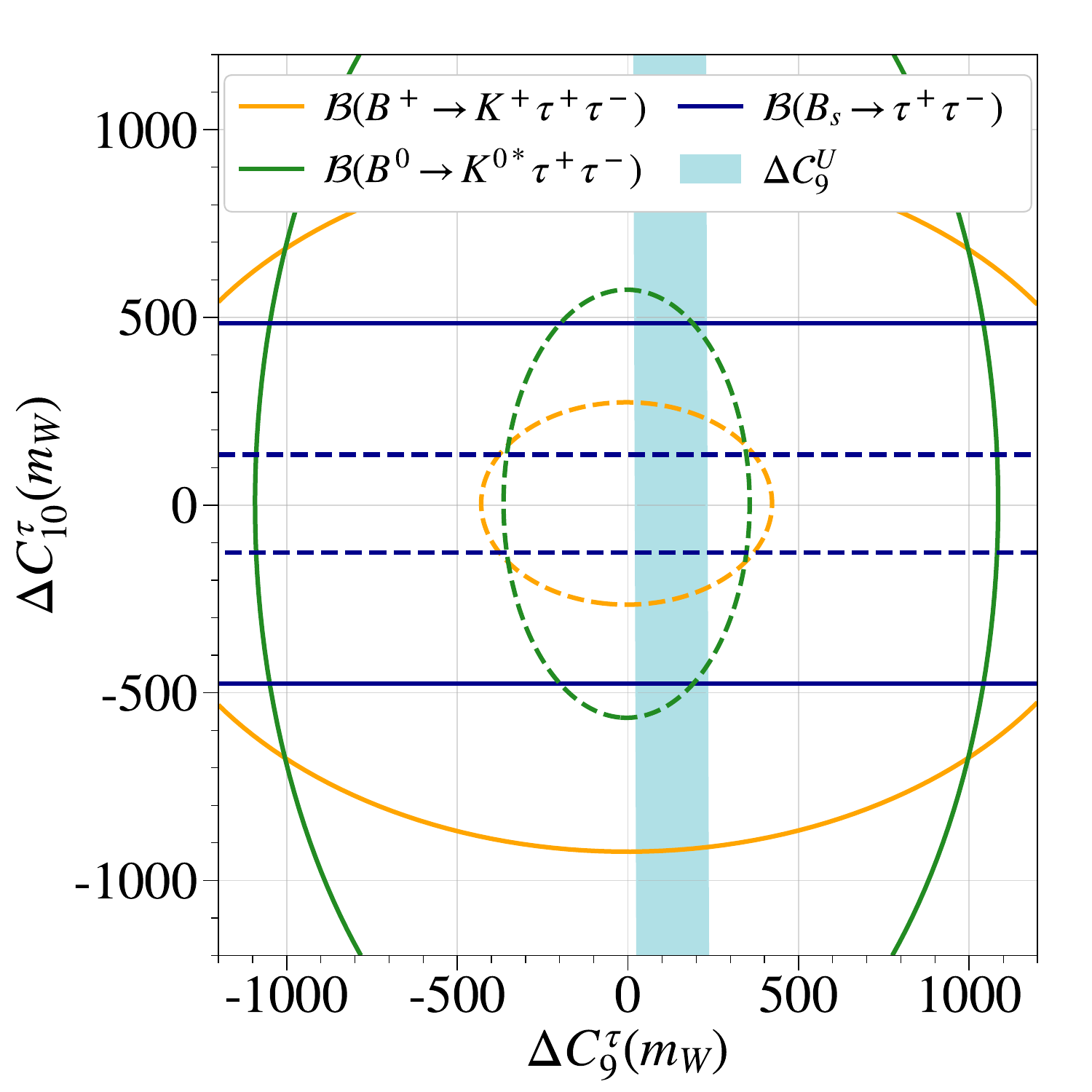}~~~\includegraphics[width=0.42\linewidth]{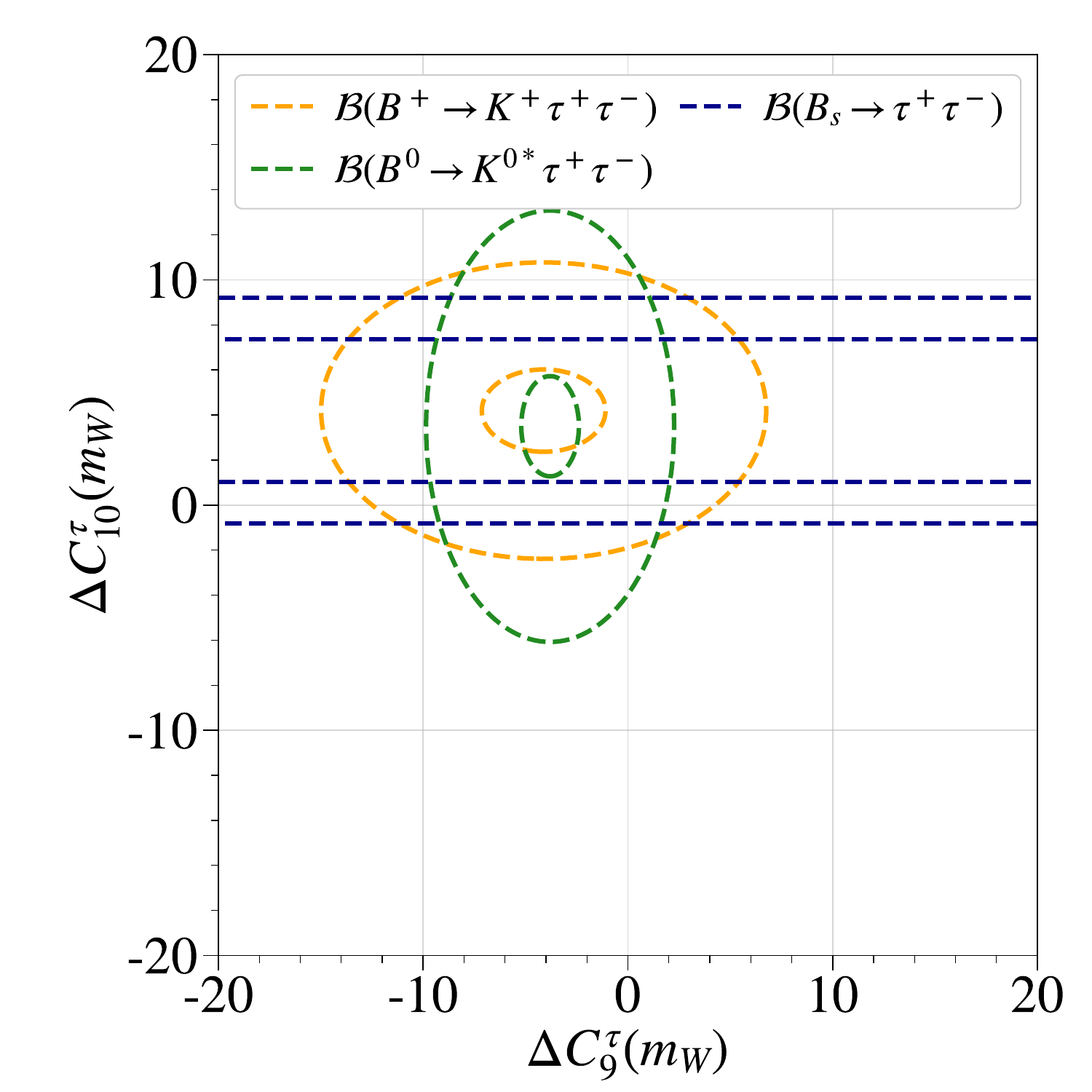}
    \caption{Indirect 95\% bounds in the $\Delta\cC_9^{\tau}-\Delta\cC_{10}^{\tau}$ plane from $b\to s \mu\mu$ measurements (light blue band), along with contours corresponding to current (solid lines) and projected (dashed lines) 95\% upper limits of $b\to s \tau\tau$ branching ratios. Left plot: current constraints and projected future sensitivities from Belle II (10ab$^{-1}$) and LHCb (full upgrade II dataset). Right plot: Projected constraints at FCC-ee, where allowed regions are between two dashed lines, and the origin is always allowed. The current $\Delta \cC_9^U$ band is outside the plotted region in the right-hand plot. In all cases we assume that new physics couples only to $\tau\tau$ at the electroweak scale.\label{fig:c9c10taucomparison}}
\end{figure}

\begin{figure}
    \centering
    \includegraphics[width=0.42\linewidth]{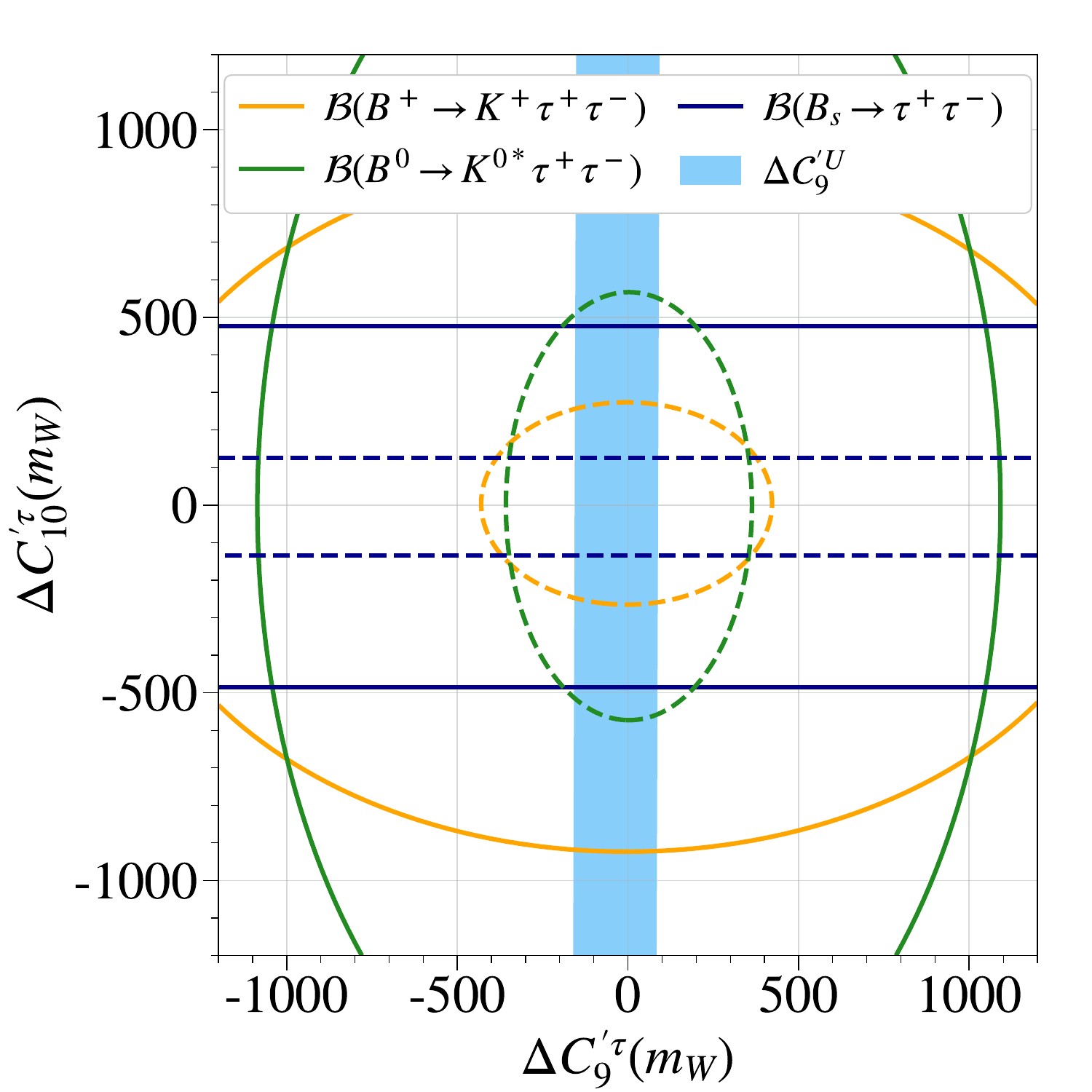}~~~\includegraphics[width=0.42\linewidth]{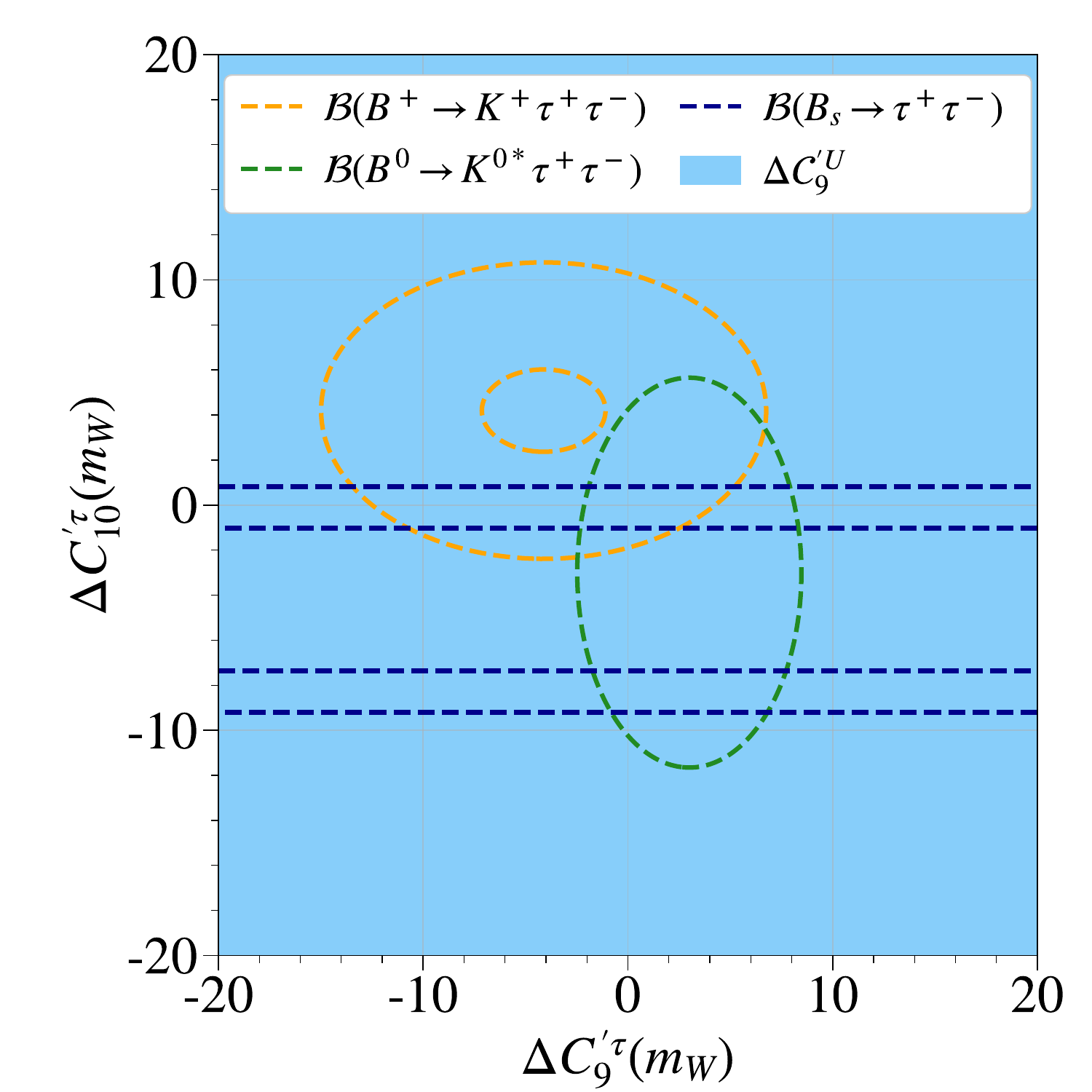}
    \caption{Same as \cref{fig:c9c10taucomparison} but now in the $\Delta\cC_9^{\prime\tau}-\Delta\cC_{10}^{\prime\tau}$ plane.}
        \label{fig:c9c10ptaucomparison}
\end{figure}

Assuming that new physics arises from above the electroweak scale, additional constraints can be put on operators arising from third generation lepton doublets via $b\to s \nu \nu$ processes. These are currently much more precisely constrained than $b\to s \tau \tau$ processes, so we now investigate to what extent the $\Delta\cC_9^{\prime U}$ bounds can provide competitive or complementary constraints to those from $b\to s \nu \nu$, when phrased in terms of SMEFT coefficients at the electroweak scale. In the space of the SMEFT coefficients matching to $bs\tau\tau$ operators, three directions give no contribution to $b\to s\nu\nu$ processes at tree level: $[\cC_{qe}]^{b\tau}_{s\tau}$, $[\cC_{ed}]^{\tau b}_{\tau s}$, and the direction for which $[\cC_{lq}^{(1)}]^{b\tau}_{s\tau}=[\cC_{lq}^{(3)}]^{b\tau}_{s\tau}$. For NP coupled to left-handed quarks, \cref{fig:SMEFTboundsbsll} (left) is drawn in the space of the two directions $[\cC_{qe}]^{b\tau}_{s\tau}$, and $[\cC_{lq}^{(1)}]^{b\tau}_{s\tau}=[\cC_{lq}^{(3)}]^{b\tau}_{s\tau}$. Here we show the current bounds from $b\to s \tau \tau$ branching ratios and $\Delta\cC_9^U$, as well as the projected constraints on $\mathcal{B}(B\to K^{(*)}\tau\tau)$ at Belle II (10 ab$^{-1}$ and 50 ab$^{-1}$) \cite{ATLAS:2025lrr} and $\mathcal{B}(B_s\to \tau\tau)$ at LHCb (300 fb$^{-1}$) \cite{LHCb:2018roe}. We also perform a combined fit to current constraints, both with and without the indirect constraint from $\Delta \cC_9^{U}$. Including the constraint from $\Delta\cC_9^{U}$, it can be seen that we should not expect any signals in Belle II $B\to K^{*}\tau\tau$ measurements with 10 ab$^{-1}$. The combined constraint from $b\to s \tau \tau$ processes without $\Delta \cC_9^{U}$ is not enough to reach this conclusion.
On the other hand, there is still some room for signals in $B\to K\tau\tau$ at Belle II and $B_s\to \tau^+ \tau^-$ at LHCb, as well as in $\mathcal{B}(B\to K^{(*)}\tau\tau)$ with 50 ab$^{-1}$ at Belle II. The direction which is unconstrained by $\Delta \cC_9^{U}$ alone corresponds to the linear combination of SMEFT Wilson coefficients which match to $\Delta \cC_{10}^{\tau}$, hence corresponding to the `parity-stabilised exception' mentioned in the previous subsection, which requires $[\cC_{lq}^{(1)}]^{\tau b}_{\tau s}=[\cC_{lq}^{(3)}]^{\tau b}_{\tau s}=-[\cC_{qe}^{(1)}]^{b \tau}_{s \tau}$. Within a single-particle UV completion, this could only be achieved by a $Z^\prime$ with equal and opposite couplings to left-handed and right-handed $\tau$s. This flat direction is also orthogonal to the flat direction of the $B_s\to \tau \tau$ branching ratio, since (in this plane) this observable depends only on $\Delta \cC_{10}^{\tau}$.

The two independent SMEFT coefficients for NP coupled to right-handed quarks in $b\to s \tau \tau$ are shown in \cref{fig:SMEFTboundsbsll} (right). $[\cC_{ld}]^{\tau b}_{\tau s}$, shown along the horizontal axis, contributes to $b\to s \nu\nu$ processes, and the corresponding $2\sigma$ regions are shown in green and teal.\footnote{The two disconnected regions away from zero for $B\to K\bar\nu\nu$ are due to the fact that this branching ratio measurement has an excess over the SM~\cite{Belle-II:2023esi}. As seen in \cref{fig:SMEFTboundsbsll} (right), some parts of the favoured region are in tension with bounds from $B\to K^{*}\bar \nu\nu$ while other parts are allowed. See Refs.~\cite{Bause:2023mfe,Allwicher:2023xba} for detailed discussion.} In combination, it is seen that bounds from $\Delta\cC_9^{\prime U}$ and $b\to s \nu\nu$ are enough to constrain right-handed currents to a level such that projected sensitivities on $b\to s \tau \tau$ processes at Belle II and LHCb are not expected to explore new parameter space at all. 
Overall, $\Delta\cC_9^{(\prime) U}$ constraints indeed provide complementary information to direct bounds on tauonic new physics. If signals are seen in upcoming searches for $B\to K^{*}\tau\tau$ at Belle II with $10$ab$^{-1}$, then they will be in tension with current constraints unless the underlying physics is light, or has a per-mille tuning between tauonic and muonic couplings at the electroweak scale such that their effects in $\Delta \cC_9^{(\prime) U}$ cancel at $\mu_b$. Any signals in $b\to s \tau\tau$ processes with full LHCb and Belle II datasets will be pointing to new physics coupling to left-handed quarks, unless similar caveats hold. 

\begin{figure}
    \centering
    \includegraphics[height=0.4\linewidth]{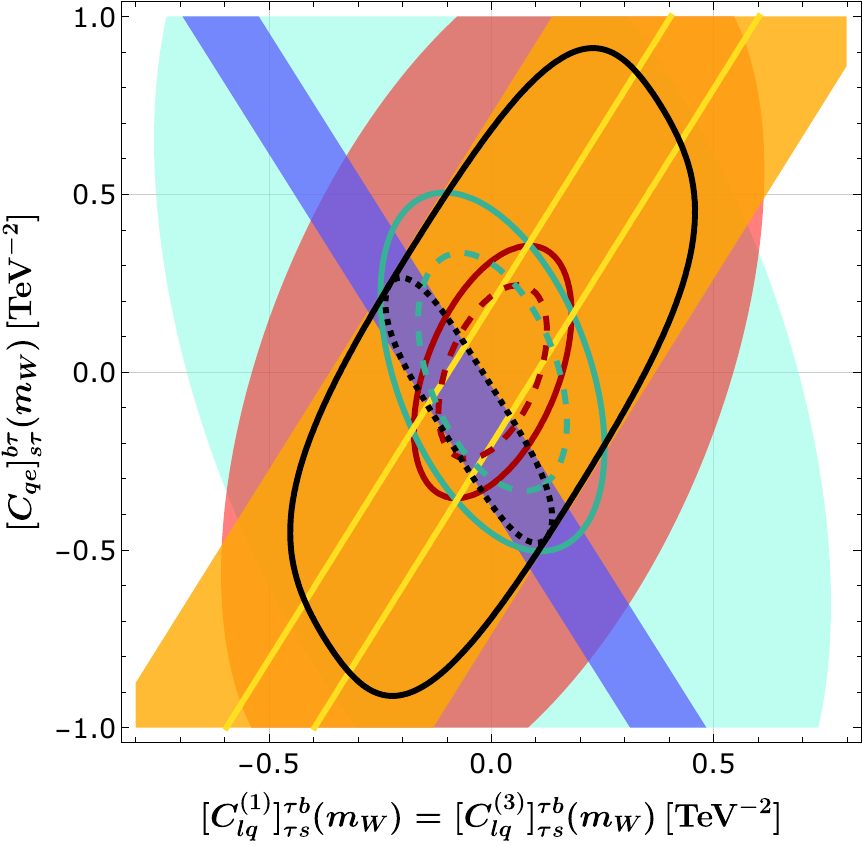}~~~\includegraphics[height=0.4\linewidth]{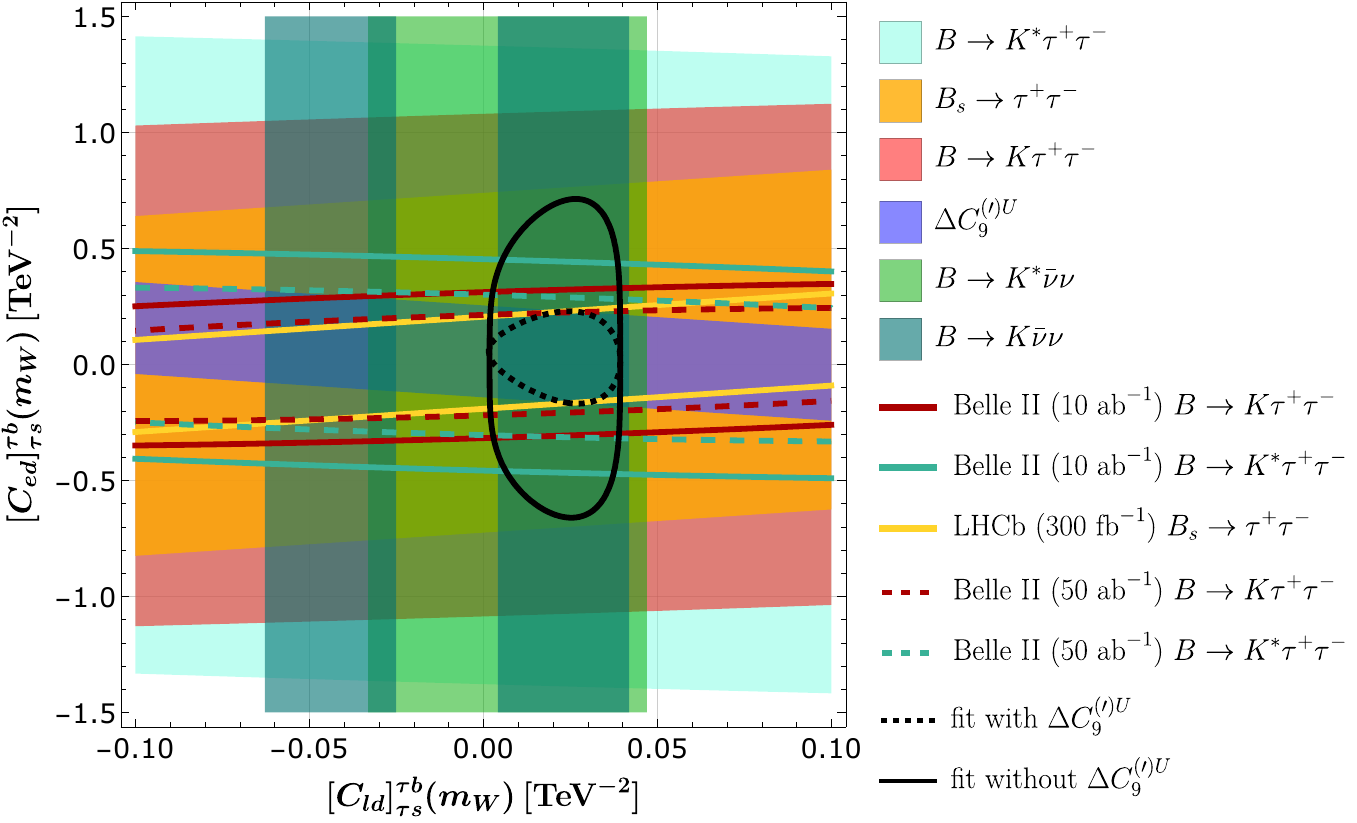}
    \caption{Current direct and indirect constraints (shaded regions) on tauonic SMEFT operators at the electroweak scale, along with projected constraints (coloured lines) on $b\to s \tau\tau$ branching ratios from Belle II 10 ab$^{-1}$ (solid) and 50 ab$^{-1}$ (dashed) and LHCb (full upgrade II dataset). The solid black contour shows the combined fit to current bounds not including the indirect limits from $\Delta \cC_9^{(\prime) U}$, while the dashed black contour shows the combined fit including $\Delta \cC_9^{(\prime) U}$. All regions are at 95\% C.L.. The left and right plots are for left- and right-handed quarks respectively, and the coefficient on the horizontal axis of the left-hand plot has been chosen to avoid tree level effects in $b\to s \nu\nu$.}
    \label{fig:SMEFTboundsbsll}
\end{figure}

Finally, in \cref{fig:constraints}, in the light blue bars we show the overall allowed size of the branching ratio of tauonic processes to be consistent at $2\sigma$ with the constraints from $\Delta \cC_9^{(\prime) U}$, assuming that the entire effect comes from a single third-lepton-generation SMEFT Wilson coefficient. These are expressed relative to the SM prediction for the corresponding branching ratio:
\begin{equation}
    \overline{\mathcal{B}}(X \to Y) \equiv \frac{\mathcal{B}(X \to Y)}{\mathcal{B}(X \to Y)_{SM}} \,.
\end{equation}
Also shown in dark blue are current constraints from the $B\to K^*\bar\nu\nu$ branching ratio upper limit. Vertical lines show current and projected direct bounds on the tauonic branching ratios. 

Summarising this subsection, for some directions in BSM parameter space, current bounds on $\Delta \cC_9^{(\prime)U}$ provide indirectly the leading constraints on new physics inducing a $b\to s \tau \tau$ transition. If excesses are seen in branching ratios of $b\to s \tau \tau$ processes at Belle II or at LHCb Upgrade II, then constraints from $\Delta \cC_9^{(\prime)U}$ would imply that it is arising from new states coupling to left-handed quarks, or it is due to light new physics. These conclusions were reached by assuming that only tauonic Wilson coefficients are non-zero at the electroweak scale; of course new physics could arise from well above the electroweak scale, or have some direct couplings to other generations, which would mean that Wilson coefficients involving light leptons have extra BSM contributions at our starting point. However, unless there is per-mille fine tuning between coefficients at the electroweak scale, these additional light lepton contributions will only further reduce the space for tauonic new physics.

\begin{figure}[h]
    \centering
    \includegraphics[width=\linewidth]{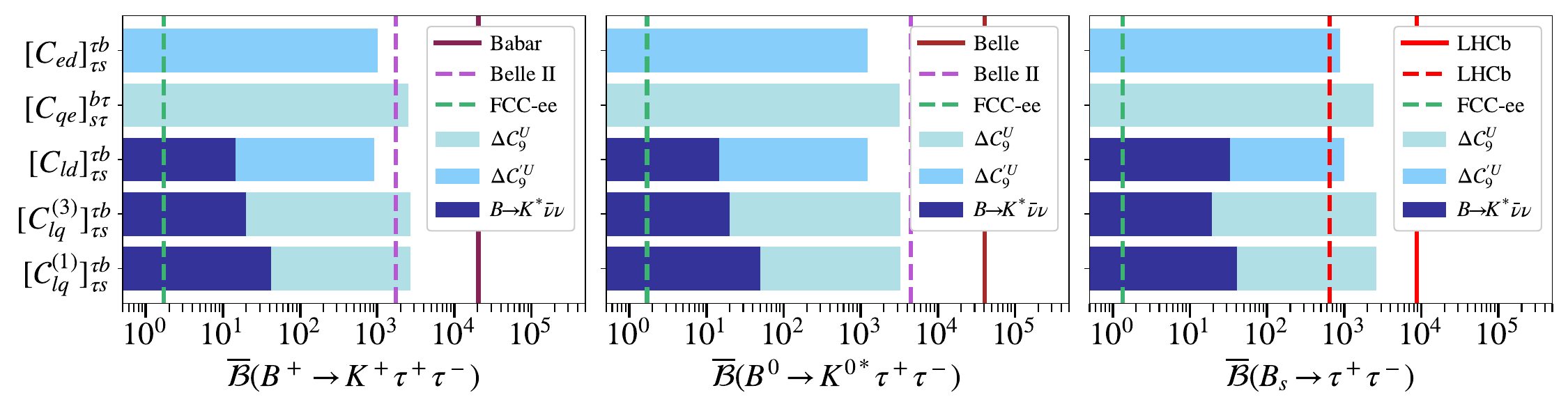}
    \caption{Comparison of upper limits on branching ratios, for processes involving a $b\to s\tau^+\tau^-$ transition, divided by their SM predictions. Horizontal light blue bars represent the allowed regions from $\Delta \cC_9^{(\prime)U}$ at 95\% confidence. Horizontal dark blue bars are regions allowed by the experimental upper limit on $\mathcal{B}(B^+\to K^*\bar\nu\nu)$. Solid vertical lines represent experimental upper limits on the branching ratio (of the process indicated by the $x$-axis label) from Babar, Belle and LHCb. Dashed lines indicate projected limits from Belle II (10 ab$^{-1}$), LHCb Upgrade II and FCC-ee.}
    \label{fig:constraints}
\end{figure}

\subsection{New constraints on four-quark operators\label{sec:fourquarkpheno}}
New physics contributions to the four-quark $\Delta F_d=1$ operators are not generally well constrained, due to large theory errors associated with fully hadronic observables, as well as the fact that the SM can also contribute at tree level to some of these observables through, e.g., $bscc$ operators. As discussed in previous sections, and as seen in \cref{fig:eigenvector_plot}, within the down-octet block there is a small admixture of four-quark operators in ADM eigenvectors which consist predominantly of semileptonic operators (and vice versa). This means that a four-quark operator will generally mix slightly into semileptonic operators between the electroweak scale and the $b$ scale. Indeed in the SM, about half of the $\mathcal{C}_9^\text{SM}$ Wilson coefficient (see \cref{eq:effHambsll} for the definition of this coefficient) is due to this mixing from $bscc$ operators which are generated at tree level from $W$ exchange. In models of new physics, in principle any four-quark operator could be the dominant contribution at the electroweak scale. We give the numerical expressions for the contribution from 4-quark SMEFT operators to $\Delta \cC_9^{(\prime)U}$ in \cref{sec:4quarkrun}.\footnote{We note here that we do not expect the vector LEFT operators to run significantly into $b\to s \gamma$ processes, since from \cref{eq:dipoleRGs} the dipoles are generated at one loop only by scalar and tensor operators. Some traditional bases for four-quark operators based on the structure of SM interactions have Fierz-transformed operators which can obscure this distinction.}

\begin{figure}
    \centering
    \includegraphics[width=0.8\linewidth]{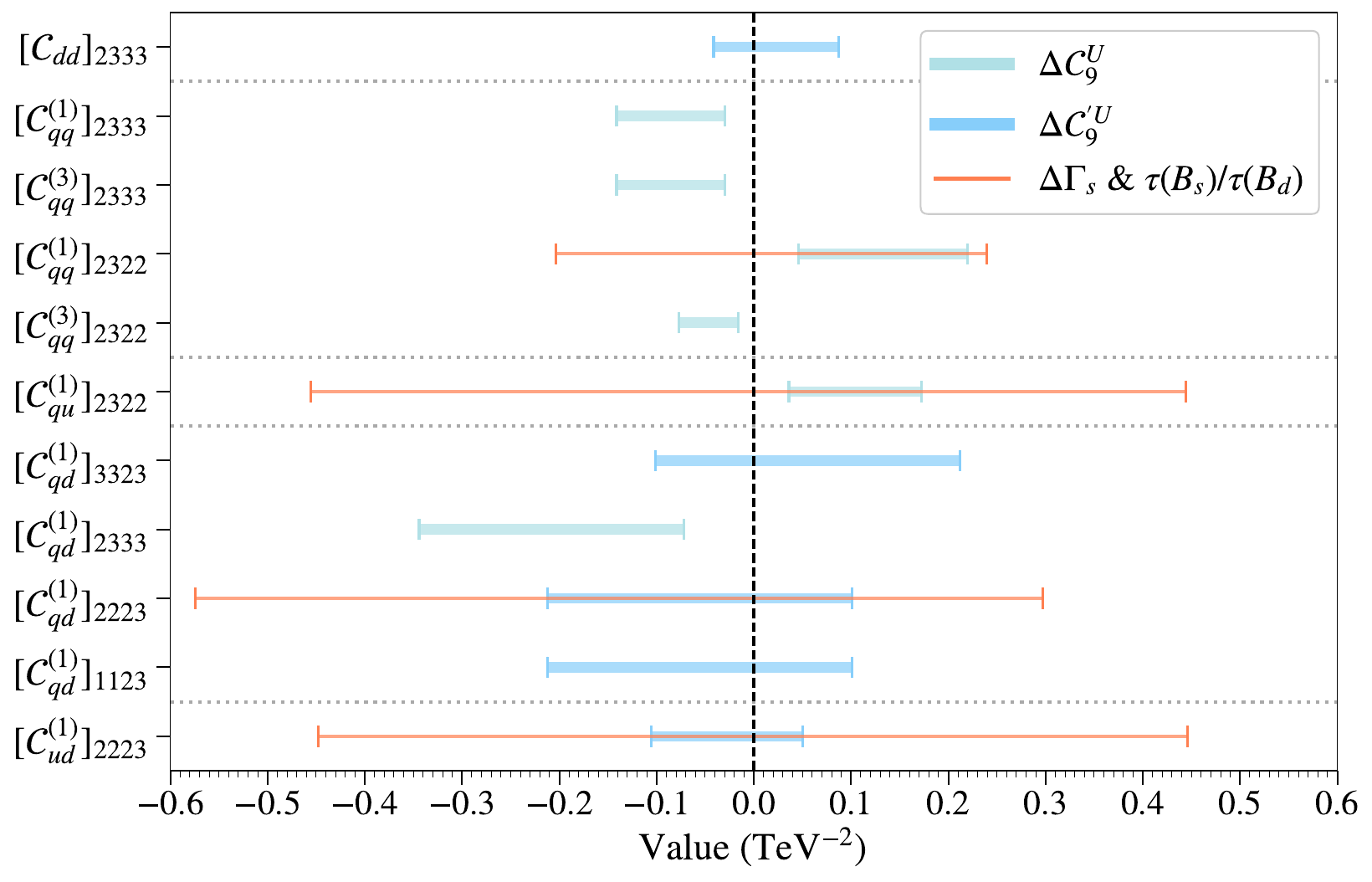}
    \caption{Indirect constraints at 95\% C.L. on individual four-quark SMEFT operators (defined at the electroweak scale) from their RG-induced effects in $\Delta \cC_9^{(\prime)U}$. Also shown are combined 95\% constraints from $\Delta \Gamma_s$ and $\tau(B_s)/\tau(B_d)$ on operators containing $bscc$ interactions. Flavour indices are in the down-aligned basis.}
    \label{fig:bsccLimitsSmall}
\end{figure}

\begin{figure}
    \centering
    \includegraphics[width=0.8\linewidth]{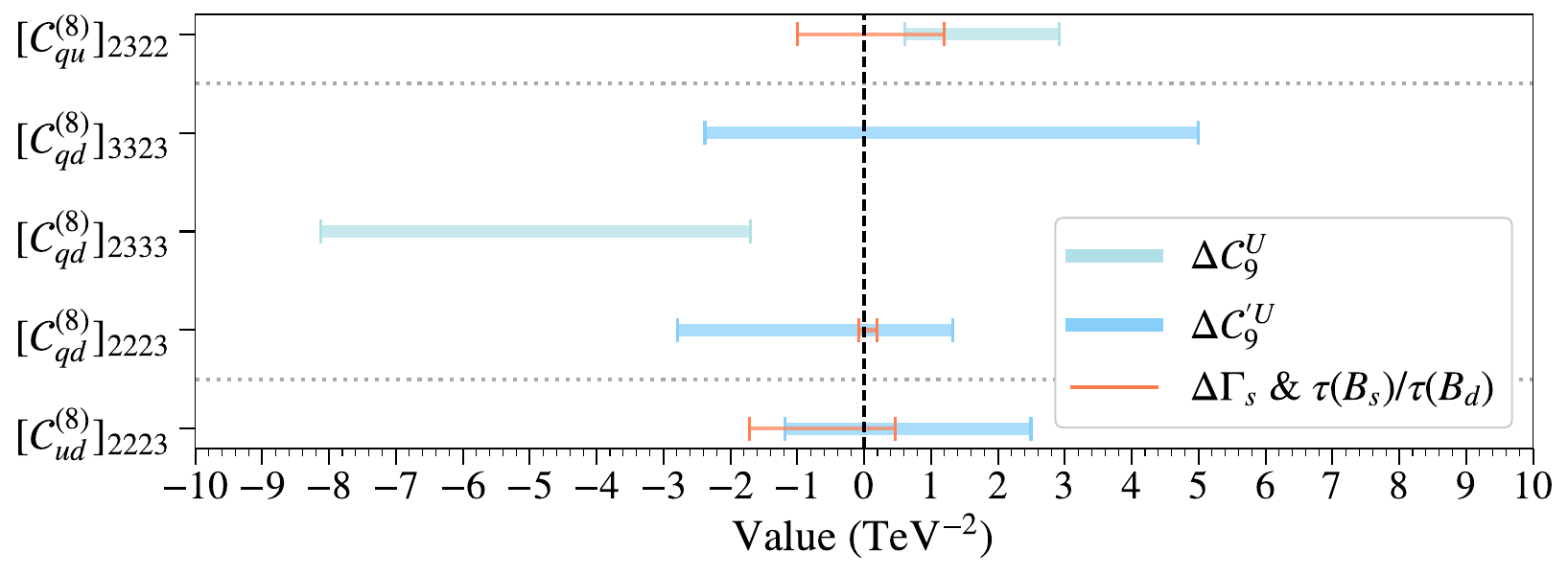}
    \caption{As in \cref{fig:bsccLimitsSmall}, but for operators with colour octet currents.}
    \label{fig:bsccLimitsLarge}
\end{figure}

In \cref{fig:bsccLimitsSmall,fig:bsccLimitsLarge}, we show the resulting indirect $2\sigma$ constraints on individual 4-quark SMEFT Wilson coefficients at the electroweak scale, from current limits on $\Delta \cC_9^{(\prime)U}$. Here, we restrict ourselves to showing operators which match to LEFT $bsqq$ operators with $q=c,b$.\footnote{Note, however, that some SMEFT operators that we study contain second generation quark doublets, in which case they match to $bsss$ as well as $bscc$ operators in the LEFT, and can contribute at tree level to $B\to K^{(*)}\phi$ decays. These could provide additional constraints~\cite{Biswas:2024bhn} which we do not include here.} This is to simplify the plots (other flavour index choices generally give comparable regions) and to avoid contributions to $B_{\{s,d\}}\to MM$ decays, where $M=\{\pi, K^{(*)}\}$.\footnote{These modes could likely provide leading constraints on $bsqq$ operators with $q=u,d$, since the SM does not contribute to them at tree level. However, the theoretical situation is a little unclear (see e.g.~\cite{Bhattacharya:2025wcq}), which may be due to underestimated hadronic effects or to new physics~\cite{Bhattacharya:2022akr,Biswas:2023pyw,Biswas:2025drz}. In fact, recent papers have found possible BSM connections between anomalies in these modes and in $b\to s ll$~\cite{Alguero:2020xca,Datta:2024zrl} or $b\to c \tau \nu$~\cite{Lizana:2023kei}.} 
We also show combined constraints from $\Delta \Gamma_s$ and $\tau(B_s)/\tau(B_d)$ for operators containing charms for which the corresponding calculations are available~\cite{Jager:2017gal,Jager:2019bgk,Lenz:2019lvd}.
It was demonstrated in Refs.~\cite{Jager:2017gal,Jager:2019bgk} that these charm-containing LEFT operators can induce important lepton flavour universal effects in $b\to s ll$ transitions (see also \cite{Crivellin:2023saq}). Despite the fact that these operators contribute at tree level to $\Delta \Gamma_s$ and $\tau(B_s)/\tau(B_d)$, we see in \cref{fig:bsccLimitsSmall} that the RGE-induced $\Delta \cC_9^{(\prime)U}$ constraints outperform them for the corresponding SMEFT operators. On the other hand, the operators shown in \cref{fig:bsccLimitsLarge} are much less well constrained by $\Delta \cC_9^{(\prime)U}$ because their running into $\Delta \cC_9^{(\prime)U}$ is colour-suppressed and occurs formally at two loop. However, models of heavy new physics will not generally match purely onto the operators in \cref{fig:bsccLimitsLarge} without also generating the operators in \cref{fig:bsccLimitsSmall}, with the notable exception of a colour octet electroweak singlet vector state~\cite{deBlas:2017xtg}.

In \cref{fig:bsccLimitsSmall,fig:bsccLimitsLarge} we have assumed a down-aligned basis for operators containing quark doublets. If we instead assume an up-aligned basis, then $bsqq$ LEFT operators can be matched to at tree level by a SMEFT operator involving at least one current of third generation quark doublets, such that the down-type flavour change is generated just through the CKM rotation. In this case we can expect correlated effects in top physics and electroweak precision observables, and we can compare the constraints we find to those found from global fits of electroweak data. This is shown in \cref{fig:topvsflavour}, where the red `Global fits' regions are individual bounds from Refs.~\cite{Celada:2024mcf} and \cite{DiNoi:2025uhu}, which include current top, Higgs and electroweak precision data in their fits. The Wilson coefficients shown here are defined in those references as:
\begin{align}
\label{eq:topfitWCs}
    c_{QQ}^1=2\mathcal{C}_{qq}^{(1)3333}-\frac{2}{3} \mathcal{C}_{qq}^{(3)3333}, ~c_{QQ}^3=\mathcal{C}_{qq}^{(3)3333},~c_{Qd}^1=\mathcal{C}_{qd}^{(1)33jj},~c_{Qd}^8=\mathcal{C}_{qd}^{(8)33jj},
\end{align}
where $j=1,\,2$ or $3$. We do not show $c_{Qd}^8$ in \cref{fig:topvsflavour} since the $\Delta \cC_9^U$ constraint is colour-suppressed and very weak. However, for the other coefficients it can be seen that the $\Delta \cC_9^U$ constraints are roughly comparable with those from top, Higgs and electroweak data, even without explicit flavour-changing indices in the theory above the electroweak scale.\footnote{As mentioned, these results rely on having up-aligned left-handed quark doublets in the SMEFT operators, such that via the CKM rotation they match to $bsqq$ operators at tree level. Operators with down-aligned doublets, or purely right-handed currents, would instead match to these operators at one loop, and directly to $bsll$ operators at two loop. The two-loop matching from down-aligned and right-handed third generation four-quark operators onto $bsll$ LEFT operators has recently been calculated in Ref.~\cite{Haisch:2024wnw}.} 

\begin{figure}
    \centering
    \includegraphics[width=0.6\linewidth]{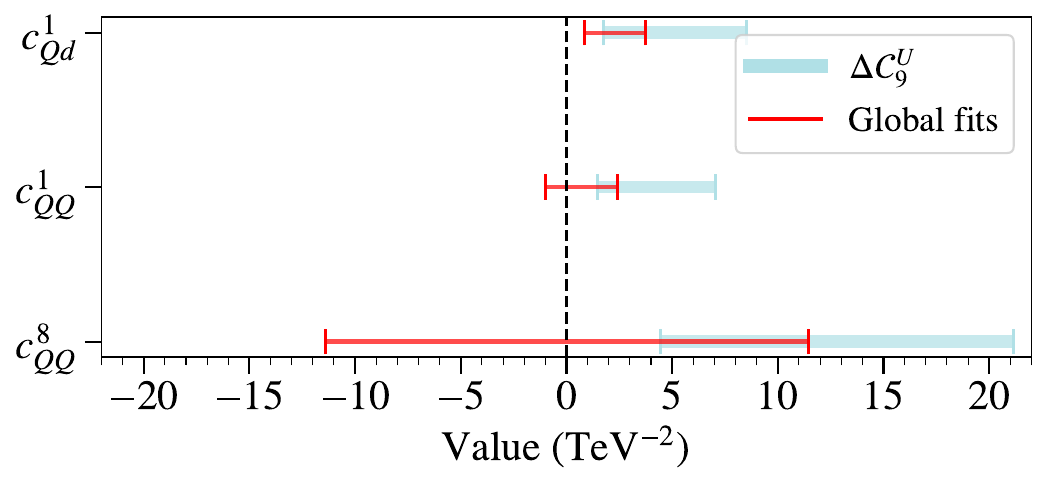}
    \caption{Comparison of indirect bounds from $\Delta \cC_9^U$ to those from fits to electroweak, Higgs and top data \cite{Celada:2024mcf,DiNoi:2025uhu}, all at $95\%$ C.L.. The Wilson coefficients are defined in \cref{eq:topfitWCs}, and are assumed to be in the up-aligned flavour basis and defined around the electroweak scale.}
    \label{fig:topvsflavour}
\end{figure}

An important simplification within the above analysis is that we study the impact of a single SMEFT operator at a time. As discussed in the text around \cref{eq:justdd1,eq:justdd2}, we have identified two linear combinations of operators which are prevented by symmetry (see \cref{sec:fullflav}) from running into the semi-leptonic down-octet operators. Interestingly, the colour structures required to lie within these closed directions are respectively achieved by a second Higgs doublet $\varphi\sim (1,2,\frac12)$ or a colour octet Higgs doublet $\Phi\sim (8,2,\frac12)$~\cite{deBlas:2017xtg}. However, in any explicit model, correlated $\Delta F_d=2$ effects can be expected. For example, given that these scalar particles generate the Wilson coefficient $[c_{dd}^{V1,LR}]^{21}_{13}$ via the coupling combination $(y^d)^*_{21}y_{13}$, then they will inevitably contribute to $K$ and $B$ mixing via coefficients proportional to $|(y^d)_{21}|^2$ and $|(y^d)_{13}|^2$ respectively. So it appears difficult to successfully `hide' four-quark new physics in these closed directions; however, it may be that effects in $bsll$ are avoidable in specific models, and this could provide an additional model discriminator if a pattern of deviations emerges.

\section{Conclusions\label{sec:conc}}

We have presented a decomposition of the four-fermion vectorial operators of the LEFT --- as described in the basis of \cite{Jenkins:2017jig} --- into operators with definite flavour and parity charges. Using the non-Abelian vectorial flavour symmetry of the fermion kinetic terms allows us to partition the thousand-odd operators into small groups, the largest of which has 23 real degrees of freedom. Note that we do not assume any flavour structure on the LEFT coefficients; we are effectively treating the Wilson coefficients as a general set of spurions, whose behaviour is usefully categorised by their non-trivial flavour and parity charges. This work leaves open the extension to other LEFT operators, and also to using the full independent non-Abelian symmetries of the left- and right-handed fields, which will further reduce the size of these groups.

We expect that this decomposition will function as a useful sanity check of calculations performed in more traditional bases of Wilson coefficients with generational indices. Moreover, it may be an efficient way to organise and execute future calculations, such as those of anomalous dimensions. It can be used to reliably extend partial results (for instance, if a partial calculation gives the running of at least one component of the $\mathbf{10}$, it must be the same for the other components). The decomposition could also be built in from the beginning, and will naturally organise diagrams into factorised kinematic, colour, and flavour parts with definite exchange symmetry. Calculating the RG should then be a matter of calculating a handful of reusable colour contractions, flavour contractions, and kinematic contractions (i.e., loop integrals) --- see the example for the one-loop RG of the SMEFT in \cite{Machado:2022ozb}.

In this paper, we used the decomposition to obtain a more global perspective on the effects of the RG flow in the LEFT. The decomposition reduces the RG into small enough blocks that it is semi-analytically soluble via a Magnus expansion method in the QED and QCD couplings, which easily gives a solution to per-mille accuracy.

In the case of operators mediating $\Delta F_d = 1$ transitions, and specifically $b \to s$ transitions, we study the eigendirections of the flow from the $W$ mass to the $b$ mass scale, highlighting in particular the nearly degenerate directions which are mixed significantly at two loop. The fact that lepton flavour non-universal operators are generally split between lepton octet and lepton singlet flavour structures means that the degree of lepton flavour non-universality is scale-dependent. The difference between the anomalous dimensions of the singlet and octet eigenvectors determines the maximum lepton non-universality ratio expected from heavy new physics, which is order $10^3$.

Having used symmetries to condense the anomalous dimension matrix down into dense blocks, practically everything runs into everything else. In the $b \to s$ block, $b\to s ll$ observables with light leptons are precisely measured, and running therefore extends this sensitivity, suppressed by a loop factor, to practically all new physics that can populate this block. We use this to point out that there is limited room for signals in future $b\to s \tau \tau$ searches at LHCb and Belle II, when taken together with other constraints. Specifically, absent per-mille level fine-tuning, $b\to s \tau \tau$ searches at current experiments will not explore any new parameter space for new physics that couples only to right-handed quarks. Four-quark operators can also be constrained in this way, with constraints that are competitive with tree level constraints from lifetime ratios and decay rate differences of $B$ mesons, and top, Higgs and electroweak observables. These insights illustrate the power of the global picture of the LEFT that we gain by exploiting the non-Abelian symmetries.

\appendix

\section{Formulae and inputs for phenomenological analysis}

\subsection{Bases and formulae for $b\to s ll$}
\label{sec:basisbsll}
The Wilson coefficients of semileptonic four-fermion SMEFT operators match at tree level to LEFT operators in the San Diego basis through
\begin{align}
    [\cC^{(1)}_{lq}]_{pqrs}+ [\cC^{(3)}_{lq}]_{pqrs} &= [c^{V,LL}_{ed}]^{qs}_{pr} \,, \label{first_matching_eqn}
    ~~~[\cC_{ed}]_{pqrs} = [c^{V,RR}_{ed}]^{qs}_{pr}\,, \\
    [\cC_{qe}]_{pqrs} &= [c^{V,LR}_{de}]^{qs}_{pr}\,, 
    ~~~[\cC_{ld}]_{pqrs} = [c^{V,LR}_{ed}]^{qs}_{pr} \,,
\end{align}
where all coefficients are evaluated at $\mu_W$. Arranging them in a matrix form
\begin{equation}
  \mathbf{c} = \begin{pmatrix}
    [c_{ed}^{V,LL}]^{ed_i}_{ed_j} & [c_{e d}^{V,LL}]^{\mu d_i}_{\mu d_j} & [c_{e d}^{V,LL}]^{\tau d_i}_{\tau d_j} \\
    [c_{ed}^{V,RR}]^{ed_i}_{ed_j}& [c_{ed}^{V,RR}]^{\mu d_i}_{\mu d_j}& [c_{ed}^{V,RR}]^{\tau d_i}_{\tau d_j}\\
    [c_{de}^{V,LR}]^{d_i e}_{d_j e} & [c_{de}^{V,LR}]^{d_i \mu}_{d_j \mu}  & [c_{de}^{V,LR}]^{ d_i \tau}_{d_j \tau}  \\
    [c_{ed}^{V,LR}]^{e d_i}_{e d_j} &[c_{ed}^{V,LR}]^{\mu d_i}_{\mu d_j} & [c_{ed}^{V,LR}]^{\tau d_i}_{\tau d_j} 
  \end{pmatrix} \, ,
\end{equation}
these can be expressed in terms of the definite flavour and parity basis at $\mu_W$ via the (inverse of)
\begin{align}
  \mathbf{c} =&
  \begin{pmatrix}
    - [c_{ed,A}^+]_{\{ \mathbf{8} , 4 \}_e} - \sqrt{\frac13} [c_{ed,A}^+]_{\{ \mathbf{8} , 6 \}_e}  &\hspace{2ex}
    [c_{ed,A}^+]_{\{ \mathbf{8} , 4 \}_e}  - \sqrt{\frac13} [c_{ed,A}^+]_{\{ \mathbf{8} , 6 \}_e} &\hspace{2ex}
    \sqrt{\frac43} [c_{ed,A}^+]_{\{ \mathbf{8} , 6 \}_e} \\
    - [c_{ed,A}^+]_{\{ \mathbf{8} , 4 \}_e} - \sqrt{\frac13} [c_{ed,A}^+]_{\{ \mathbf{8} , 6 \}_e} &\hspace{2ex}
    [c_{ed,A}^+]_{\{ \mathbf{8} , 4 \}_e} - \sqrt{\frac13} [c_{ed,A}^+]_{\{ \mathbf{8} , 6 \}_e} &\hspace{2ex}
    \sqrt{\frac43} [c_{ed,A}^+]_{\{ \mathbf{8} , 6 \}_e} \\
     - [c_{ed,B}^+]_{\{ \mathbf{8} , 4 \}_e} - \sqrt{\frac13} [c_{ed,B}^+]_{\{ \mathbf{8} , 6 \}_e}  &\hspace{2ex}
    [c_{ed,B}^+]_{\{ \mathbf{8} , 4 \}_e} - \sqrt{\frac13} [c_{ed,B}^+]_{\{ \mathbf{8} , 6 \}_e} &\hspace{2ex}
    \sqrt{\frac43} [c_{ed,B}^+]_{\{ \mathbf{8} , 6 \}_e} \\
    - [c_{ed,B}^+]_{\{ \mathbf{8} , 4 \}_e} - \sqrt{\frac13} [c_{ed,B}^+]_{\{ \mathbf{8} , 6 \}_e} &\hspace{2ex}
    [c_{ed,B}^+]_{\{ \mathbf{8} , 4 \}_e} - \sqrt{\frac13} [c_{ed,B}^+]_{\{ \mathbf{8} , 6 \}_e} &\hspace{2ex}
    \sqrt{\frac43} [c_{ed,B}^+]_{\{ \mathbf{8} , 6 \}_e}
  \end{pmatrix} \notag\\[5pt]
  &+
  \sqrt{\frac23}
  \begin{pmatrix}
    [c_{ed,A}^+]_{\{ \mathbf{1} , 1 \}_e} &
    [c_{ed,A}^+]_{\{ \mathbf{1} , 1 \}_e} &
    [c_{ed,A}^+]_{\{ \mathbf{1} , 1 \}_e} \\
    [c_{ed,A}^+]_{\{ \mathbf{1} , 1 \}_e} &
    [c_{ed,A}^+]_{\{ \mathbf{1} , 1 \}_e} &
    [c_{ed,A}^+]_{\{ \mathbf{1} , 1 \}_e} \\
    [c_{ed,B}^+]_{\{ \mathbf{1} , 1 \}_e} &
    [c_{ed,B}^+]_{\{ \mathbf{1} , 1 \}_e} &
    [c_{ed,B}^+]_{\{ \mathbf{1} , 1 \}_e} \\
    [c_{ed,B}^+]_{\{ \mathbf{1} , 1 \}_e} &
    [c_{ed,B}^+]_{\{ \mathbf{1} , 1 \}_e} &
    [c_{ed,B}^+]_{\{ \mathbf{1} , 1 \}_e} 
  \end{pmatrix} 
  +
  \begin{pmatrix}
    (-1) \times \left[c^+ \rightarrow c^- \right] \\
    (+1) \times \left[c^+ \rightarrow c^- \right] \\
    (-1) \times \left[c^+ \rightarrow c^- \right] \\
    (+1) \times \left[c^+ \rightarrow c^- \right] 
  \end{pmatrix} \, , \label{eq:bsllDecomp}
\end{align}
and then run to the scale $\mu_b$.

The traditional $b\to s ll$ coefficients of the effective Hamiltonian \cref{eq:effHambsll} are then related to the San Diego basis coefficients through:\begin{align}
	\Delta\mathcal{C}_{9}^{\ell} &= \frac{\pi v^2}{\alpha V_{ts}^*V_{tb}} \left( [c_{de}^{LR}]^{b \ell}_{s \ell} + [c_{ed}^{LL}]^{\ell b}_{\ell s} \right), \\
 	\Delta\mathcal{C}_{10}^{\ell} &= \frac{\pi v^2}{\alpha V_{ts}^*V_{tb}} \left( [c_{de}^{LR}]^{b \ell}_{s \ell} - [c_{ed}^{LL}]^{\ell b}_{\ell s} \right), \\
    \Delta\mathcal{C}_{9}^{\ell\prime} &= \frac{\pi v^2}{\alpha V_{ts}^*V_{tb}} \left( [c_{ed}^{LR}]^{\ell b}_{\ell s} + [c_{ed}^{RR}]^{\ell b}_{\ell s} \right), \\
    \Delta\mathcal{C}_{10}^{\ell\prime} &= \frac{\pi v^2}{\alpha V_{ts}^*V_{tb}} \left( -[c_{ed}^{LR}]^{\ell b}_{\ell s} + [c_{ed}^{RR}]^{\ell b}_{\ell s} \right) .
\end{align}
We take the expressions for the branching ratios for the processes $B^+ \to K^+\tau^+\tau^-$ and $B^0 \to K^{0*}\tau^+\tau^-$ in terms of these Wilson coefficients from \cite{Capdevila:2017iqn}, and the branching ratio of $B_s \to \tau^+\tau^-$ from \cite{Bobeth:2011st}. The list of experimental sensitivities and theoretical predictions we use is given in \cref{tab:bstautauobs}.\footnote{Tera-Z prospects for $B_s\to \phi \bar{\nu}\nu$, not included in this table, have been studied in Ref.~\cite{Li:2022tov}.}

\begin{table}[h!]
    \centering
    \scalebox{0.7}{
    \begin{tabular}{l|l | l | l | l}
       Observable  & SM prediction & Current value & Belle II/LHCb proj. & FCC-ee proj.\\
       \hline
$\mathcal{B}(B^+ \to K^+\tau^+\tau^-)$ &   $(1.4\pm0.2)\times10^{-7}$ \cite{Cornella:2021sby} & $< 2.25 \times 10^{-3}$ \cite{BaBar:2016wgb} & $< 1.9 \times 10^{-4}$  \cite{ATLAS:2025lrr} & $\pm 35$\% \cite{Miralles:2024,Allwicher:2025bub}\\
$\mathcal{B}(B^0 \to K^{0*}\tau^+\tau^-)$ & $(0.98\pm0.09)\times10^{-7}$ \cite{Capdevila:2017iqn} & $< 3.1 \times 10^{-3}$ \cite{Belle:2021ecr} & $< 3.4 \times 10^{-4}$ \cite{ATLAS:2025lrr}& $\pm 35$\% \cite{Miralles:2024,Allwicher:2025bub}\\
$\mathcal{B}(B_s \to \tau^+\tau^-)$ &   $(7.73\pm0.49)\times10^{-7}$ \cite{Bobeth:2013uxa} & $< 6.8 \times 10^{-3}$ (95\% C.L.) \cite{LHCb:2017myy}  & $< 5.0 \times 10^{-4}$ (95\% C.L.) \cite{LHCb:2018roe} &  $\pm 17.5$\% \cite{Miralles:2024,Allwicher:2025bub}\\
\hline
$\mathcal{B}(B^+ \to K^+\bar\nu\nu)$ & $(4.44\pm0.30)\times10^{-6}$ \cite{Allwicher:2023xba} &$(2.3\pm0.7)\times10^{-5}$ \cite{Belle-II:2023esi} & $\pm 14$\% \cite{ATLAS:2025lrr} & $\pm 3$\% \cite{Amhis:2023mpj,Allwicher:2025bub}\\
$\mathcal{B}(B^0 \to K^{*0}\bar\nu\nu)$ &$(9.8\pm1.4)\times10^{-6}$ \cite{Allwicher:2023xba} & $<1.8\times 10^{-5}$ \cite{Belle:2017oht} & $\pm 33$\% \cite{ATLAS:2025lrr} & $\pm 3$\% \cite{Amhis:2023mpj,Allwicher:2025bub}\\
    \end{tabular}}
    \caption{Current and projected future sensitivities on branching ratios of $B$ and $B_s$ mesons to final states involving third generation leptons. Upper bounds are at $90\%$ C.L.~unless otherwise stated; uncertainties are at $68\%$. Percentage uncertainties are relative to their SM central value. All projections in the fourth column are for Belle II assuming 10 ab$^{-1}$ integrated luminosity, except $\mathcal{B}(B_s \to \tau^+\tau^-)$, which is for LHCb after Upgrade II.}
    \label{tab:bstautauobs}
\end{table}

\subsection{Four-quark operator running expressions}
\label{sec:4quarkrun}
We collect here numerical expressions for conventional weak effective theory coefficients, at the $\mu_b$ scale, in terms of the coefficients of four-quark SMEFT operators, which are understood to be evaluated at $\mu_W\sim m_W$, and are in the basis in which the down-quark Yukawa matrices are diagonal.
\begin{align}
\Delta \mathcal{C}_{9}^{U}(\mu_b) &= v^2\left(48 [\mathcal{C}^{(1)}_{qd}]_{23ii}
+ 2.0 [\mathcal{C}^{(8)}_{qd}]_{23ii}
+ 21 [\mathcal{C}^{(1)}_{qq}]_{1321}
- 96 [\mathcal{C}^{(1)}_{qq}]_{1123} \right.\nonumber\\
&- 75 [\mathcal{C}^{(1)}_{qq}]_{2223}
+ 120 [\mathcal{C}^{(1)}_{qq}]_{2333}
- 74 [\mathcal{C}^{(3)}_{qq}]_{1321}
+ 290 [\mathcal{C}^{(3)}_{qq}]_{1123} \nonumber\\
&\left.+ 210 [\mathcal{C}^{(3)}_{qq}]_{2223}
+ 120[\mathcal{C}^{(3)}_{qq}]_{2333}
- 95 [\mathcal{C}^{(1)}_{qu}]_{23jj}
- 5.6[\mathcal{C}^{(8)}_{qu}]_{23jj}\right), \\
\Delta \mathcal{C}_{10}^{U}(\mu_b) &= v^2\left(0.17 [\mathcal{C}^{(1)}_{qd}]_{23ii}
+ 0.09 [\mathcal{C}^{(1)}_{qq}]_{1321}
- 0.34 [\mathcal{C}^{(1)}_{qq}]_{1123}
- 0.25 [\mathcal{C}^{(1)}_{qq}]_{2223}\right. \nonumber\\
&+ 0.43 [\mathcal{C}^{(1)}_{qq}]_{2333}
- 0.30 [\mathcal{C}^{(3)}_{qq}]_{1321}
+ 1.0 [\mathcal{C}^{(3)}_{qq}]_{1123}
+ 0.73 [\mathcal{C}^{(3)}_{qq}]_{2223} \nonumber\\
&+ \left.0.43 [\mathcal{C}^{(3)}_{qq}]_{2333}
- 0.34 [\mathcal{C}^{(1)}_{qu}]_{23jj}\right), \\
\Delta \mathcal{C}_{9}^{\prime U}(\mu_b) &= v^2\left(
48 [\mathcal{C}^{(1)}_{qd}]_{3323}
- 48 [\mathcal{C}^{(1)}_{qd}]_{jj23}
+ 2.0 [\mathcal{C}^{(8)}_{qd}]_{3323}
- 3.6 [\mathcal{C}^{(8)}_{qd}]_{jj23} \right.\nonumber\\
&+ 95 [\mathcal{C}_{dd}]_{1123}
+ 21 [\mathcal{C}_{dd}]_{1321}
+ 120 [\mathcal{C}_{dd}]_{2223}
+ 120 [\mathcal{C}_{dd}]_{2333}\nonumber\\
&\left.- 95 [\mathcal{C}^{(1)}_{ud}]_{jj23}
+ 4.0 [\mathcal{C}^{(8)}_{ud}]_{jj23}\right), \\
\Delta \mathcal{C}_{10}^{\prime U}(\mu_b) &= v^2\left(- 0.17 [\mathcal{C}^{(1)}_{qd}]_{3323}
+ 0.17 [\mathcal{C}^{(1)}_{qd}]_{jj23} -0.34 [\mathcal{C}_{dd}]_{1123}
- 0.090 [\mathcal{C}_{dd}]_{1321}\right. \nonumber\\
&\left.- 0.43 [\mathcal{C}_{dd}]_{2223}
- 0.43 [\mathcal{C}_{dd}]_{2333}
+ 0.34 [\mathcal{C}^{(1)}_{ud}]_{jj23}\right) \,,
\end{align}
where the repeated index `$i$' signifies summation over indices $i=1,2,3,$ and repeated `$j$' indices signify summation over $j=1,2$. We calculated the above using the second-order Magnus expansion method in \cref{sec:magnus}. The numerical coefficients in the expressions for $\Delta \mathcal{C}_{9}^{(\prime) U}(\mu_b)$ all agree to within 20\% with those found with \texttt{wilson}~\cite{Aebischer:2018bkb}. We were unable to compare $\Delta \mathcal{C}_{10}^{(\prime) U}(\mu_b)$ since \texttt{wilson} does not include QED resummation, without which no contribution to these coefficients is generated by running from the four-quark operators.

\section{One- and two-loop down-octet matrices\label{app:onetwoloopexplicit}}

Here we give the QCD and QED pieces of the $12\times 12$ anomalous dimension matrix of the parity-even down-octet block ($\db_d=8$, $\db_{u,e}=1$, $P=+1$), as defined in \cref{eq:tDependenceGamma}, and as calculated in the San Diego basis in \cite{Jenkins:2017dyc} and \cite{Naterop:2025cwg}.
\begin{equation}
\scalebox{0.8}{$\hat{\gamma}^{(1)}_e = \left(
\begin{array}{cccccccccccc}
 \frac{76}{27} & \frac{4 \sqrt{5}}{27} & \frac{10}{27} & \frac{5}{27} & \frac{2 \sqrt{5}}{27} & \frac{\sqrt{5}}{27} & \frac{\sqrt{\frac{5}{3}}}{3} &
   \frac{\sqrt{\frac{5}{3}}}{3} & -\frac{8 \sqrt{10}}{27} & -\frac{4 \sqrt{10}}{27} & -\frac{8 \sqrt{10}}{27} & -\frac{4 \sqrt{10}}{27} \\
 \frac{8 \sqrt{5}}{27} & \frac{40}{27} & \frac{2 \sqrt{5}}{27} & \frac{\sqrt{5}}{27} & \frac{2}{27} & \frac{1}{27} & \frac{1}{3 \sqrt{3}} & \frac{1}{3 \sqrt{3}} &
   -\frac{8 \sqrt{2}}{27} & -\frac{4 \sqrt{2}}{27} & -\frac{8 \sqrt{2}}{27} & -\frac{4 \sqrt{2}}{27} \\
 \frac{80}{27} & \frac{8 \sqrt{5}}{27} & -\frac{16}{27} & \frac{10}{27} & \frac{4 \sqrt{5}}{27} & \frac{2 \sqrt{5}}{27} & \frac{2 \sqrt{\frac{5}{3}}}{3} & \frac{2
   \sqrt{\frac{5}{3}}}{3} & -\frac{16 \sqrt{10}}{27} & -\frac{8 \sqrt{10}}{27} & -\frac{16 \sqrt{10}}{27} & -\frac{8 \sqrt{10}}{27} \\
 \frac{80}{27} & \frac{8 \sqrt{5}}{27} & \frac{20}{27} & -\frac{26}{27} & \frac{4 \sqrt{5}}{27} & \frac{2 \sqrt{5}}{27} & \frac{2 \sqrt{\frac{5}{3}}}{3} & \frac{2
   \sqrt{\frac{5}{3}}}{3} & -\frac{16 \sqrt{10}}{27} & -\frac{8 \sqrt{10}}{27} & -\frac{16 \sqrt{10}}{27} & -\frac{8 \sqrt{10}}{27} \\
 \frac{16 \sqrt{5}}{27} & \frac{8}{27} & \frac{4 \sqrt{5}}{27} & \frac{2 \sqrt{5}}{27} & -\frac{32}{27} & \frac{2}{27} & \frac{2}{3 \sqrt{3}} & \frac{2}{3 \sqrt{3}} &
   -\frac{16 \sqrt{2}}{27} & -\frac{8 \sqrt{2}}{27} & -\frac{16 \sqrt{2}}{27} & -\frac{8 \sqrt{2}}{27} \\
 \frac{16 \sqrt{5}}{27} & \frac{8}{27} & \frac{4 \sqrt{5}}{27} & \frac{2 \sqrt{5}}{27} & \frac{4}{27} & -\frac{34}{27} & \frac{2}{3 \sqrt{3}} & \frac{2}{3 \sqrt{3}} &
   -\frac{16 \sqrt{2}}{27} & -\frac{8 \sqrt{2}}{27} & -\frac{16 \sqrt{2}}{27} & -\frac{8 \sqrt{2}}{27} \\
 \frac{32 \sqrt{\frac{5}{3}}}{3} & \frac{16}{3 \sqrt{3}} & \frac{8 \sqrt{\frac{5}{3}}}{3} & \frac{4 \sqrt{\frac{5}{3}}}{3} & \frac{8}{3 \sqrt{3}} & \frac{4}{3 \sqrt{3}}
   & 8 & 4 & -\frac{32 \sqrt{\frac{2}{3}}}{3} & -\frac{16 \sqrt{\frac{2}{3}}}{3} & -\frac{32 \sqrt{\frac{2}{3}}}{3} & -\frac{16 \sqrt{\frac{2}{3}}}{3} \\
 \frac{32 \sqrt{\frac{5}{3}}}{3} & \frac{16}{3 \sqrt{3}} & \frac{8 \sqrt{\frac{5}{3}}}{3} & \frac{4 \sqrt{\frac{5}{3}}}{3} & \frac{8}{3 \sqrt{3}} & \frac{4}{3 \sqrt{3}}
   & 4 & 0 & -\frac{32 \sqrt{\frac{2}{3}}}{3} & -\frac{16 \sqrt{\frac{2}{3}}}{3} & -\frac{32 \sqrt{\frac{2}{3}}}{3} & -\frac{16 \sqrt{\frac{2}{3}}}{3} \\
 -\frac{32 \sqrt{10}}{27} & -\frac{16 \sqrt{2}}{27} & -\frac{8 \sqrt{10}}{27} & -\frac{4 \sqrt{10}}{27} & -\frac{8 \sqrt{2}}{27} & -\frac{4 \sqrt{2}}{27} & -\frac{4
   \sqrt{\frac{2}{3}}}{3} & -\frac{4 \sqrt{\frac{2}{3}}}{3} & -\frac{8}{27} & \frac{32}{27} & \frac{64}{27} & \frac{32}{27} \\
 -\frac{32 \sqrt{10}}{27} & -\frac{16 \sqrt{2}}{27} & -\frac{8 \sqrt{10}}{27} & -\frac{4 \sqrt{10}}{27} & -\frac{8 \sqrt{2}}{27} & -\frac{4 \sqrt{2}}{27} & -\frac{4
   \sqrt{\frac{2}{3}}}{3} & -\frac{4 \sqrt{\frac{2}{3}}}{3} & \frac{64}{27} & -\frac{40}{27} & \frac{64}{27} & \frac{32}{27} \\
 -\frac{32 \sqrt{10}}{27} & -\frac{16 \sqrt{2}}{27} & -\frac{8 \sqrt{10}}{27} & -\frac{4 \sqrt{10}}{27} & -\frac{8 \sqrt{2}}{27} & -\frac{4 \sqrt{2}}{27} & -\frac{4
   \sqrt{\frac{2}{3}}}{3} & -\frac{4 \sqrt{\frac{2}{3}}}{3} & \frac{64}{27} & \frac{32}{27} & \frac{136}{27} & \frac{32}{27} \\
 -\frac{32 \sqrt{10}}{27} & -\frac{16 \sqrt{2}}{27} & -\frac{8 \sqrt{10}}{27} & -\frac{4 \sqrt{10}}{27} & -\frac{8 \sqrt{2}}{27} & -\frac{4 \sqrt{2}}{27} & -\frac{4
   \sqrt{\frac{2}{3}}}{3} & -\frac{4 \sqrt{\frac{2}{3}}}{3} & \frac{64}{27} & \frac{32}{27} & \frac{64}{27} & \frac{104}{27} \\
\end{array}
\right)$}
\end{equation}
\begin{equation}
\scalebox{0.8}{$\hat{\gamma}^{(1)}_g = \left(
\begin{array}{cccccccccccc}
 \frac{46}{9} & -\frac{2 \sqrt{5}}{9} & \frac{5}{36} & -\frac{5}{36} & \frac{\sqrt{5}}{36} & -\frac{\sqrt{5}}{36} & 0 & 0 & \frac{\sqrt{\frac{5}{2}}}{9} &
   -\frac{\sqrt{\frac{5}{2}}}{9} & \frac{\sqrt{\frac{5}{2}}}{9} & -\frac{\sqrt{\frac{5}{2}}}{9} \\
 -\frac{4 \sqrt{5}}{9} & -\frac{68}{9} & -\frac{\sqrt{5}}{18} & \frac{\sqrt{5}}{18} & -\frac{1}{18} & \frac{1}{18} & 0 & 0 & -\frac{\sqrt{2}}{9} & \frac{\sqrt{2}}{9} &
   -\frac{\sqrt{2}}{9} & \frac{\sqrt{2}}{9} \\
 \frac{40}{9} & -\frac{8 \sqrt{5}}{9} & -\frac{97}{9} & \frac{25}{9} & \frac{\sqrt{5}}{9} & -\frac{\sqrt{5}}{9} & 0 & 0 & \frac{2 \sqrt{10}}{9} & -\frac{2 \sqrt{10}}{9}
   & \frac{2 \sqrt{10}}{9} & -\frac{2 \sqrt{10}}{9} \\
 -\frac{80}{9} & \frac{16 \sqrt{5}}{9} & \frac{158}{9} & -\frac{14}{9} & -\frac{2 \sqrt{5}}{9} & \frac{2 \sqrt{5}}{9} & 0 & 0 & -\frac{4 \sqrt{10}}{9} & \frac{4
   \sqrt{10}}{9} & -\frac{4 \sqrt{10}}{9} & \frac{4 \sqrt{10}}{9} \\
 \frac{8 \sqrt{5}}{9} & -\frac{8}{9} & \frac{\sqrt{5}}{9} & -\frac{\sqrt{5}}{9} & -\frac{101}{9} & \frac{29}{9} & 0 & 0 & \frac{2 \sqrt{2}}{9} & -\frac{2 \sqrt{2}}{9} &
   \frac{2 \sqrt{2}}{9} & -\frac{2 \sqrt{2}}{9} \\
 -\frac{16 \sqrt{5}}{9} & \frac{16}{9} & -\frac{2 \sqrt{5}}{9} & \frac{2 \sqrt{5}}{9} & \frac{166}{9} & -\frac{22}{9} & 0 & 0 & -\frac{4 \sqrt{2}}{9} & \frac{4
   \sqrt{2}}{9} & -\frac{4 \sqrt{2}}{9} & \frac{4 \sqrt{2}}{9} \\
 0 & 0 & 0 & 0 & 0 & 0 & 0 & 0 & 0 & 0 & 0 & 0 \\
 0 & 0 & 0 & 0 & 0 & 0 & 0 & 0 & 0 & 0 & 0 & 0 \\
 \frac{8 \sqrt{10}}{9} & -\frac{8 \sqrt{2}}{9} & \frac{\sqrt{10}}{9} & -\frac{\sqrt{10}}{9} & \frac{\sqrt{2}}{9} & -\frac{\sqrt{2}}{9} & 0 & 0 & \frac{52}{9} &
   \frac{20}{9} & \frac{4}{9} & -\frac{4}{9} \\
 -\frac{16 \sqrt{10}}{9} & \frac{16 \sqrt{2}}{9} & -\frac{2 \sqrt{10}}{9} & \frac{2 \sqrt{10}}{9} & -\frac{2 \sqrt{2}}{9} & \frac{2 \sqrt{2}}{9} & 0 & 0 & -\frac{68}{9}
   & -\frac{76}{9} & -\frac{8}{9} & \frac{8}{9} \\
 \frac{8 \sqrt{10}}{9} & -\frac{8 \sqrt{2}}{9} & \frac{\sqrt{10}}{9} & -\frac{\sqrt{10}}{9} & \frac{\sqrt{2}}{9} & -\frac{\sqrt{2}}{9} & 0 & 0 & \frac{4}{9} &
   -\frac{4}{9} & -\frac{98}{9} & \frac{26}{9} \\
 -\frac{16 \sqrt{10}}{9} & \frac{16 \sqrt{2}}{9} & -\frac{2 \sqrt{10}}{9} & \frac{2 \sqrt{10}}{9} & -\frac{2 \sqrt{2}}{9} & \frac{2 \sqrt{2}}{9} & 0 & 0 & -\frac{8}{9} &
   \frac{8}{9} & \frac{160}{9} & -\frac{16}{9} \\
\end{array}
\right)$}
\end{equation}
\begin{equation}
\scalebox{0.65}{$\hat{\gamma}^{(2)}_e = \left(
\begin{array}{cccccccccccc}
 -\frac{12598}{2187} & \frac{344 \sqrt{5}}{2187} & -\frac{940}{2187} & -\frac{470}{2187} & -\frac{188 \sqrt{5}}{2187} & -\frac{94 \sqrt{5}}{2187} & \frac{590
   \sqrt{\frac{5}{3}}}{243} & \frac{50 \sqrt{\frac{5}{3}}}{243} & -\frac{40 \sqrt{10}}{2187} & -\frac{20 \sqrt{10}}{2187} & -\frac{2920 \sqrt{10}}{2187} & -\frac{1460
   \sqrt{10}}{2187} \\
 \frac{688 \sqrt{5}}{2187} & -\frac{15694}{2187} & -\frac{188 \sqrt{5}}{2187} & -\frac{94 \sqrt{5}}{2187} & -\frac{188}{2187} & -\frac{94}{2187} & \frac{590}{243
   \sqrt{3}} & \frac{50}{243 \sqrt{3}} & -\frac{40 \sqrt{2}}{2187} & -\frac{20 \sqrt{2}}{2187} & -\frac{2920 \sqrt{2}}{2187} & -\frac{1460 \sqrt{2}}{2187} \\
 -\frac{4640}{2187} & -\frac{464 \sqrt{5}}{2187} & -\frac{1790}{2187} & \frac{860}{2187} & \frac{344 \sqrt{5}}{2187} & \frac{172 \sqrt{5}}{2187} & \frac{244
   \sqrt{\frac{5}{3}}}{243} & \frac{1108 \sqrt{\frac{5}{3}}}{243} & -\frac{5552 \sqrt{10}}{2187} & -\frac{2776 \sqrt{10}}{2187} & -\frac{944 \sqrt{10}}{2187} & -\frac{472
   \sqrt{10}}{2187} \\
 -\frac{4640}{2187} & -\frac{464 \sqrt{5}}{2187} & \frac{1720}{2187} & -\frac{2650}{2187} & \frac{344 \sqrt{5}}{2187} & \frac{172 \sqrt{5}}{2187} & \frac{244
   \sqrt{\frac{5}{3}}}{243} & \frac{1108 \sqrt{\frac{5}{3}}}{243} & -\frac{5552 \sqrt{10}}{2187} & -\frac{2776 \sqrt{10}}{2187} & -\frac{944 \sqrt{10}}{2187} & -\frac{472
   \sqrt{10}}{2187} \\
 -\frac{928 \sqrt{5}}{2187} & -\frac{464}{2187} & \frac{344 \sqrt{5}}{2187} & \frac{172 \sqrt{5}}{2187} & -\frac{3166}{2187} & \frac{172}{2187} & \frac{244}{243 \sqrt{3}}
   & \frac{1108}{243 \sqrt{3}} & -\frac{5552 \sqrt{2}}{2187} & -\frac{2776 \sqrt{2}}{2187} & -\frac{944 \sqrt{2}}{2187} & -\frac{472 \sqrt{2}}{2187} \\
 -\frac{928 \sqrt{5}}{2187} & -\frac{464}{2187} & \frac{344 \sqrt{5}}{2187} & \frac{172 \sqrt{5}}{2187} & \frac{344}{2187} & -\frac{3338}{2187} & \frac{244}{243 \sqrt{3}}
   & \frac{1108}{243 \sqrt{3}} & -\frac{5552 \sqrt{2}}{2187} & -\frac{2776 \sqrt{2}}{2187} & -\frac{944 \sqrt{2}}{2187} & -\frac{472 \sqrt{2}}{2187} \\
 \frac{3904 \sqrt{\frac{5}{3}}}{243} & \frac{1952}{243 \sqrt{3}} & -\frac{3056 \sqrt{\frac{5}{3}}}{243} & -\frac{1528 \sqrt{\frac{5}{3}}}{243} & -\frac{3056}{243
   \sqrt{3}} & -\frac{1528}{243 \sqrt{3}} & -\frac{14}{3} & -\frac{2248}{81} & \frac{14240 \sqrt{\frac{2}{3}}}{243} & \frac{7120 \sqrt{\frac{2}{3}}}{243} & -\frac{18016
   \sqrt{\frac{2}{3}}}{243} & -\frac{9008 \sqrt{\frac{2}{3}}}{243} \\
 -\frac{9920 \sqrt{\frac{5}{3}}}{243} & -\frac{4960}{243 \sqrt{3}} & \frac{1264 \sqrt{\frac{5}{3}}}{243} & \frac{632 \sqrt{\frac{5}{3}}}{243} & \frac{1264}{243 \sqrt{3}}
   & \frac{632}{243 \sqrt{3}} & -\frac{1816}{81} & \frac{914}{27} & -\frac{18592 \sqrt{\frac{2}{3}}}{243} & -\frac{9296 \sqrt{\frac{2}{3}}}{243} & \frac{11360
   \sqrt{\frac{2}{3}}}{243} & \frac{5680 \sqrt{\frac{2}{3}}}{243} \\
 \frac{7616 \sqrt{10}}{2187} & \frac{3808 \sqrt{2}}{2187} & -\frac{544 \sqrt{10}}{2187} & -\frac{272 \sqrt{10}}{2187} & -\frac{544 \sqrt{2}}{2187} & -\frac{272
   \sqrt{2}}{2187} & \frac{880 \sqrt{\frac{2}{3}}}{243} & -\frac{2792 \sqrt{\frac{2}{3}}}{243} & -\frac{20416}{2187} & \frac{13120}{2187} & -\frac{12928}{2187} &
   -\frac{6464}{2187} \\
 \frac{7616 \sqrt{10}}{2187} & \frac{3808 \sqrt{2}}{2187} & -\frac{544 \sqrt{10}}{2187} & -\frac{272 \sqrt{10}}{2187} & -\frac{544 \sqrt{2}}{2187} & -\frac{272
   \sqrt{2}}{2187} & \frac{880 \sqrt{\frac{2}{3}}}{243} & -\frac{2792 \sqrt{\frac{2}{3}}}{243} & \frac{26240}{2187} & -\frac{33536}{2187} & -\frac{12928}{2187} &
   -\frac{6464}{2187} \\
 -\frac{1600 \sqrt{10}}{2187} & -\frac{800 \sqrt{2}}{2187} & \frac{2336 \sqrt{10}}{2187} & \frac{1168 \sqrt{10}}{2187} & \frac{2336 \sqrt{2}}{2187} & \frac{1168
   \sqrt{2}}{2187} & -\frac{2864 \sqrt{\frac{2}{3}}}{243} & \frac{1240 \sqrt{\frac{2}{3}}}{243} & -\frac{17536}{2187} & -\frac{8768}{2187} & -\frac{57136}{2187} &
   \frac{13120}{2187} \\
 -\frac{1600 \sqrt{10}}{2187} & -\frac{800 \sqrt{2}}{2187} & \frac{2336 \sqrt{10}}{2187} & \frac{1168 \sqrt{10}}{2187} & \frac{2336 \sqrt{2}}{2187} & \frac{1168
   \sqrt{2}}{2187} & -\frac{2864 \sqrt{\frac{2}{3}}}{243} & \frac{1240 \sqrt{\frac{2}{3}}}{243} & -\frac{17536}{2187} & -\frac{8768}{2187} & \frac{26240}{2187} &
   -\frac{70256}{2187} \\
\end{array}
\right)$}
\end{equation}
\begin{equation}
\scalebox{0.65}{$\hat{\gamma}^{(2)}_{eg} = \left(
\begin{array}{cccccccccccc}
 -\frac{7538}{729} & \frac{250 \sqrt{5}}{729} & \frac{685}{972} & \frac{5}{4} & \frac{137 \sqrt{5}}{972} & \frac{\sqrt{5}}{4} & -\frac{85 \sqrt{\frac{5}{3}}}{81} &
   -\frac{85 \sqrt{\frac{5}{3}}}{81} & -\frac{557 \sqrt{\frac{5}{2}}}{243} & -\frac{203 \sqrt{\frac{5}{2}}}{243} & -\frac{281 \sqrt{\frac{5}{2}}}{243} & -\frac{479
   \sqrt{\frac{5}{2}}}{243} \\
 \frac{68 \sqrt{5}}{729} & \frac{2188}{729} & \frac{197 \sqrt{5}}{486} & \frac{71 \sqrt{5}}{486} & \frac{197}{486} & \frac{71}{486} & -\frac{46}{81 \sqrt{3}} &
   -\frac{46}{81 \sqrt{3}} & -\frac{35 \sqrt{2}}{27} & -\frac{221 \sqrt{2}}{243} & -\frac{173 \sqrt{2}}{81} & -\frac{17 \sqrt{2}}{243} \\
 \frac{760}{243} & \frac{88 \sqrt{5}}{27} & \frac{6757}{729} & \frac{3911}{729} & \frac{563 \sqrt{5}}{729} & \frac{145 \sqrt{5}}{729} & -\frac{124 \sqrt{\frac{5}{3}}}{81}
   & -\frac{124 \sqrt{\frac{5}{3}}}{81} & -\frac{842 \sqrt{10}}{729} & -\frac{1990 \sqrt{10}}{729} & -\frac{1922 \sqrt{10}}{729} & -\frac{910 \sqrt{10}}{729} \\
 \frac{4640}{243} & -\frac{224 \sqrt{5}}{243} & \frac{6724}{729} & \frac{4988}{729} & -\frac{124 \sqrt{5}}{729} & \frac{652 \sqrt{5}}{729} & -\frac{184
   \sqrt{\frac{5}{3}}}{81} & -\frac{184 \sqrt{\frac{5}{3}}}{81} & -\frac{3548 \sqrt{10}}{729} & \frac{1436 \sqrt{10}}{729} & -\frac{956 \sqrt{10}}{729} & -\frac{1156
   \sqrt{10}}{729} \\
 \frac{152 \sqrt{5}}{243} & \frac{88}{27} & \frac{563 \sqrt{5}}{729} & \frac{145 \sqrt{5}}{729} & \frac{4505}{729} & \frac{3331}{729} & -\frac{124}{81 \sqrt{3}} &
   -\frac{124}{81 \sqrt{3}} & -\frac{842 \sqrt{2}}{729} & -\frac{1990 \sqrt{2}}{729} & -\frac{1922 \sqrt{2}}{729} & -\frac{910 \sqrt{2}}{729} \\
 \frac{928 \sqrt{5}}{243} & -\frac{224}{243} & -\frac{124 \sqrt{5}}{729} & \frac{652 \sqrt{5}}{729} & \frac{7220}{729} & \frac{2380}{729} & -\frac{184}{81 \sqrt{3}} &
   -\frac{184}{81 \sqrt{3}} & -\frac{3548 \sqrt{2}}{729} & \frac{1436 \sqrt{2}}{729} & -\frac{956 \sqrt{2}}{729} & -\frac{1156 \sqrt{2}}{729} \\
 \frac{4768 \sqrt{\frac{5}{3}}}{81} & \frac{3872}{81 \sqrt{3}} & \frac{1244 \sqrt{\frac{5}{3}}}{81} & \frac{916 \sqrt{\frac{5}{3}}}{81} & \frac{1244}{81 \sqrt{3}} &
   \frac{916}{81 \sqrt{3}} & -16 & 0 & -\frac{5792 \sqrt{\frac{2}{3}}}{81} & -\frac{2848 \sqrt{\frac{2}{3}}}{81} & -\frac{5504 \sqrt{\frac{2}{3}}}{81} & -\frac{3136
   \sqrt{\frac{2}{3}}}{81} \\
 \frac{4768 \sqrt{\frac{5}{3}}}{81} & \frac{3872}{81 \sqrt{3}} & \frac{1244 \sqrt{\frac{5}{3}}}{81} & \frac{916 \sqrt{\frac{5}{3}}}{81} & \frac{1244}{81 \sqrt{3}} &
   \frac{916}{81 \sqrt{3}} & 0 & 16 & -\frac{5792 \sqrt{\frac{2}{3}}}{81} & -\frac{2848 \sqrt{\frac{2}{3}}}{81} & -\frac{5504 \sqrt{\frac{2}{3}}}{81} & -\frac{3136
   \sqrt{\frac{2}{3}}}{81} \\
 -\frac{752 \sqrt{10}}{81} & -\frac{112 \sqrt{2}}{243} & -\frac{736 \sqrt{10}}{729} & -\frac{1040 \sqrt{10}}{729} & -\frac{736 \sqrt{2}}{729} & -\frac{1040 \sqrt{2}}{729}
   & \frac{128 \sqrt{\frac{2}{3}}}{81} & \frac{128 \sqrt{\frac{2}{3}}}{81} & \frac{21119}{729} & \frac{7993}{729} & \frac{6944}{729} & \frac{7264}{729} \\
 \frac{2048 \sqrt{10}}{243} & -\frac{832 \sqrt{2}}{81} & -\frac{532 \sqrt{10}}{729} & \frac{196 \sqrt{10}}{729} & -\frac{532 \sqrt{2}}{729} & \frac{196 \sqrt{2}}{729} &
   \frac{608 \sqrt{\frac{2}{3}}}{81} & \frac{608 \sqrt{\frac{2}{3}}}{81} & \frac{9470}{729} & \frac{10066}{729} & \frac{5312}{729} & -\frac{2624}{729} \\
 -\frac{112 \sqrt{10}}{27} & -\frac{1936 \sqrt{2}}{243} & -\frac{1708 \sqrt{10}}{729} & -\frac{500 \sqrt{10}}{729} & -\frac{1708 \sqrt{2}}{729} & -\frac{500
   \sqrt{2}}{729} & -\frac{16 \sqrt{\frac{2}{3}}}{81} & -\frac{16 \sqrt{\frac{2}{3}}}{81} & \frac{9536}{729} & \frac{8128}{729} & \frac{24629}{729} & -\frac{3725}{729} \\
 -\frac{448 \sqrt{10}}{243} & \frac{128 \sqrt{2}}{27} & \frac{1412 \sqrt{10}}{729} & -\frac{884 \sqrt{10}}{729} & \frac{1412 \sqrt{2}}{729} & -\frac{884 \sqrt{2}}{729} &
   \frac{896 \sqrt{\frac{2}{3}}}{81} & \frac{896 \sqrt{\frac{2}{3}}}{81} & \frac{1856}{729} & -\frac{6080}{729} & -\frac{51766}{729} & -\frac{5594}{729} \\
\end{array}
\right)$}
\end{equation}
\begin{equation}
\scalebox{0.65}{$\hat{\gamma}^{(2)}_g = \left(
\begin{array}{cccccccccccc}
 \frac{29306}{243} & -\frac{229 \sqrt{5}}{243} & -\frac{12805}{1944} & -\frac{8255}{1944} & -\frac{2561 \sqrt{5}}{1944} & -\frac{1651 \sqrt{5}}{1944} & 0 & 0 & \frac{3217
   \sqrt{\frac{5}{2}}}{486} & \frac{563 \sqrt{\frac{5}{2}}}{486} & -\frac{2561 \sqrt{\frac{5}{2}}}{486} & -\frac{1651 \sqrt{\frac{5}{2}}}{486} \\
 -\frac{242 \sqrt{5}}{243} & \frac{17684}{243} & -\frac{1273 \sqrt{5}}{972} & -\frac{347 \sqrt{5}}{972} & -\frac{1273}{972} & -\frac{347}{972} & 0 & 0 & \frac{725}{243
   \sqrt{2}} & \frac{1327}{243 \sqrt{2}} & -\frac{1273}{243 \sqrt{2}} & -\frac{347}{243 \sqrt{2}} \\
 -\frac{100}{243} & -\frac{4084 \sqrt{5}}{243} & -\frac{4571}{243} & \frac{8405}{243} & \frac{1957 \sqrt{5}}{486} & -\frac{337 \sqrt{5}}{486} & 0 & 0 & -\frac{851
   \sqrt{10}}{243} & -\frac{1201 \sqrt{10}}{243} & \frac{1957 \sqrt{10}}{243} & -\frac{337 \sqrt{10}}{243} \\
 -\frac{35440}{243} & \frac{3632 \sqrt{5}}{243} & \frac{45565}{243} & \frac{14159}{243} & \frac{68 \sqrt{5}}{243} & \frac{1228 \sqrt{5}}{243} & 0 & 0 & -\frac{2024
   \sqrt{10}}{243} & \frac{296 \sqrt{10}}{243} & \frac{136 \sqrt{10}}{243} & \frac{2456 \sqrt{10}}{243} \\
 -\frac{20 \sqrt{5}}{243} & -\frac{4084}{243} & \frac{1957 \sqrt{5}}{486} & -\frac{337 \sqrt{5}}{486} & -\frac{8485}{243} & \frac{9079}{243} & 0 & 0 & -\frac{851
   \sqrt{2}}{243} & -\frac{1201 \sqrt{2}}{243} & \frac{1957 \sqrt{2}}{243} & -\frac{337 \sqrt{2}}{243} \\
 -\frac{7088 \sqrt{5}}{243} & \frac{3632}{243} & \frac{68 \sqrt{5}}{243} & \frac{1228 \sqrt{5}}{243} & \frac{45293}{243} & \frac{9247}{243} & 0 & 0 & -\frac{2024
   \sqrt{2}}{243} & \frac{296 \sqrt{2}}{243} & \frac{136 \sqrt{2}}{243} & \frac{2456 \sqrt{2}}{243} \\
 0 & 0 & 0 & 0 & 0 & 0 & \frac{368}{9} & 0 & 0 & 0 & 0 & 0 \\
 0 & 0 & 0 & 0 & 0 & 0 & 0 & \frac{368}{9} & 0 & 0 & 0 & 0 \\
 \frac{6460 \sqrt{10}}{243} & -\frac{2788 \sqrt{2}}{243} & -\frac{1121 \sqrt{\frac{5}{2}}}{243} & -\frac{1147 \sqrt{\frac{5}{2}}}{243} & -\frac{1121}{243 \sqrt{2}} &
   -\frac{1147}{243 \sqrt{2}} & 0 & 0 & \frac{22400}{243} & \frac{6922}{243} & -\frac{2242}{243} & -\frac{2294}{243} \\
 -\frac{4928 \sqrt{10}}{243} & \frac{9248 \sqrt{2}}{243} & -\frac{796 \sqrt{10}}{243} & \frac{148 \sqrt{10}}{243} & -\frac{796 \sqrt{2}}{243} & \frac{148 \sqrt{2}}{243} &
   0 & 0 & -\frac{3517}{243} & \frac{9349}{243} & -\frac{3184}{243} & \frac{592}{243} \\
 -\frac{20 \sqrt{10}}{243} & -\frac{4084 \sqrt{2}}{243} & \frac{1957 \sqrt{\frac{5}{2}}}{243} & -\frac{337 \sqrt{\frac{5}{2}}}{243} & \frac{1957}{243 \sqrt{2}} &
   -\frac{337}{243 \sqrt{2}} & 0 & 0 & -\frac{1702}{243} & -\frac{2402}{243} & -\frac{11099}{486} & \frac{17147}{486} \\
 -\frac{7088 \sqrt{10}}{243} & \frac{3632 \sqrt{2}}{243} & \frac{68 \sqrt{10}}{243} & \frac{1228 \sqrt{10}}{243} & \frac{68 \sqrt{2}}{243} & \frac{1228 \sqrt{2}}{243} & 0
   & 0 & -\frac{4048}{243} & \frac{592}{243} & \frac{45497}{243} & \frac{12931}{243} \\
\end{array}
\right)$}
\end{equation}

\acknowledgments

We are grateful to Peter Stoffer for helpful comments on the manuscript, and to Matthew Kirk for discussions on Ref.~\cite{Jager:2019bgk}. SR and DS are supported by the STFC under grant ST/X000605/1. SR is supported by UKRI Stephen Hawking Fellowship EP/W005433/1. BS is supported by a Lord Kelvin Adam Smith scholarship from the University of Glasgow.

\bibliographystyle{JHEP}
\bibliography{biblio.bib}
\end{document}